\documentclass[showpacs,preprintnumbers,amsmath,amssymb,nofootinbib]{revtex4}

\usepackage{graphicx}% Include figure files
\usepackage{dcolumn}% Align table columns on decimal point
\usepackage{bm}% bold math

\def\simge{\mathrel{%
       \rlap{\raise 0.511ex \hbox{$>$}}{\lower 0.511ex \hbox{$\sim$}}}}
\def\simle{\mathrel{
       \rlap{\raise 0.511ex \hbox{$<$}}{\lower 0.511ex \hbox{$\sim$}}}}
\newcommand{\Vsi}{V^{\bf 1}}
\newcommand{\Vad}{V^{\bf 8}}
\newcommand{\Vsy}{V^{\bf 6}}
\newcommand{\Van}{V^{{\bf 3}^*}}
\newcommand{\Vm} {V^R}
\newcommand{\Osi}{{\it \Omega}^{\bf 1}}
\newcommand{\Oad}{{\it \Omega}^{\bf 8}}
\newcommand{\Osy}{{\it \Omega}^{\bf 6}}
\newcommand{\Oan}{{\it \Omega}^{{\bf 3}^*}}
\newcommand{\Om} {{\it \Omega}^R}
\newcommand{\tr}{{\rm tr}}
\begin{document}

%%%%%%%%%%%%%%%%%%%%%%%%%%%%%%%%%%%%%%%
%\baselineskip 16pt plus 1pt minus 1pt%
%%%%%%%%%%%%%%%%%%%%%%%%%%%%%%%%%%%%%%%

\preprint{\sf TKYNT-09-18, UTHEP-589, \ \ 2009/September}

\title{Equation of State and Heavy-Quark Free Energy 
at Finite Temperature and Density 
in Two Flavor Lattice QCD with Wilson Quark Action}

\author{S.~Ejiri$^1$, Y.~Maezawa$^2$, N.~Ukita$^3$, S.~Aoki$^{4,5}$, 
T.~Hatsuda$^6$, N.~Ishii$^6$, K.~Kanaya$^4$ and T.~Umeda$^7$ \\
(WHOT-QCD Collaboration)}

\affiliation{
$^1$Physics Department,
Brookhaven National Laboratory, Upton, New York 11973, USA \\
$^2$En'yo Laboratory, Nishina Accelerator Research Center, RIKEN, 
Wako 351-0198, Japan \\
$^3$Center for Computational Sciences,
University of Tsukuba, Tsukuba, Ibaraki 305-8577, Japan \\
$^4$Graduate School of Pure and Applied Sciences, 
University of Tsukuba, Tsukuba, Ibaraki 305-8571, Japan \\
$^5$RIKEN BNL Research Center,
Brookhaven National Laboratory, Upton, New York 11973, USA \\
$^6$Department of Physics, The University of Tokyo, 
Tokyo 113-0033, Japan \\
$^7$Graduate School of Education, Hiroshima University, 
Hiroshima 739-8524, Japan
}

%\date{\today}
\date{June 30, 2010}

\begin{abstract}
We study the equation of state at finite temperature and density
in two-flavor QCD with the RG-improved gluon action
and the clover-improved Wilson quark action on a $ 16^3 \times 4$ lattice.
Along the lines of constant physics at $m_{\rm PS}/m_{\rm V} = 0.65$ and
0.80,
we compute 
the second and forth derivatives of the grand canonical partition
function with respect to the quark chemical potential $\mu_q = (\mu_u+\mu_d)/2$ and the isospin chemical potential $\mu_I = (\mu_u-\mu_d)/2$ at vanishing chemical potentials, and study the behaviors of thermodynamic quantities at
finite $\mu_q$ using these derivatives for the case $\mu_I=0$.
In particular, we study density fluctuations at nonezero temperature and
density
by calculating the quark number and isospin susceptibilities and their
derivatives with respect to $\mu_q$.
To suppress statistical fluctuations, we also examine new techniques applicable at low densities.
We find a large enhancement in the fluctuation of quark number when the
density increased near the pseudo-critical temperature, suggesting a
critical point at finite $\mu_q$ 
terminating the first order transition line between hadronic and quark
gluon plasma phases.
This result agrees with the previous
results using staggered-type quark actions qualitatively.
Furthermore, we study heavy-quark free energies and Debye screening masses at
finite density by measuring the first and second derivatives
of these quantities for various color channels of heavy quark-quark and
quark-anti-quark pairs.
The results suggest that, to the leading order of $\mu_q$, the interaction between two quarks
becomes stronger at finite densities, while that between quark and anti-quark becomes weaker.
\end{abstract}


\maketitle

\section{Introduction}
\label{sec:intro}

Heavy-ion collision experiments are taking place at BNL
aiming at the experimental studies of a new state of matter,
the quark-gluon plasma \cite{YHM}.
In order to extract unambiguous signals for the QCD phase transition
from the heavy-ion collision experiments, quantitative calculations
directly from the first principles of QCD are indispensable.
At present, the lattice QCD simulation is the only systematic method to
do so.
Various computational techniques have been developed to study the nature
of quark matter at finite temperature $(T)$ and at small chemical
potentials $\mu_u$ and $\mu_d$ \cite{DeTarlat08,Ejilat08}.
From intensive studies for the isosymmetric case $\mu_u = \mu_d = \mu_q$,
it turned out that accurate zero-temperature simulations are important
to set the scale
% etc.
to achieve high %quantitative
precision results at finite $T$ and $\mu_q$.

Most of the lattice QCD studies at finite $\mu_q$ so far have been
performed using
staggered-type quark actions with the fourth-root trick for the quark
determinant.
However, 
the fourth-root trick makes the theory non-local and thus the universality arguments fragile.
%Here,
It should be kept in mind that the staggered-type quarks for two-favor QCD
does not show the scaling properties at finite $T$ expected from
the three-dimensional O(4) spin model \cite{Pisarski,Rajagopal}.
This may suggest large lattice artifacts to
the results of staggered-type quarks near the transition point.
Moreover, problems in the staggered quark formulation at finite density are pointed out in \cite{Golterman06}.
Since the theoretical base for the fourth-root trick is not clear,
it is indispensable to carry out simulations adopting different
lattice quark actions to control and estimate systematic errors
due to the lattice discretization.

Several years ago, the CP-PACS Collaboration has studied
finite-temperature QCD
using the clover-improved Wilson quark action coupled with the
RG-improved Iwasaki
action for gluons \cite{cp1,cp2}.
With two flavors of dynamical quarks, the phase structure, the
transition temperature and the equation of
state have been investigated.
In contrast to the case of the staggered-type quarks, both the standard
Wilson quark
action \cite{Iwasaki}
and the clover-improved Wilson quark action \cite{cp1} reproduce the
expected universality
around the critical point of the chiral phase transition:
the subtracted chiral condensate shows the scaling behavior with the
critical exponents and scaling function of the three-dimensional O(4)
spin model.
Moreover, extensive calculations of major physical quantities such as
the light hadron masses
have been carried out at $T=0$ using the same action \cite{cp3,cp4}.
Therefore, it is worth revisiting this action armed with recent
techniques for finite $\mu_q$.

In the $(T, \mu_q)$ plane, phenomenological studies suggest
the existence of a critical point at which the first order phase
transition line separating the hadronic phase and the quark-gluon-plasma
phase terminates \cite{AY,Bard,SRS}.
Because the critical point has second order characteristics,
the fluctuation of the net quark number will diverge as we approach to the
critical point
in the $(T,\mu_q)$ plane, while the fluctuation in the isospin number
will remain finite \cite{Kunihiro,HS}. Such
hadronic fluctuations may be experimentally examined in heavy-ion
collisions by an event-by-event analysis.
The Bielefeld-Swansea Collaboration reported lattice results for the
quark number susceptibility (the second
derivative of the thermodynamic grand canonical potential
$\omega/T^4 = -(VT^3)^{-1} \ln {\cal Z}$, which is proportional to the
pressure of the system)
by the Taylor expansion method using a p4-improved staggered quark
action \cite{BS03,BS05,isen06}:
From a calculation of the Taylor expansion coefficients of $\omega/T^4$ up
to $O[(\mu_q/T)^6]$,
they found that the quark number fluctuation increases rapidly
as $\mu_q$ increases in the region near the transition temperature.
This suggests indirectly the existence of the nearby critical point in
the $(T,\mu_q)$ plane.
Moreover, 2+1 flavor simulations in staggered quarks with almost
physical quark masses have recently been performed and the same
behaviors in the fluctuations have been found at finite density
\cite{milc07,rbcb08}. 
Therefore, it is important to confirm the result using the Wilson-type quarks.

In this paper, we study thermodynamic properties of QCD at finite
temperature and density
with two flavors of clover-improved Wilson quarks coupled with the
RG-improved Iwasaki gluons.
The simulations are performed along lines of constant physics
corresponding to the pion and rho meson mass ratio, $m_{\rm PS}/m_{\rm
V} = 0.65$ and 0.80 at $T=0$.
We calculate the Taylor coefficients for the pressure in terms of
$\mu_q/T$ up to the fourth order, and study the quark number and isospin
susceptibilities at finite $\mu_q$.
Since the odd derivatives vanish at $\mu_q=0$, the fourth derivative is
the leading contribution to the $\mu_q$-dependence of susceptibilities.
We find that Wilson-type quarks require much more statistics than
staggered-type quarks to
obtain the susceptibilities with a comparable quality.
To overcome this problem, we introduce a couple of tricks in the
evaluation of the Taylor expansion coefficients.
Furthermore, we adopt a hybrid method of Taylor expansion and spectral
reweighting 
in which $\omega/T^4$ for the reweighting is approximated by a
truncated Taylor expansion \cite{BS02,BS05}.
%We estimate the quark determinant, which is needed in the reweighting method,
%by a Taylor series up to $O(\mu_q^4)$. This is considerably cheaper than a
%calculation of the full determinant.
%Moreover,
Since the applicable range of the reweighting method is narrow
due to the sign problem, we introduce the Gaussian method 
% to eliminate the sign problem
proposed in \cite{eji07}.
Using these techniques, we compute the quark number density and
the susceptibility in a relatively wide range of $\mu_q/T$,
and compare the results with those with staggered-type quarks.

We also extend our previous study of heavy-quark free energies
in various color channels at $\mu_q=0$ \cite{Maezawa:2007fc} 
to finite $\mu_q$.
% and discuss how the properties found at $\mu_{q} =0$ is modified 
% at finite $\mu_{q}$.
At $T > T_{pc}$, where $T_{pc}$ is the pseudo-critical temperature, we
calculate the Taylor expansion coefficients for the heavy-quark free
energies between a static quark ($Q$) and an antiquark ($\bar{Q}$) and
those between $Q$ and $Q$,
for all color channels up to the second order in $\mu_q/T$.
By comparing the expansion coefficients of the free energies,
we find that the inter-quark interaction between $Q$ and $\bar{Q}$
becomes weaker,
whereas that between $Q$ and $Q$ becomes stronger as $\mu_q$ increases.
The expansion coefficients of
the effective running coupling $\alpha_{\rm eff} (T,\mu_{q})$
and the Debye screening mass $m_D (T,\mu_{q})$ are also extracted
by fitting the numerical results with a screened Coulomb form; we find that
the heavy-quark free energies are well reproduced by
the channel dependent Casimir factor and the channel independent
$\alpha_{\rm eff} (T,\mu_{q})$ and $m_D (T,\mu_{q})$ at $T \simge 2 T_{pc}$.
Magnitude of the second order coefficient of $m_D (T,\mu_{q})$ does not
agree with
that of the leading-order calculation in the thermal perturbation theory.

In Sec.~\ref{sec:simu}, we summarize our lattice action and simulation
parameters, and determine the pseudo-critical temperature.
In Sec.~\ref{sec:eostay}, we calculate the Taylor expansion coefficients
of the thermodinamic grand canonical potenital in terms of 
the quark chemical potentials $\mu_u$ and $\mu_d$ 
and evaluate them for the isosymmetric case $\mu_u=\mu_d=\mu_q$ 
at $\mu_q=0$ up to $O(\mu_q^4)$.
In Sec.~\ref{sec:eosrew}, we adopt the hybrid method combined with the
Gaussian method, to improve the calculation.
The static quark free energies and the Debye screening mass are
discussed in Sec.~\ref{sec:hqfe}.
Conclusions and discussions are given in Sec.~\ref{sec:conc}.
We summarize properties of the pressure and the quark number
susceptibility in the free gas limit in Appendix \ref{ap:free}.
Appendix \ref{ap:gaussian} is devoted to a description of detailed
derivations of formulae for the Gaussian method.
Results of the fits of heavy-quark free energies are summarized in
Appendix \ref{ap:fit}.

%%%%%%%%%%%%%%%%%%%%%%%%%%%%%%%%%%%%%%%%%%%%%%%%%%%%
\section{Phase structure and lines of constant physics at $\mu_q=0$}
\label{sec:simu}

\subsection{Lattice action}
\label{sec:action}

First, we summarize our simulation details. 
We adopt the same lattice actions as in our previous study at $\mu_q=0$ \cite{Maezawa:2007fc}.
We use the RG-improved Iwasaki gauge action \cite{rg} and the
$N_f=2$ clover-improved Wilson quark action \cite{cl} defined by
\begin{eqnarray}
  S   &=& S_g + S_q, \\
  S_g &=& 
  -{\beta}\sum_x\left(
   c_0\sum_{\mu<\nu;\mu,\nu=1}^{4}W_{\mu\nu}^{1\times1}(x) 
   +c_1\sum_{\mu\ne\nu;\mu,\nu=1}^{4}W_{\mu\nu}^{1\times2}(x)\right), \\
  S_q &=& \sum_{f=u,d}\sum_{x,y}\bar{\psi}_x^f M_{x,y}\psi_y^f,
  \label{eq:action}
\end{eqnarray}
where $\beta=6/g^2$, $c_1=-0.331$, $c_0=1-8c_1$ and
\begin{eqnarray}
 M_{x,y} &=& \delta_{xy}
   -{K}\sum_{i=1}^3 \{(1-\gamma_{i})U_{x,i}\delta_{x+\hat{i},y}
    +(1+\gamma_{i})U_{y,i}^{\dagger}\delta_{x,y+\hat{i}}\}
       \nonumber \\ &&
   -{K} \{e^{\mu}(1-\gamma_{4})U_{x,4}\delta_{x+\hat{4},y}
    +e^{-\mu}(1+\gamma_{4})U_{y,4}^{\dagger}\delta_{x,y+\hat{4}}\}
   -\delta_{xy}{c_{SW}}{K}\sum_{\mu<\nu}\sigma_{\mu\nu}F_{\mu\nu}.
\label{eq:fermact}
\end{eqnarray}
Here $K$ is the hopping parameter, $\mu \equiv \mu_q a$ is the quark chemical 
potential in lattice unit and 
$F_{\mu\nu}$ is the lattice field strength,
$F_{\mu\nu} = (f_{\mu\nu}-f^{\dagger}_{\mu\nu})/(8i)$,
 with $f_{\mu\nu}$ the standard clover-shaped combination of gauge links.
For the clover coefficient $c_{SW}$, we adopt a mean field value using
$W^{1\times 1}$ calculated in the one-loop perturbation theory
\cite{rg}: 
$ {c_{SW}}=(W^{1\times 1})^{-3/4}=(1-0.8412\beta^{-1})^{-3/4}$.
We denote the spatial and temporal lattice size as $N_s$ and $N_t$ respectively.
At $\mu_q=0$, the phase diagram of this action in the $(\beta,K)$ plane has been
obtained by the CP-PACS Collaboration \cite{cp1,cp2}. 

%as shown in Fig.~\ref{fig:phdiag} for $N_t=4$. 
%The solid line $K_c(T=0)$
%with filled circles is the chiral limit where pseudo-scalar
%mass vanishes at zero temperature. Above the $K_c(T=0)$ line, the parity-flavor
%symmetry is spontaneously broken \cite{Aoki:1983qi}.
%%,Aoki:1986xr,Aoki:1987us}.
%% At finite temperatures, the cusp of the
%%parity-broken phase retracts from the large $\beta$ limit to a finite
%%$\beta$ \cite{Aoki:1995yf,Aoki:1996pw}. 
%The solid line $K_c(T>0)$ connecting open symbols
%represents the boundary of the parity-broken phase.
%The region below $K_c$ corresponds to the two-flavor QCD with finite quark mass.
%We perform simulations in this region after investigating the relation 
%between the simulation parameters $(\beta, K)$ and the physical parameters, 
%e.g. quark mass and lattice spacing.
%
%The dashed line $K_t$ with filled diamond represents the pseudo-critical 
%line of the finite temperature deconfinement transition determined
%from the peak of Polyakov loop susceptibility. This line separates the hot
%phase (the quark-gluon plasma phase) and the cold phase (the hadron phase).
%The crossing of the $K_t$ and the $K_c(T=0)$ lines is the chiral
%phase transition point.
%
%\begin{figure}[tbp]
%  \begin{center}
%    \begin{tabular}{c}
%    \includegraphics[width=75mm]{figureA/phdiag.eps}
%    \end{tabular}
%    \caption{Phase diagram at $\mu_q=0$ for $N_t=4$ lattices.}
%    \label{fig:phdiag}
%  \end{center}
%\end{figure}

\begin{figure}[tbp]
  \begin{center}
    \begin{tabular}{c}
    \includegraphics[width=80mm]{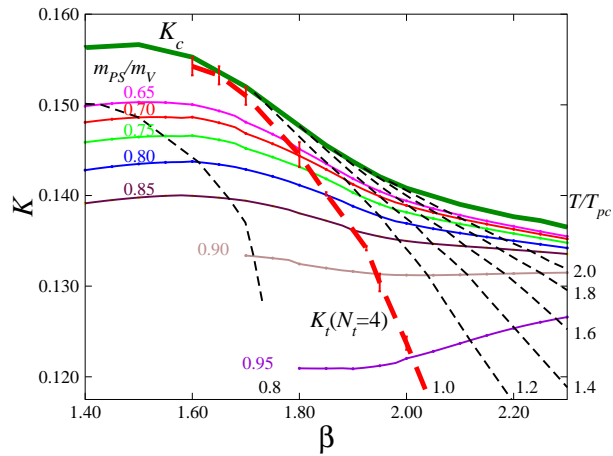}
    \end{tabular}
    \caption{
    Lines of constant physics (LCP) determined by $m_{\rm PS}/m_{\rm V}$ at $T=0$ (solid lines)
  %  in the ($\beta,K$) plane 
    for $m_{\rm PS}/m_{\rm V} = 0.65$,
     0.70, 0.75, 0.80, 0.85, 0.90 and 0.95.
    $K_c$ is the chiral limit, i.e. $m_{\rm PS}/m_{\rm V}=0$.
    Dashed lines represent lines of constant $T/T_{pc}$ on $N_t=4$ lattices, 
    where $T_{pc}$ is the pseudo-critical temperature corresponding to
    $K_t(N_t=4)$ shown by the thick dashed line. }
    \label{fig:simpara}
  \end{center}
\end{figure}

For phenomenological applications, we need to investigate the
temperature dependence of thermodynamic observables in a given physical system. 
On the lattice, ``a given physical system'' corresponds to a given set of values of dimension-less ratios of physical observables at $T=0$ and $\mu_q=0$.
Assuming the scaling, this forms a line in the coupling parameter space, called the line of constant physics (LCP), along which the lattice scale (lattice spacing $a$) is varied for a given physical system.
On a finite-temperature lattice with fixed $N_t$, the temperature, $T=1/N_t a$, is varied along a LCP according to the variation of $a$.
In this study, we determine LCP by $m_{\rm PS}/m_{\rm V}$ (the ratio of pseudo-scalar and vector meson masses at $T=0$ and $\mu_q=0$). 
The bold solid line denoted as $K_c$ in Fig.~\ref{fig:simpara} represents 
the chiral limit, i.e. $m_{\rm PS}/m_{\rm V}=0$.
Above the $K_c$ line, the parity-flavor symmetry is spontaneously broken 
\cite{Aoki:1983qi}.
The region below $K_c$ corresponds to the two-flavor QCD with finite quark mass.
We perform simulations in this region.
The lines of constant $m_{\rm PS}/m_{\rm V}$ are investigated 
in Refs.~\cite{cp2,Maezawa:2007fc}, which is shown as thin solid lines 
in Fig.~\ref{fig:simpara}, corresponding to 
$m_{\rm PS}/m_{\rm V}=0.65$, 0.70, 0.75, 0.80, 0.85, 0.90 and 0.95. 

The temperature $T$ is estimated by the zero-temperature vector meson
mass $m_{\rm V}a(\beta,K)$ using
\begin{eqnarray}
 \frac{T}{m_{\rm V}}(\beta,K)=\frac{1}{N_t \times m_{\rm V}a(\beta,K)}.
\end{eqnarray}
The lines of constant $T/T_{pc}$ is determined by the ratio of
${T}/{m_{\rm V}}$ to ${T_{pc}}/{m_{\rm V}}$ where
$T_{pc}/m_{\rm V}$ is obtained by $T/m_{\rm V}$ at $K_t$ on the same LCP.
We use an interpolation function,  
$ T_{pc}/m_{\rm V} = 
A (1+B( m_{\rm PS}/ m_{\rm V})^2)/(1+C(m_{\rm PS}/m_{\rm V})^2) $
with $A=0.2253(71)$, $B=-0.933(17)$ and $C=-0.820(39)$,
obtained in Ref.~\cite{cp2} to evaluate $T_{pc}/m_{\rm V}$ 
for each $m_{\rm PS}/m_{\rm V}$.
The bold dashed line denoted as $K_t(N_t=4)$ in Fig.~\ref{fig:simpara} 
represents the pseudo-critical line $T/T_{pc}=1$.
The thin dashed lines represent the results for 
$T/T_{pc}=0.8$, 1.2, 1.4, 1.6, 1.8, 2.0 at $N_t=4$.

We perform finite temperature simulations on a lattice with a
temporal extent $N_t=4$ and a spatial extent $N_s=16$ along the LCP's at 
$m_{\rm PS}/m_{\rm V}=0.65$ and 0.80. 
The standard hybrid Monte Carlo algorithm is employed to generate full QCD
configurations with two flavors of dynamical quarks. The length of one
trajectory is unity and the step size of 
 the molecular dynamics is tuned to
achieve an acceptance rate greater than 70\%.
Runs are carried out in the range $\beta=1.50$--2.40 at thirteen values of
$T/T_{pc}\sim 0.82$--4.0 for $m_{\rm PS}/m_{\rm V}=0.65$ and twelve values of
$T/T_{pc}\sim 0.76$--3.0 for $m_{\rm PS}/m_{\rm V}=0.80$.
Our simulation
parameters and the corresponding temperatures are summarized 
in Table \ref{tab:parameter}.
Because the determination of the pseudo critical line is 
more difficult than the calculation of $T/m_{\rm V}$, the dominant source  
for the error of $T/T_{pc}$ in Table~\ref{tab:parameter}
 is the overall factor $T_{pc}/m_{\rm V}$.
The number of trajectories for each run after thermalization is
5000--6000. We measure physical quantities at every 10 trajectories.
The study of heavy quark free energies at $\mu_q=0$ using the same configurations 
have been already published in Ref.~\cite{Maezawa:2007fc}.

\begin{table}[tbp]
 \begin{center}
 \caption{Simulation parameters for $m_{\rm PS}/m_{\rm V}=0.65$ (left) 
 and $m_{\rm PS}/m_{\rm V}=0.80$ (right) on a $16^3 \times 4$ lattice.}
 \label{tab:parameter}
 {\renewcommand{\arraystretch}{1.2} \tabcolsep = 3mm
 \newcolumntype{a}{D{.}{.}{2}}
 \newcolumntype{b}{D{.}{.}{6}}
 \newcolumntype{d}{D{.}{.}{0}}
 \begin{tabular}{|abbd|c|abbd|}
 \cline{1-4} \cline{6-9}
 \multicolumn{1}{|c}{$\beta$} &
 \multicolumn{1}{c} {$K$}    & 
 \multicolumn{1}{c} {$T/T_{pc}$} & 
 \multicolumn{1}{c|}{Traj.} & 
 \multicolumn{1}{c} {} & 
 \multicolumn{1}{|c}{$\beta$} &
 \multicolumn{1}{c} {$K$}    & 
 \multicolumn{1}{c} {$T/T_{pc}$} & 
 \multicolumn{1}{c|}{Traj.} \\
 \cline{1-4} \cline{6-9}
 1.50 & 0.150290 & 0.82(3)  & 5000 & & 1.50 & 0.143480 & 0.76(4)  & 5500 \\
 1.60 & 0.150030 & 0.86(3)  & 5000 & & 1.60 & 0.143749 & 0.80(4)  & 6000 \\
 1.70 & 0.148086 & 0.94(3)  & 5000 & & 1.70 & 0.142871 & 0.84(4)  & 6000 \\
 1.75 & 0.146763 & 1.00(4)  & 5000 & & 1.80 & 0.141139 & 0.93(5)  & 6000 \\
 1.80 & 0.145127 & 1.07(4)  & 5000 & & 1.85 & 0.140070 & 0.99(5)  & 6000 \\
 1.85 & 0.143502 & 1.18(4)  & 5000 & & 1.90 & 0.138817 & 1.08(5)  & 6000 \\
 1.90 & 0.141849 & 1.32(5)  & 5000 & & 1.95 & 0.137716 & 1.20(6)  & 6000 \\
 1.95 & 0.140472 & 1.48(5)  & 5000 & & 2.00 & 0.136931 & 1.35(7)  & 5000 \\
 2.00 & 0.139411 & 1.67(6)  & 5000 & & 2.10 & 0.135860 & 1.69(8)  & 5000 \\
 2.10 & 0.137833 & 2.09(7)  & 5000 & & 2.20 & 0.135010 & 2.07(10) & 5000 \\
 2.20 & 0.136596 & 2.59(9)  & 5000 & & 2.30 & 0.134194 & 2.51(13) & 5000 \\
 2.30 & 0.135492 & 3.22(12) & 5000 & & 2.40 & 0.133395 & 3.01(15) & 5000 \\
 2.40 & 0.134453 & 4.02(15) & 5000 & &      &          &          & \\
 \cline{1-4} \cline{6-9}
 \end{tabular}}
 \end{center}
\end{table}

\subsection{Critical temperature}
\label{sec:tc}

\begin{figure}[t]
\begin{center}
\includegraphics[width=6.5cm]{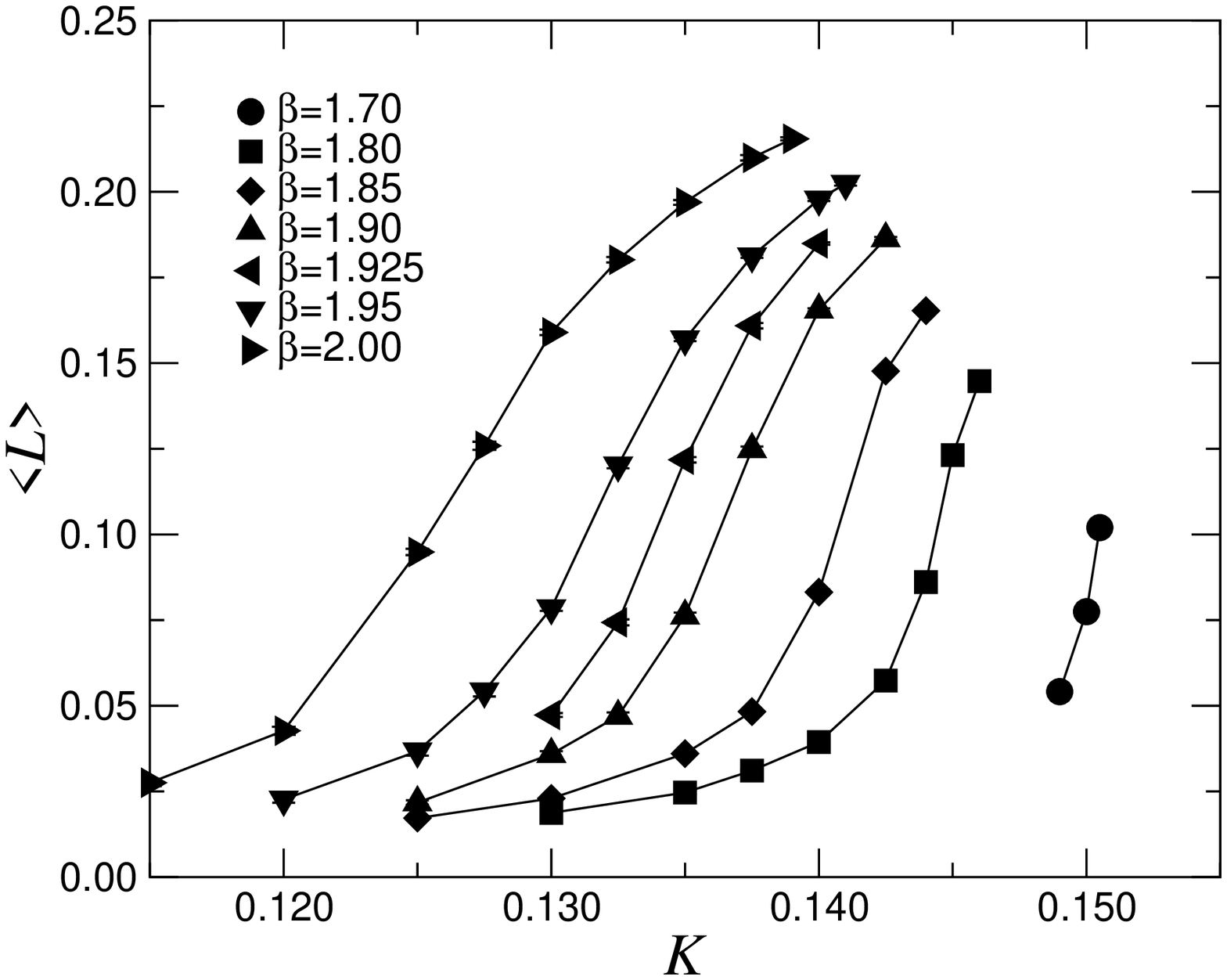}
\hskip 0.5cm
\includegraphics[width=6.5cm]{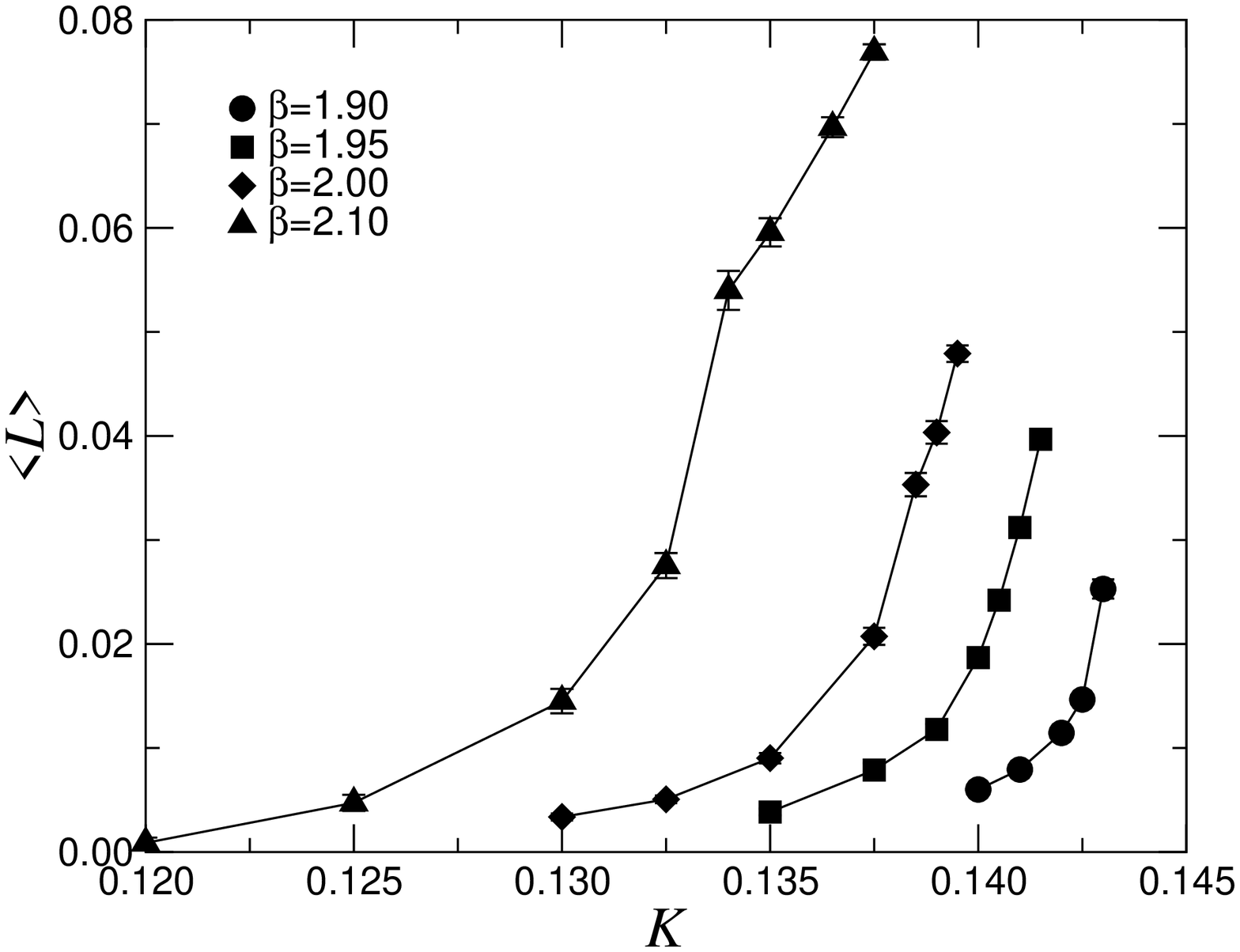}
\vskip -0.2cm
\caption{$K$-dependence of Polyakov loop 
for $N_t=4$ (left) and $6$ (right).
Data at $\beta=1.7$ and $1.8$ for $N_t=4$ and $1.9$ and $1.95$ 
for $N_t=6$ are renewed from Refs.~\cite{cp1,cp2}. 
}
\label{fig:pol}
\end{center}
%\vskip -0.3cm
%\setcounter{figure}{1}
\end{figure} 

\begin{figure}[t]
\begin{center}
\includegraphics[width=6.5cm]{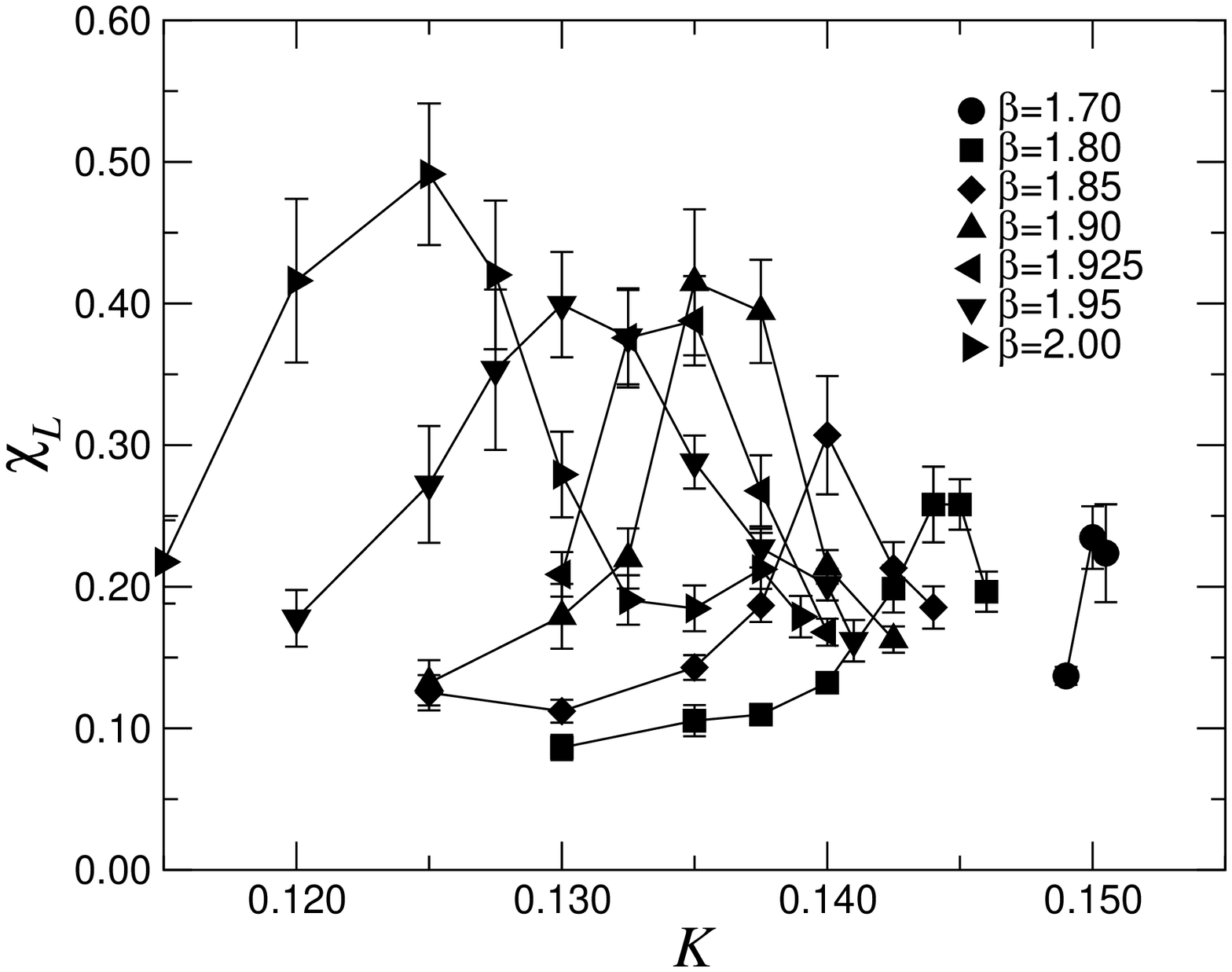}
\hskip 0.5cm
\includegraphics[width=6.5cm]{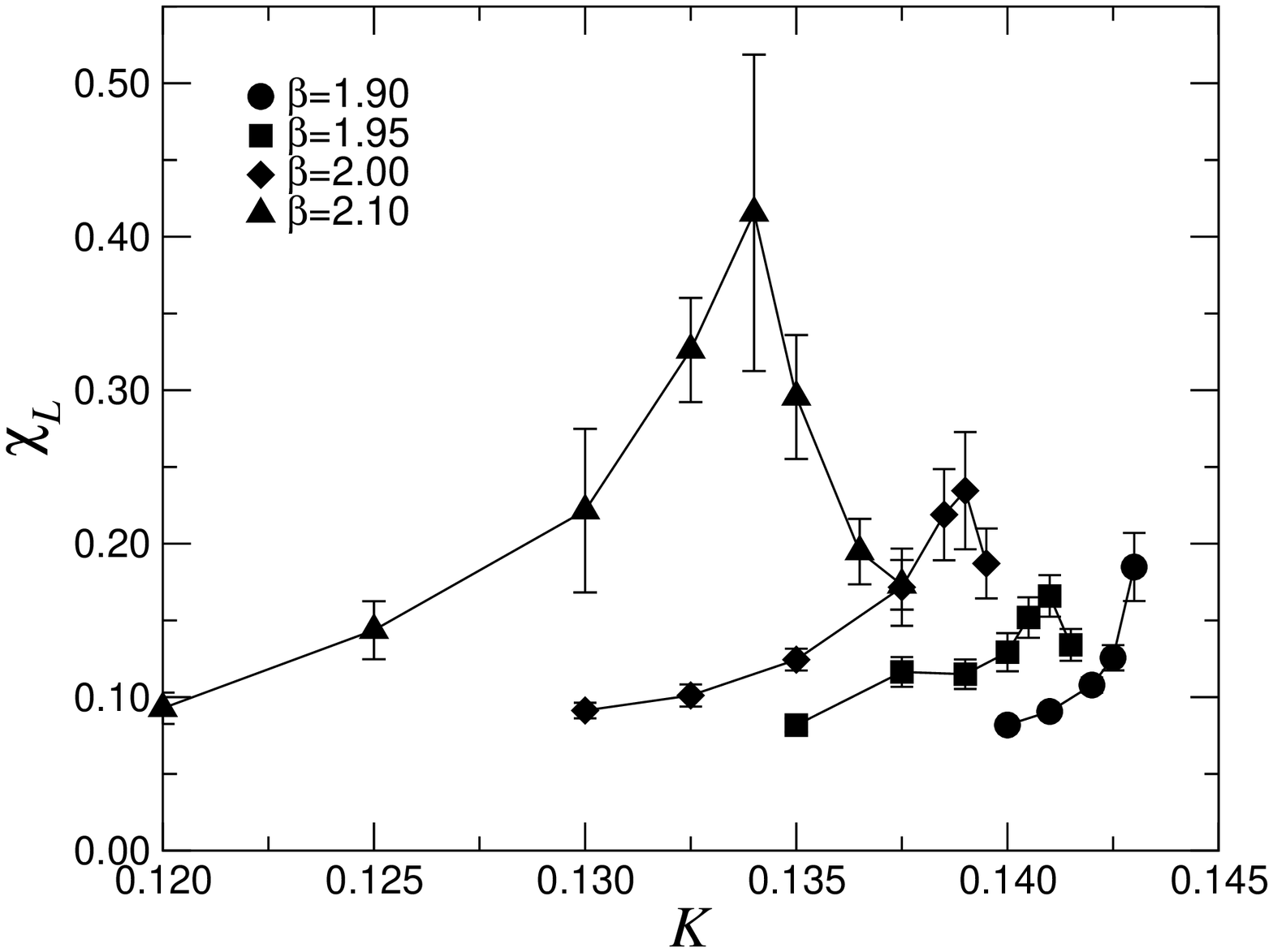}
\vskip -0.2cm
\caption{$K$-dependence of Polyakov loop susceptibility
for $N_t=4$ (left) and $6$ (right).}
\label{fig:plsus}
\end{center}
%\vskip -0.3cm
%\setcounter{figure}{1}
\end{figure} 

\begin{table}[hbt]
\begin{center}
\caption{Finite temperature transition/crossover point $K_t$
for $N_t=4$ and $6$. 
Results for 
$m_{\rm PS}(T=0)/m_{\rm V}(T=0)$, $m_{\rm PS} a (T=0)$, 
$m_q^{\rm AWI} a (T>0)$, 
$T_{pc}/m_{\rm V}(T=0)$, $T_{pc}/\sqrt{\sigma}$, $T_{pc} r_0$, 
and $m_{\rm PS} r_0$, 
are interpolated to the $K_t$ line, 
}
\label{tab:tc}
\begin{tabular}{|llllllllll|}
\hline
$\beta$ & $K_t$ & $K_c$ & $m_{\rm PS}/m_{\rm V}$ & $m_{\rm PS} a$ 
& $m_q^{\rm AWI} a$ & $T_{pc}/m_{\rm V}$ & 
$T_{pc}/\sqrt{\sigma}$ & $T_{pc} r_0$ & $m_{\rm PS} r_0$ \\ \hline
\multicolumn{10}{|c|}{$N_t=4$} \\
\hline
1.700 & 0.15014(33) & 0.151987(22) & 0.509(35)  & 0.579(51) &            & 0.2197(47) & & & \\
1.800 & 0.14425(16) & 0.147678(15) & 0.7070(79) & 0.849(18) & 0.1107(77) & 0.2083(21) & 0.4204(29) & 0.4716(42) & 1.601(37) \\
1.850 & 0.14019(18) & 0.145526(58) & 0.7905(60) & 1.031(15) & 0.1864(72) & 0.1917(20) & 0.4359(60) & 0.484(11) & 1.994(55) \\
1.900 & 0.13621(15) & 0.143737(48) & 0.8525(39) & 1.183(11) & 0.2464(49) & 0.1801(12) & 0.4382(70) & 0.484(16) & 2.290(79) \\
1.925 & 0.13417(23) &              &            &           & 0.2725(67) & & & & \\
1.950 & 0.13040(97) & 0.142072(14) & 0.9051(64) & 1.440(66) & 0.363(25)  & 0.1572(62) & & & \\
2.000 & 0.12371(73) & 0.140811(55) & 0.9450(36) & 1.689(39) & 0.500(18)  & 0.1398(29) & & & \\
2.100 & 0.10921(43) & 0.139020(21) & 0.9790(13) & 2.196(18) &            & 0.1114(9) & & & \\ 
%1.700 & 0.15014(33) & & 0.509(35)  & & 0.2197(47) & 0.3517(83) & 0.3982(170) \\
%1.950 & 0.13040(97) & & 0.9051(64) & & 0.1572(62) & 0.4009(167)& 0.4422(260) \\
\hline
\multicolumn{10}{|c|}{$N_t=6$} \\
\hline
1.950 & 0.14090(13) & 0.142072(14) & 0.591(21)  & 0.448(24) & 0.0451(51) & 0.2202(44) & 0.4336(40) & 0.4973(58) & 1.336(73) \\
2.000 & 0.13861(21) & 0.140811(55) & 0.725(16)  & 0.580(27) & 0.080(10)  & 0.2086(53) & 0.4639(77) & 0.530(13) & 1.842(98) \\
2.100 & 0.13365(40) & 0.139020(21) & 0.8635(78) & 0.821(34) & 0.175(13)  & 0.1753(58) & 0.491(12)  & 0.570(13) & 2.81(13) \\
2.200 & 0.12539(25) & 0.137658(53) & 0.9481(19) & 1.240(16) & 0.3607(67) & 0.1275(15) & & & \\
2.300 & 0.11963(15) & 0.136513(85) & 0.9724(12) & 1.454(8)  & 0.4813(39) & 0.1114(6)  & & & \\
%2.200 & 0.12539(25) & & 0.9481(19) & & 0.1275(15) & 0.4440(67) & 0.5167(129)\\
\hline
\end{tabular}
\end{center}
\end{table}

\begin{table}[hbt]
\begin{center}
\caption{The critical point $(\beta_{ct})$ and critical temperature $(T_c)$ 
in the chiral limit obtained by various fitting procedures. 
The fit range for $\beta$ is written in ``$\beta$ range". 
$T_c$ in a physical unit is estimated from the vector meson mass 
$m_{\rm V}=m_{\rho}=770 {\rm MeV}$.}
\label{tab:bcfit}
\begin{tabular}{|llllll|}
\hline
$N_t$ \ & $h \sim m_q a$ \hspace{7mm} & $\beta$ range \hspace{4mm} & 
$\beta_{ct}$ \hspace{10mm} & $T_c$ ($m_{\rm V}$-input) \ & $T_c r_0$ \\ 
\hline
4 & $1/K_t-1/K_c$      & 1.70--1.95 & 1.619(10) & 180(3) MeV & \\
4 & $1/K_t-1/K_c$      & 1.70--1.90 & 1.611(12) & 179(3) MeV & \\
4 & $(m_{\rm PS} a)^2$ & 1.70--1.95 & 1.559(16) & 172(3) MeV & \\
4 & $(m_{\rm PS} a)^2$ & 1.70--1.90 & 1.552(16) & 171(3) MeV & \\
4 & $m_q^{\rm AWI}a$   & 1.80--1.90 & 1.601(20) & 177(4) MeV & \\
4 & $m_q^{\rm AWI}a$   & 1.80--1.95 & 1.596(18) & 176(3) MeV & \\
6 & $1/K_t-1/K_c$      & 1.95--2.20 & 1.870(6)  & 184(5) MeV & 0.434(9) \\
6 & $1/K_t-1/K_c$      & 1.95--2.10 & 1.840(14) & 171(4) MeV & 0.401(16) \\
6 & $(m_{\rm PS} a)^2$ & 1.95--2.20 & 1.835(9)  & 170(4) MeV & 0.396(12) \\
6 & $(m_{\rm PS} a)^2$ & 1.95--2.10 & 1.786(25) & 160(9) MeV & 0.350(23) \\
6 & $m_q^{\rm AWI}a$   & 1.95--2.20 & 1.835(10) & 170(4) MeV & 0.396(12) \\
6 & $m_q^{\rm AWI}a$   & 1.95--2.10 & 1.810(19) & 167(4) MeV & 0.372(20) \\
\hline                                                       
\end{tabular}
\end{center}
\end{table}

%The critical temperature ($T_c$) is one of the most fundamental quantities 
%in the QCD thermodynamics. 
We update the analysis of the pseudo critical temperature done 
in Refs.~\cite{cp1,cp2}, performing additional simulations 
at $\beta=6/g^2=1.7$ and $1.8$ on an $N_s^3 \times N_t = 16 \times 4$ lattice
and at $1.9$ and $1.95$ on $N_s^3 \times N_t = 16^3 \times 6$.
The number of trajectories for each new run is 1050--4200 
after thermalization.
We add the new data to the data in Refs.~\cite{cp1,cp2} 
and determine the pseudo-critical hopping parameters $K_t$ defined from 
the peak of the Polyakov loop susceptibility on $16^3 \times 4$ and 
$16^3 \times 6$ lattices, as a function of $\beta$.
%Because the Polyakov loop is an indicator of the deconfinement, 
%this $K_t$ is the transition point for confinement properties. 
Figures \ref{fig:pol} and \ref{fig:plsus} are the results of 
the Polyakov loop $\langle L \rangle$ and Polyakov loop susceptibility 
$\chi_L$, respectively.
We find a pronounced peak in the Polyakov loop susceptibility 
except for $\beta=1.90$ at $N_t=6$.
The peak position of the susceptibility $(K_t)$ is determined by 
fitting three or four data near the peak with the Gaussian form.
%using the same way done in \cite{cp1,cp2}.
The results are summarized in Table \ref{tab:tc} together with values 
of some quantities at $K_t$ to set a physical scale. 

We use the data of the pseudo-scalar and vector meson masses at $T=0$, 
$m_{\rm PS}$ and $m_{\rm V}$, summarized in the Table IV of Ref.~\cite{cp2}, and 
interpolate them following the method discussed in  Refs.~\cite{cp1} and \cite{cp2}.
We also calculate the current quark mass defined through an axial vector 
Ward-Takahashi identity, $\nabla_\mu A_\mu = 2m_q^{\rm AWI} P + O(a),$ 
where $P$ is the pseudo-scalar density and $A_\mu$ the $\mu$-th
component of the local axial vector current \cite{Itoh,Bochicchio}.  
Because the $T$-dependence in $m_q^{\rm AWI}$ is small, 
we use the data of $m_q^{\rm AWI}$ obtained in finite temperature 
simulations at $N_t=4$ and $6$ \cite{cp1,cp2} .
%In practice we make a simultaneous fit of the two-point functions 
%$\langele A_\mu(t)P(0) \rangle$ and $\langle P(t)P(0) \rangle$ 
%to extract the pion mass $m_\pi$ 
%and the amplitudes $\langle 0|P|\pi(\vec{p}=0)\rangle$, 
%$\langle 0|A_\mu|\pi(\vec{p}=0)\rangle$, from which we compute
%\begin{eqnarray}
%m_q^{\rm AWI} =-m_\pi \frac{\langle\,0\,|\,A_\mu\,|\,\pi(\vec{p}=0)\,\rangle}
%	         {\langle\,0\,|\,P\,|\,\pi(\vec{p}=0)\,\rangle} .
%\label{eq:mq}
%\end{eqnarray}
%We choose $\mu=4$ for zero temperature simulations, and $\mu=3$ 
%at finite temperatures for which the screening masses are determined 
%along the $z$ axis.
In Table \ref{tab:tc}, $m_q^{\rm AWI}$ on the $K_t$ line are obtained using a cubic spline 
interpolation for each $\beta$. 
A straight line interpolation leads to almost identical results within statistical errors.
The values of the string tension $\sigma$ and the Sommer scale $r_0$ 
\cite{Sommer} are estimated by interpolating or extrapolating 
the data at $\beta=1.80,1.95,2.10$ and $2.20$ %given in Table V of  Ref.~
\cite{cp4} in the $(\beta, 1/K-1/K_c)$ parameter plane.

\begin{figure}[t]
\begin{center}
\includegraphics[width=7.5cm]{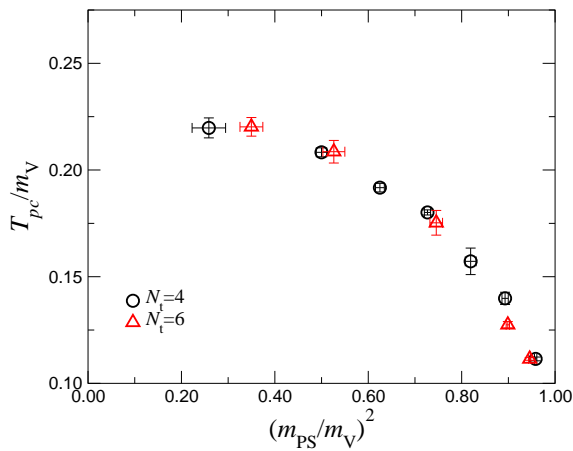}
\vskip -0.2cm
\caption{$T_{pc}/m_{\rm V}$ vs. $m_{\rm PS}/m_{\rm V}$ 
for $N_t=4$ (circle) and $6$ (triangle).
The lightest two points for $N_t=4$ and the lightest one point for $N_t=6$ 
are updated from Ref.~\cite{cp2}.
}
\label{fig:tpcmv}
\end{center}
%\vskip -0.3cm
%\setcounter{figure}{1}
\end{figure}

The results of the pseudo-critical temperature are also shown 
in Table \ref{tab:tc}.
We plot $T_{pc}/m_{\rm V}$ as a function of 
$(m_{\rm PS}/m_{\rm V})^2$ in Fig.~\ref{fig:tpcmv}, and 
find that the results of $N_t=4$ and 6 agree with each other. 
Note that $T_{pc}/m_{\rm V}$ vanishes in the heavy quark limit  
$m_{\rm PS}/m_{\rm V}=1$.
%, $T_{pc}/m_{\rm V}$ increases as 
% $m_{\rm PS}/m_{\rm V}$ decreases. 
Figure \ref{fig:tpcmv} suggests $T_{pc}/m_{\rm V} \sim 0.22$ 
($T_{pc} \sim 170 {\rm MeV}$) in the chiral limit.

We denote the critical temperature in the chiral limit as $T_c$.
As discussed in \cite{cp1,Iwasaki}, 
the subtracted chiral condensate \cite{Bochicchio}
satisfies the scaling behavior with the critical exponents and 
scaling function of the 3-dimensional O(4) spin model.
For the reduced temperature $t$ and external magnetic field $h$, we adopt
$t \sim \beta-\beta_{ct}$ and $h \sim m_q$, 
where $\beta_{ct}$ is the critical transition point in the chiral limit. 
For a precise determination of $T_c$, we need to deduce $\beta_{ct}$ from the data.
In this study, we perform critical scaling fits 
assuming that the pseudo-critical temperature $t_{pc}$ from the 
Polyakov loop susceptibility, as well as that from the chiral condensate, 
follows the scaling law 
$t_{pc} \sim h^{1/y}$ with the O(4) critical exponent $1/y \equiv 1/(\beta \delta) = 0.537(7)$. 
In practice, we fit the data of $\beta_{pc}(K)$, i.e. the inverse function of 
$K_t(\beta)$ in Table~\ref{tab:tc}, by
\begin{eqnarray}
\beta_{pc}=\beta_{ct}+Ah^{1/y}
\end{eqnarray}
with two free parameters, $\beta_{ct}$ and $A$.

\begin{figure}[t]
\begin{center}
\begin{tabular}{cc}
\includegraphics[width=55mm]{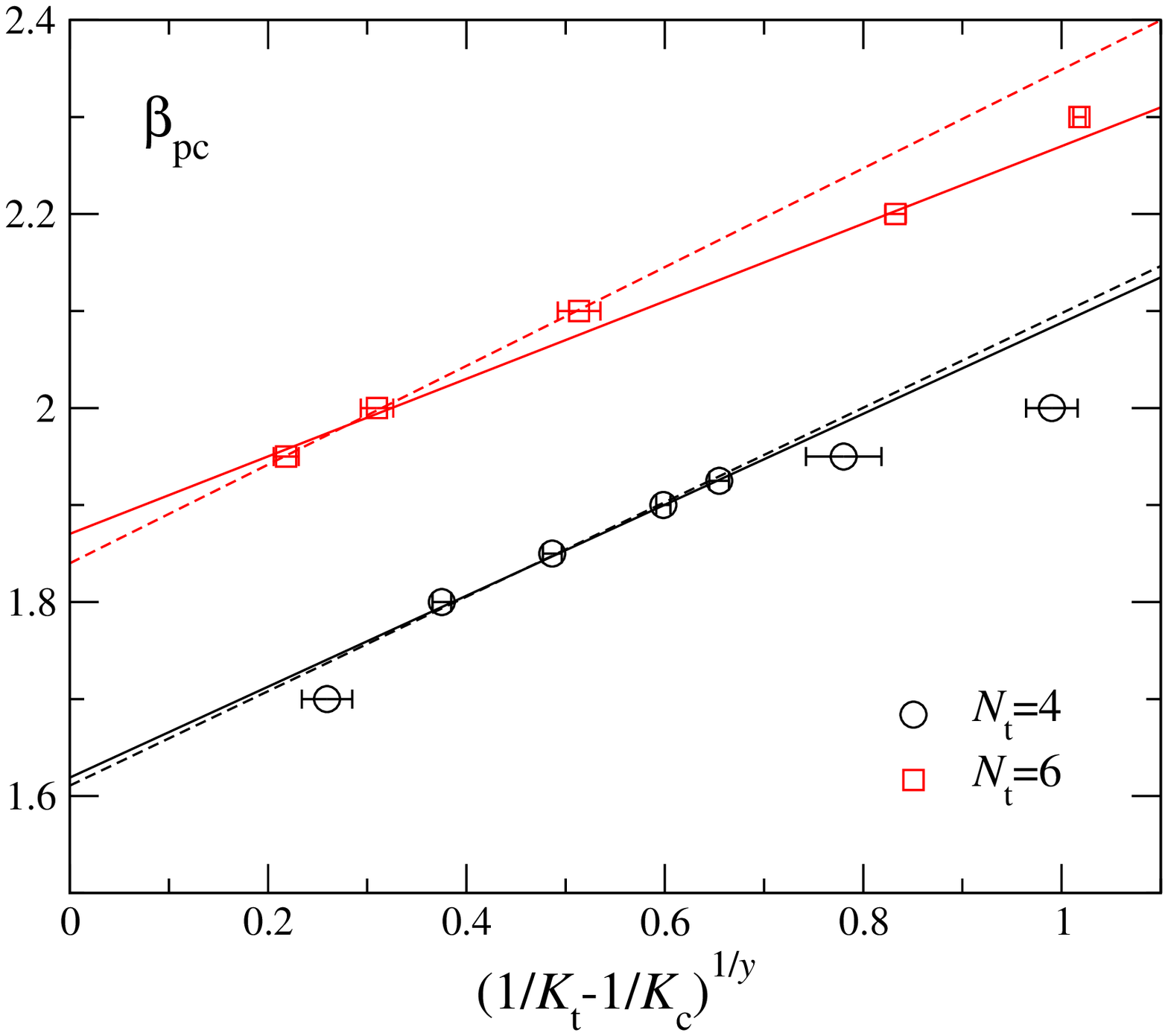} 
\includegraphics[width=55mm]{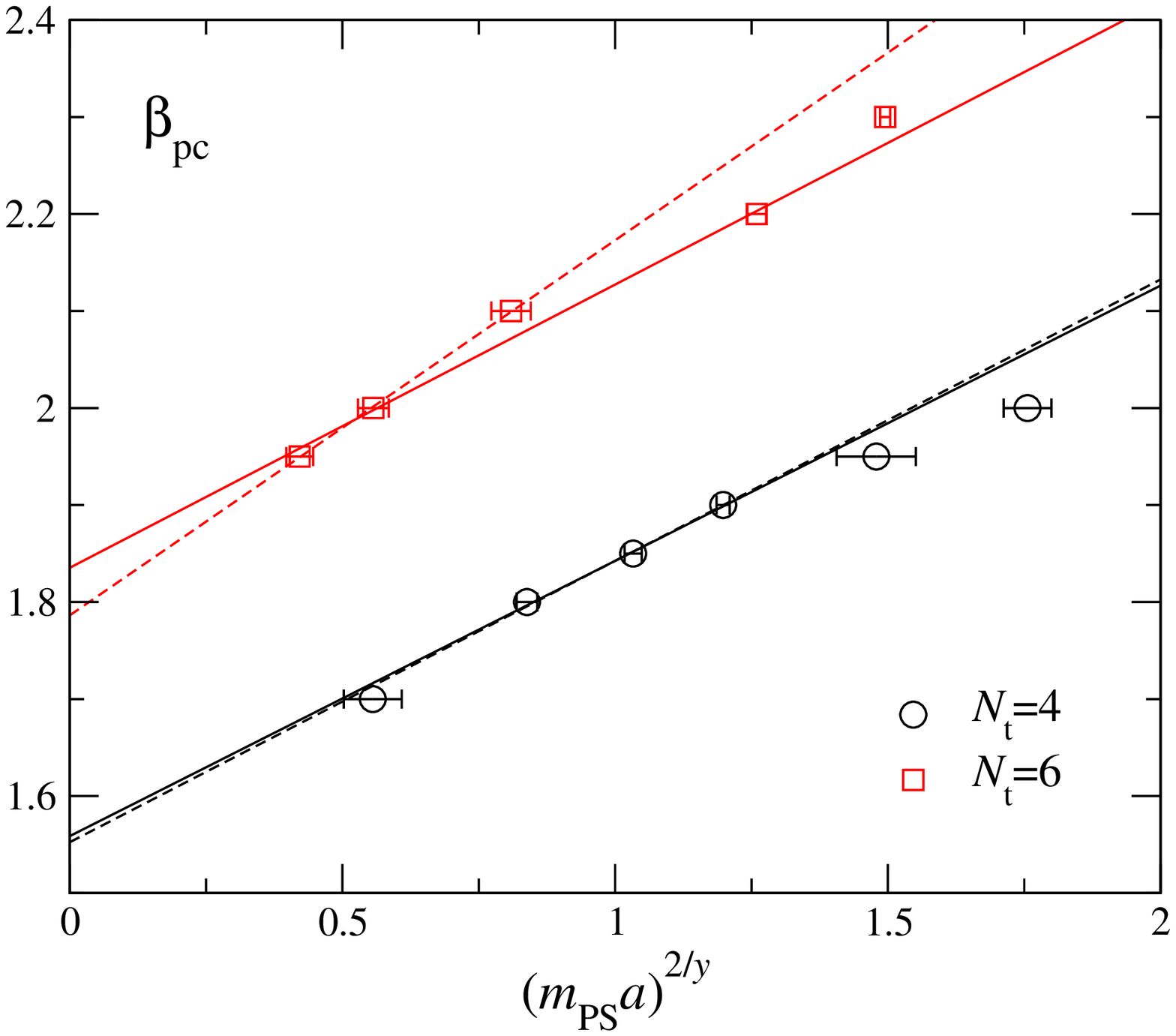} 
\includegraphics[width=55mm]{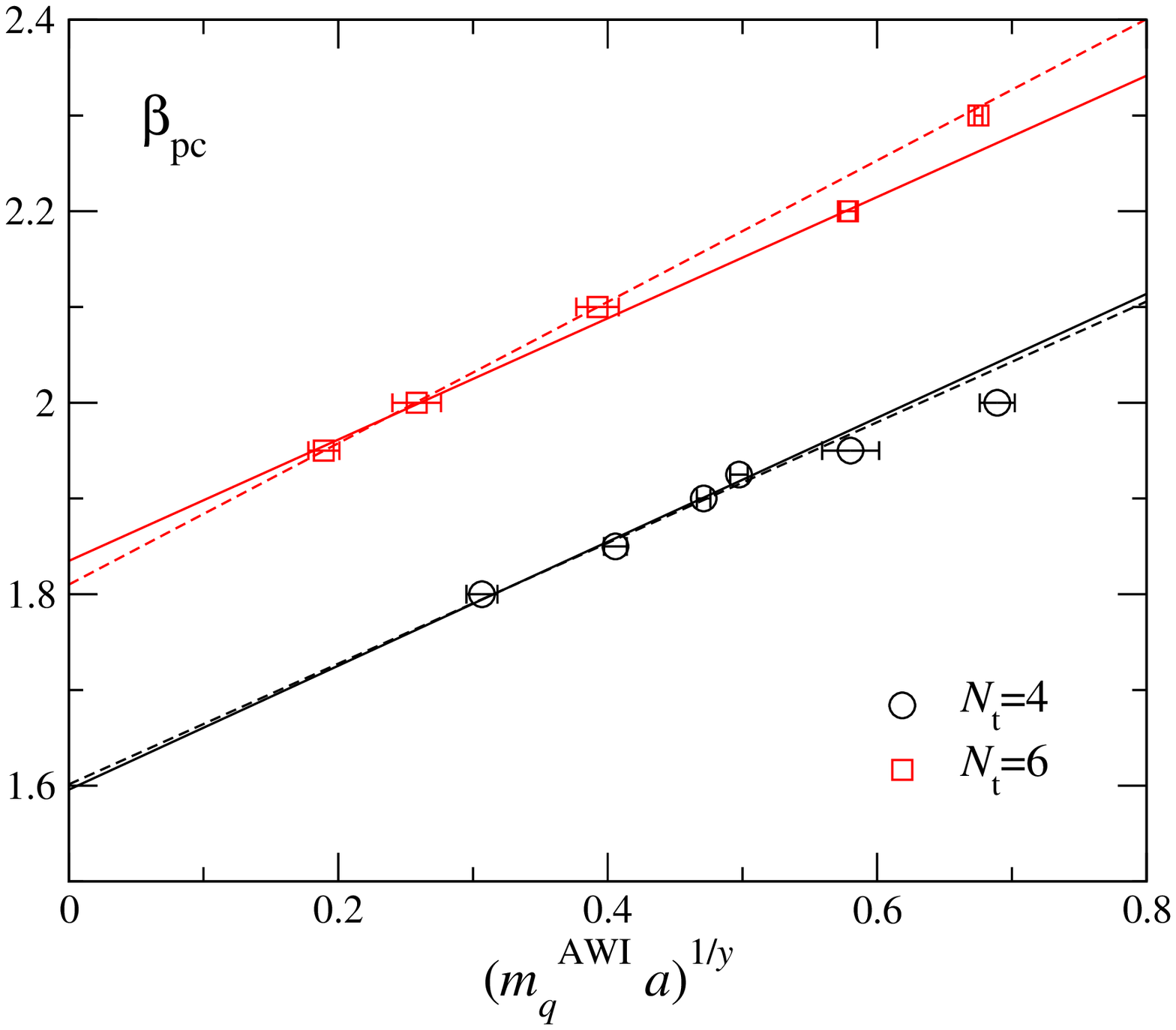} 
\end{tabular}
\vspace*{-2mm}
\caption{The pseudo-critical point $\beta_{pc}$ 
as a function of $(m_q a)^{1/y}$ with $m_q a \sim 1/K - 1/K_c$(left), 
$(m_{\rm PS}^2 a)^2$(center) and  $m_q^{\rm AWI} a$(right) 
for $N_t=4$ (circle) and $N_t=6$ (square).
We fit the data in two fit ranges. 
The solid and dashed lines are the fit results with the long 
and short fit ranges, respectively.
}
\label{fig:bpcmq}
\end{center}
%\vspace{-0.5cm}
\end{figure}

For the quark mass $m_q \sim h$ in the scaling fits, we test three variants. 
The first is $m_q a \sim 1/K - 1/K_c$, where $K_c$ is the chiral point 
at which the pion mass vanishes at $T=0$ for each $\beta$. 
The second is $m_q a \sim (m_{\rm PS} a)^2$. 
The third is $m_q^{\rm AWI} a$, i.e. the quark mass defined 
by the axial vector Ward-Takahashi identity.
We plot $\beta_{pc}$ as a function of $1/K - 1/K_c$ (left),  
$(m_{\rm PS} a)^2$ (center) and $m_q^{\rm AWI} a$ (right) 
in Fig.~\ref{fig:bpcmq}.
The results of $\beta_{ct}$ and $T_c$ are summarized in Table \ref{tab:bcfit}, 
where $T_c$ in MeV is calculated by $T_c=1/[N_t a(\beta_{ct})]$
with $a$ from the vector meson mass 
$m_{\rm V}(T=0)=m_{\rho}=770$ MeV at $\beta_{ct}$ on $K_c$. 
We test two fit ranges of $\beta$ for each extrapolation, which is denoted 
in Table \ref{tab:bcfit} as ``$\beta$ range".
We note that these O(4) fits reproduce the data of $\beta_{pc}$ 
much better than a naive linear fit $\beta_{pc}=\beta_{ct}+Ah$. 
A tentative conclusion is that the critical temperature in the chiral limit 
is in the range 171--180 MeV for $N_t=4$ and 160--184 MeV for $N_t=6$. 
There is still a large uncertainty from the choice of the fit ansatz 
and the fit range. 
To remove this, further simulations at lighter quark masses are necessary.

For a comparison with other groups, we estimate 
$T_c$ in units of the Sommer scale 
$r_0$ \cite{Sommer} at $\beta_{ct}$ in the chiral limit for $N_t=6$. 
Using the data of $r_0/a$ in the chiral limit at $\beta=1.80$, 1.95 and 
2.10 \cite{cp4}, we interpolate $a/r_0$ by 
a quadratic function and calculate $T_c r_0 = (N_t a/r_0)^{-1}$. 
The estimates are about $T_c r_0 \approx 0.40$, as listed 
in Table \ref{tab:bcfit}.
These values are close to $T_c r_0=0.402(29)$ obtained by the MILC Collaboration 
using the asqtad quark action in 2+1 flavor QCD
%in the chiral limit 
\cite{MILC05} 
\footnote{Originally, $T_c$ is given in units of $r_1$ in 
Ref.~\cite{MILC05}. 
The scale of $T_c$ has been converted to $r_0$ using 
$r_0/r_1=1.4795$ \cite{Cheng06}.}%
.
On the other hand, the RBC-Bielefeld Collaboration obtained $T_c r_0=0.444(6)^{+12}_{-3}$ 
using a 2+1 flavor p4fat3 improved staggered quark action \cite{Cheng06}.
From a simulation of 2 flavor QCD using a clover 
improved Wilson action and the standard one-plaquette gauge action, the DIK Collaboration obtained 
$T_c r_0 = 0.438(6)^{+13}_{-7}$ at the physical pion mass point, and the value in the chiral limit 
is $2 \%$ smaller than this value \cite{Bornya07}.
Our result is somewhat smaller than these values.
Finally, the Budapest-Wuppertal group used a stout-link improved staggered fermion action and 
fixed the scale by the pion decay constant $f_{\pi}$.
They found that $T_c$ determined by the chiral susceptibility 
is $T_c=151(3)(3)$ MeV and that by the renormalized Polyakov 
loop is $T_c = 176(3)(4)$ MeV in the continuum limit 
at the physical point \cite{YAoki06}. 
Our result is close to their result defined by the Polyakov loop.
For further discussions, see Refs. \cite{Karlat07,Fodlat07,HatsuQM06}.

% A care is in order when 
% we convert  $T_c r_0$ ($T_{pc} r_0$) to $T_c$ ($T_{pc}$)
% in MeV using a physical value of $r_0$. 
% Because the phenomenological estimate of $r_0$ has large 
% theoretical uncertainties, 
% it looks convenient to adopt a lattice result.
% Unfortunately, lattice results for $r_0$ suffer from sizable ambiguities yet.
% The RBC-Bielefeld Collaboration adopted the value $r_0 = 0.469(7)$ fm, 
% which was obtained by Gray {\it et al.} \cite{Gray:2005ur} 
% from a bottomonium mass \
% splitting using the AsqTad-improved staggered quark action.
% On the other hand, 
% the CP-PACS+JLQCD Collaboration found $r_0=0.516(21)$ fm from a study 
% of light hadron spectrum 
% using the clover-improved Wilson quark action \cite{Ishikawa:2006ws}. 
% This leads to about 10 \%
% difference in the value of $T_c$ and makes it difficult 
% to naively compare results from different groups.

%Next, we compare these results with those of a staggered quark action.
%We plot the results of the pseudo-critical temperature ($T_{pc}$) 
%in unit of Sommer scale $(r_0)$ as a 
%function of $m_{\rm PS} r_0$ in Fig.~\ref{fig:tcr0} together with 
%those by the RBC-Bielefeld Collaboration using 2+1 flavor p4-improved staggered 
%quark action \cite{Che06}. 
%As seen in this figure, results of $T_{pc}$ obtained by different quark 
%actions seem to approach the same function of $m_{\rm PS} r_0$ 
%as $N_t$ increases.

%\begin{figure}[t]
%\begin{center}
%\begin{tabular}{cc}
%\includegraphics[width=2.4in]{figureA/tcr0vsmpir0_QM06.eps}
%\end{tabular}
%\vspace*{-2mm}
%\caption{Comparison of $T_{pc}$ scaled by $r_0$ between 
%the staggered quark action (open symbol) \cite{Che06} and 
%the Wilson quark action (filled symbol) for $N_t=4$ and 6.
%}
%\label{fig:tcr0}
%\end{center}
%%\vspace{-0.5cm}
%\end{figure}

%%%%%%%%%%%%%%%%%%%%%%%%%%%%%%%%%%%%%%%%%%%%%%%%%%%%
\section{Equation of state at finite densities by the Taylor expansion method}
\label{sec:eostay}

The main difficulty in a study of QCD at finite density is that 
the Boltzmann weight is complex for nonzero $\mu_q$. 
The quark matrix at zero density have the $\gamma_5$ Hermiticity $M^{\dagger} = \gamma_5 M \gamma_5$
which guarantees that the quark determinant is real. 
However, at $\mu_q \ne 0$, we have only 
\begin{eqnarray}
M^{\dagger}(\mu_q) = \gamma_5 M(-\mu_q) \gamma_5, 
\label{eq:gamma5conj}
\end{eqnarray}
from Eq.~(\ref{eq:fermact}). 
Therefore, the quark determinant is complex for $\mu_q \ne 0$. 

Because configurations cannot be generated with a complex probability, 
the conventional Monte Carlo method is not applicable at $\mu_q \ne 0$. 
At present, there are three methods to study finite density QCD, all of which are applicable for small $\mu_q$ regions.
The simplest is the method based on a Taylor expansion 
in terms of $\mu_q/T$ around $\mu_q=0$ \cite{BS02,BS03,Miya02,GG}. 
%One performs a Taylor expansion of physical quantities around $\mu_q=0$ 
%and computes expansion coefficients, i.e. derivatives with respect 
%to $\mu_q/T$, at $\mu_q=0$. 
Because the simulations at $\mu_q=0$ is free from the complex weight problem, 
the expansion coefficients, i.e. derivatives of physical quantities with respect to $\mu_q/T$, can be evaluated by a conventional Monte Carlo simulation.
%After measuring the convergence region of the Taylor expansion, 
%the physical quantities can be calculated in the application range of $\mu_q$, 
%using the expansion coefficients up to an appropriate order of $\mu_q$.
The second approach is the reweighting method \cite{FK,Gibbs86,Hase92,Bar97}. 
Performing simulations at $\mu_q=0$, expectation values at finite $\mu_q$ are computed adopting a corrected Boltzmann weight.
For the correction, quark determinant at finite $\mu_q$ is estimated numerically. 
%This method is applicable for the calculation in the low density region on a relatively small lattice. 
Because fluctuations in the complex phase of the determinant are large at large $\mu_q$ and/or large lattice volume, 
a reliable calculation of expectation value becomes gradually difficult off the small $\mu_q$ and small lattice volume region due to the sign problem \cite{Ejiri04,Spli06}. 
%In the reweighting method, the expectation values of complex operators 
%must be computed. 
%If the complex phase of the operator changes its sign frequently, 
%a serious problem arises, which is called ``sign problem''. 
The third approach is the analytic continuation from simulations with 
imaginary chemical potentials \cite{dFP,dEL}. 
Since the equation (\ref{eq:gamma5conj}) is generalized to 
$M^{\dagger}(\mu_q) = \gamma_5 M(-\mu_q^*) \gamma_5$ for complex $\mu_q$, 
the Boltzmann weight is real and simulations are possible 
when the chemical potential is purely imaginary.
Using results by the imaginary chemical potential simulations, 
information at a real chemical potential can be obtained by an analytic continuation.
The analytic continuation is usually based on a Taylor expansion in 
terms of $\mu_q$ around $\mu_q=0$ for the study in the low density region, 
and improvements of the analytic continuation have been also discussed 
in \cite{Cea07,Sakai09,Yoneyama09} to obtain reliable results 
in a wide range of real $\mu_q$.
%High precision simulations for wide range of the imaginary chemical 
%potential is required for a reliable analytic continuation.

In this section, we adopt the Taylor expansion method to study the effects of $\mu_q$ in the equation of state. Most of thermodynamic quantities, such as energy density, quark number, order parameters and various susceptibilities, are given by derivatives of the thermodynamic grand canonical potential $\omega/T^4 \equiv -(\ln {\cal Z}) / (VT^3)$. 
Also, pressure, which is given by $\omega$ itself, is evaluated by integrating a derivative of $\omega$ in current studies of the equation of state. 
Therefore, the calculations of the derivative of $\omega$ is basic for the study of QCD thermodynamics by lattice simulations, and the Taylor expansion method calculating higher order derivatives in $\mu_q$ is the most natural extension from the study at $\mu_q=0$ to finite $\mu_q$.

%%%%%%%%%%%%%%%%%%%%%%%%%%%%%%%%%%%%%%%%%%%%%%%%%%%%
\subsection{Taylor expansion of the grand canonical potential}
\label{sec:taygcp}

%\begin{figure}[t]
%\begin{center}
%\includegraphics[width=3.0in]{figureB/prs12fit3t41123i.eps}
%\vskip -0.2cm
%\caption{Pressure at $\mu_q=0$ on $16^3 \times 4$ lattice 
%as a function of $T/T_{pc}$ for various $m_{\rm PS}/m_{\rm V}$ obtained by 
%CP-PACS \cite{cp2}.
%The dashed curve shows the pressure for pure gauge theory with 
%the RG-improved action \cite{CPPACS99}.}
%\label{fig:prs0mu164}
%\end{center}
%\vskip -0.3cm
%%\setcounter{figure}{1}
%\end{figure} 

We study pressure $p$ and quark number densities $n_u$ and $n_d$ defined by derivatives of the partition function ${\cal Z}(T,\mu_u,\mu_d)$:
\begin{eqnarray}
\frac{p}{T^4} = \frac{1}{VT^3} \ln {\cal Z} \equiv -\frac{\omega}{T^4},
\hspace{8mm}
\frac{n_f}{T^3} = 
\frac{1}{VT^3} \frac{\partial \ln {\cal Z}}{\partial (\mu_f/T)} 
= \frac{\partial (p/T^4)}{\partial (\mu_f/T)}, 
\hspace{5mm}
(f=u,\,d)
\label{eq:qnd}
\end{eqnarray}
where $\mu_u$ and $\mu_d$ are the chemical potentials for the u and d quarks.
Let us define the quark chemical potential $\mu_q = (\mu_u + \mu_d)/2$ and the isospin chemical potential $\mu_I = (\mu_u - \mu_d)/2$.
Taylor expansion coefficients of physical quantities are given by derivatives of them in terms of $\mu_u$ and $\mu_d$, or equivalently $\mu_q$ and $\mu_I$.
We evaluate these coefficients at $\mu_u=\mu_d=0$ and study the physical quantities as functions of $T$ and $\mu_q$ in the isosymmetric case $\mu_u=\mu_d = \mu_q$ (i.e.\ $\mu_I=0$).

We define the susceptibility of quark number by
\begin{eqnarray}
\frac{\chi_q}{T^2} 
= \left( \frac{\partial}{\partial (\mu_u/T)} 
+ \frac{\partial}{\partial (\mu_d/T)} \right) 
\frac{n_u + n_d}{T^3}, 
\end{eqnarray}
and the susceptibility of isospin number by
\begin{eqnarray}
\frac{\chi_I}{T^2} 
= \left( \frac{\partial}{\partial (\mu_u/T)} 
- \frac{\partial}{\partial (\mu_d/T)} \right) 
\frac{n_u - n_d}{T^3}. 
\end{eqnarray}
These susceptibilities correspond to the fluctuations of baryon number and isospin number in the medium, respectively \cite{Gottlieb}.
They are expected to behave quite differently near the critical point in the $(T, \mu_q)$ plane. 
%Finally, the chiral condensate is defined by a derivative of $\ln {\cal Z}$ with respect to the quark mass.

We define the Taylor expansion coefficients of the pressure 
$p(T, \mu_q)$ for the case $\mu_u=\mu_d = \mu_q$ as 
\begin{equation}
\frac{p}{T^4} =
\sum_{n=0}^\infty c_n(T) \left(\frac{\mu_q}{T}\right)^n,
%\label{eq:p}
\hspace{8mm}
c_n (T)= 
%\frac{1}{n!} \left. -\frac{\partial^n (\omega/T^4)}{\partial(\mu_q/T)^n}
%\right|_{\mu_q=0}=
\frac{1}{n!} \frac{N_t^{3}}{N_s^3} \left.
\frac{\partial^n \ln{\cal Z}}{\partial(\mu_q/T)^n} \right|_{\mu_q=0}.
\label{eq:cn}
\end{equation}
Here, $c_0(T)$ is the pressure at $\mu_q=0$ and has been 
computed by the CP-PACS Collaboration with the same action 
on $16^3 \times 4$ and $16^3 \times 6$ lattices \cite{cp1,cp2}. 
Its value in the quenched limit is given in \cite{CPPACS99}.
%Figure \ref{fig:prs0mu164} is the results of $p/T^4$ as a function of 
%$T/T_{pc}$ for various $m_{\rm PS}/m_{\rm V}$ 
%on the $16^3 \times 4$ lattice \cite{cp2}, 
%which corresponds to $c_0(T)$ in our new calculation for finite density. 

We also expand the quark number and isospin susceptibilities for 
the case $\mu_u=\mu_d=\mu_q$:
\begin{eqnarray}
\frac{\chi_q(T,\mu_q)}{T^2}
=2c_2+12c_4\left(\frac{\mu_q}{T}\right)^2+\cdots ,
\hspace{5mm}
\frac{\chi_I(T,\mu_q)}{T^2}
=2c^I_2+12c^I_4\left(\frac{\mu_q}{T}\right)^2+\cdots ,
\end{eqnarray}
where
\begin{equation}
c^I_n= \left. \frac{1}{n!}\frac{N_t^{3}}{N_s^3}
\frac{\partial^n \ln {\cal Z}(T,\mu_q+\mu_I,\mu_q-\mu_I)}
{\partial(\mu_I/T)^2 \partial(\mu_q/T)^{n-2}}
\right|_{\mu_q=0,\mu_I=0}, 
%\hspace{5mm}
%\mu_I=\frac{\mu_u - \mu_d}{2}.
\end{equation}

\subsubsection{Free quark-gluon gas at high temperature}
We expect QCD in the high temperature limit is described as free gas 
of quark and gluon. 
The pressure of the free gas in the continuum theory is given by
\begin{eqnarray}
\label{eq:psb}
\frac{p}{T^4} = \frac{8 \pi^2}{45} + \sum_{f=u,d} \left[ \frac{7\pi^2}{60}
+ \frac{1}{2} \left( \frac{\mu_f}{T} \right)^2
+ \frac{1}{4 \pi^2} \left( \frac{\mu_f}{T} \right)^4 \right]. 
\end{eqnarray}
Note that the $\mu_q$-dependence appears only through terms of $\mu_q^2$ and $\mu_q^4$.
The quark number density is a cubic function of $\mu_q$ too. 
The quark number and isospin susceptibilities are the same for the free quark-gluon gas 
and are given by a quadratic function 
\begin{eqnarray}
\frac{\chi_q}{T^2}=\frac{\chi_I}{T^2}=
N_{\rm f} \left[ 1+ \frac{3}{\pi^2} \left( \frac{\mu_q}{T} \right)^2 \right]. 
\end{eqnarray}
Therefore, the Taylor expansion will converge well in the high temperature region. 

\subsubsection{Hadron resonace gas at low temperature}
On the other hand, QCD at low temperature may be modeled by free gas of 
hadron resonances \cite{KRT}. 
The partition function of the hadron resonance gas consists of mesonic 
and baryonic contributions,   
\begin{eqnarray}
\ln{\cal Z} (T,V,\mu_q) 
=\sum_{i\in\;{\rm mesons}}\hspace{-3mm} \ln{\cal Z}^{M}_{m_i}(T,V,\mu_q)
+\hspace{-3mm} 
\sum_{i\in\;{\rm baryons}}\hspace{-3mm} \ln{\cal Z}^{B}_{m_i}(T,V,\mu_q)\; ,
\label{eq:ZHRG}
\end{eqnarray}
where 
\begin{equation}
\ln{\cal Z}^{M/B}_{m_i}(T,V,\mu_q)
=\mp \frac{V}{2\pi^2} \int_0^\infty dk k^2
\ln(1\mp z_ie^{-\varepsilon_i/T}) \quad ,
\label{eq:ZMB}
\end{equation}
with energies $\varepsilon_i^2=k^2+m_i^2$ and fugacities 
\begin{equation}
z_i=\exp\left((3B_i\mu_q)/T\right) \quad .
\label{eq:fuga}
\end{equation}
Here $B_i$ is the baryon number:
$B_i=1, -1$ and $0$ for baryons, anti-baryons and mesons, respectively.
The upper sign in Eq.~(\ref{eq:ZMB}) is for bosons, while the lower sign for fermions. 
Note that ${\cal Z}^{M}_{m_i}$ is actually independent of $\mu_q$. 
Expanding the logarithms in powers 
of fugacity, the integration over momenta, $k$, can be carried out:
\begin{equation}
\ln{\cal Z}^{M/B}_{m_i}=\frac{VTm_i^2}{2\pi^2} \sum_{l=1}^\infty
\left\{
\begin{array}{c}
1 \\ (-1)^{l+1}
\end{array}
\right\} l^{-2} K_2\left(\frac{l m_i}{
T}\right) z_i^{l} \quad , 
\label{eq:Zmi}
\end{equation}
where $K_2$ is a modified Bessel function. 
For $m_i \gg T$, the Bessel function can be approximated by 
$K_2(x)\sim\sqrt{\pi/2x}\; {\rm e}^{-x} (1+15/8x +{\cal O}(x^{-2}))$. 
Terms with $\ell\geq2$ in the series given in Eq.~(\ref{eq:Zmi}) thus are 
exponentially suppressed. 

Let us study the $\mu_q$-dependence of the partition function. 
The mesonic sector has no $\mu_q$-dependence because $B_i=0$ for mesons. On the other hand, the baryonic sector can be approximated by the leading term in the expansion of $z_i$, since all baryons are heavier than a typical temperature scale. 
We obtain 
\begin{equation}
\frac{p(T,\mu_q)}{T^4} - \frac{p(T,0)}{T^4} 
= \frac{1}{VT^3} \left[ \ln{\cal Z}(T,\mu_q) - \ln{\cal Z}(T,0) \right] 
\simeq F(T) \left[ \cosh \left( \frac{3\mu_q}{T} \right) -1 \right] ,
\label{eq:pHRG}
\end{equation}
with 
\begin{eqnarray}
F(T) = \frac{1}{\pi^2} \sum_{i\in\;{\rm baryons}}\left( \frac{m_i}{T} \right)^2
K_2 \left( \frac{m_i}{T} \right) .
\label{eq:Phip}
\end{eqnarray}
Note that each term in the sum for $F$ now counts both baryons and
anti-baryons. 
The quark number susceptibility is then given by 
\begin{equation}
\frac{\chi_q}{T^2} = 9F(T) \cosh \left( \frac{3\mu_q}{T} \right) .
\label{eq:chiHRG}
\end{equation}
From Eq.~(\ref{eq:pHRG}), the ratios of the expansion coefficients of $p/T^4$ in $\mu_q/T$ are derived, 
\begin{equation}
\frac{c_{2n+2}}{c_{2n}} = \frac{9}{(2n+2)(2n+1)}. 
\end{equation}
The ratio decreases as the order becomes higher. 
This means that the contribution from the higher order terms of $\mu_q/T$ is small in the region of $\mu_q/T \simle O(1)$.

\subsubsection{Numerical study near the transition temperature}
The behavior near the transition temperature is non-trivial. 
We expect a critical point at finite $\mu_q$. The Taylor expansion must break down at that point. 
We perform numerical simulations to study the expansion coefficients near the transition point.
Using $\mu \equiv \mu_q a$,
the explicit forms of the Taylor expansion coefficients are
\begin{eqnarray}
&& \hspace{-13mm}
c_2 = \frac{N_t}{2N_s^3} {\cal A}_2 ,
\hspace{3mm}
c_4 = \frac{1}{4! N_s^3 N_t} ({\cal A}_4 -3 {\cal A}_2^2) , 
\hspace{3mm}
c_2^I = \frac{N_t}{2N_s^3} {\cal B}_2 ,
\hspace{3mm}
c_4^I = \frac{1}{4! N_s^3 N_t} ({\cal B}_4 - {\cal B}_2 {\cal A}_2) , 
\label{eq:cncnI}
\end{eqnarray}
\begin{eqnarray}
{\cal A}_2 &=&
\left\langle {\cal D}_2 \right\rangle 
+\left\langle {\cal D}_1^2 \right\rangle, 
\hspace{5mm}
{\cal A}_4 =
\left\langle {\cal D}_4 \right\rangle 
+4\left\langle {\cal D}_3 {\cal D}_1 \right\rangle 
+3\left\langle {\cal D}_2^2 \right\rangle 
+6\left\langle {\cal D}_2 {\cal D}_1^2 \right\rangle 
+\left\langle {\cal D}_1^4 \right\rangle , \nonumber \\
{\cal B}_2 &=&
\left\langle {\cal D}_2 \right\rangle ,
\hspace{5mm}
{\cal B}_4 =
\left\langle {\cal D}_4 \right\rangle 
+2 \left\langle {\cal D}_3 {\cal D}_1 \right\rangle 
+ \left\langle {\cal D}_2^2 \right\rangle 
+ \left\langle {\cal D}_2 {\cal D}_1^2 \right\rangle ,
\label{eq:AB}
\end{eqnarray}
with
\begin{eqnarray}
{\cal D}_n = N_f \frac{\partial^n \ln \det M}{\partial \mu^n}, 
\label{eq:calDn}
\end{eqnarray}
i.e.,
\begin{eqnarray}
{\cal D}_1
%N_f \frac{\partial \ln \det M}{\partial \mu} 
&=& N_f {\rm tr} \left( M^{-1} \frac{\partial M}{\partial \mu} \right) ,
\hspace{10mm}
{\cal D}_2
%N_f \frac{\partial^2 \ln \det M}{\partial \mu^2} 
\ = \ N_f \left[ {\rm tr} \left( M^{-1} \frac{\partial^2 M}{\partial \mu^2} \right)
 - {\rm tr} \left( M^{-1} \frac{\partial M}{\partial \mu}
                   M^{-1} \frac{\partial M}{\partial \mu} \right) \right] , 
\nonumber \\
{\cal D}_3
%N_f \frac{\partial^3 \ln \det M}{\partial \mu^3} 
&=& N_f \left[ {\rm tr} \left( M^{-1} \frac{\partial^3 M}{\partial \mu^3} \right)
 -3 {\rm tr} \left( M^{-1} \frac{\partial M}{\partial \mu}
              M^{-1} \frac{\partial^2 M}{\partial \mu^2} \right)
 +2 {\rm tr} \left( M^{-1} \frac{\partial M}{\partial \mu}
        M^{-1} \frac{\partial M}{\partial \mu}
        M^{-1} \frac{\partial M}{\partial \mu} \right) \right] ,
\nonumber \\
{\cal D}_4
% N_f\frac{\partial^4 \ln \det M}{\partial \mu^4} 
&=& N_f \left[{\rm tr} \left( M^{-1} \frac{\partial^4 M}{\partial \mu^4} \right)
 -4 {\rm tr} \left( M^{-1} \frac{\partial M}{\partial \mu}
              M^{-1} \frac{\partial^3 M}{\partial \mu^3} \right)
 -3 {\rm tr} \left( M^{-1} \frac{\partial^2 M}{\partial \mu^2}
        M^{-1} \frac{\partial^2 M}{\partial \mu^2} \right) \right.
 \nonumber \\ && \left.
 +12 {\rm tr} \left( M^{-1} \frac{\partial M}{\partial \mu}
        M^{-1} \frac{\partial M}{\partial \mu}
        M^{-1} \frac{\partial^2 M}{\partial \mu^2} \right) 
 -6 {\rm tr} \left( M^{-1} \frac{\partial M}{\partial \mu}
        M^{-1} \frac{\partial M}{\partial \mu}
        M^{-1} \frac{\partial M}{\partial \mu}
        M^{-1} \frac{\partial M}{\partial \mu} \right) \right] .
\label{eq:dermu}
\end{eqnarray}
The derivative of the fermion matrix $M$ at $\mu=0$ is 
\begin{eqnarray}
\left( \frac{\partial^{n} M}{\partial \mu^{n}} \right)_{x,y}
= \left\{ 
\begin{array}{l}
-K \left(
 (1-\gamma_4) U_4(x)~\delta_{x+\hat{4},y} 
 -(1+\gamma_4) {U_4}^\dagger(x-\hat{4})~\delta_{x-\hat{4},y} \right)
\ \ {\rm for} \ n: {\rm odd.}
 \\
-K \left(
 (1-\gamma_4) U_4(x)~\delta_{x+\hat{4},y} 
 +(1+\gamma_4) {U_4}^\dagger(x-\hat{4})~\delta_{x-\hat{4},y} \right) 
\ \ {\rm for} \ n: {\rm even.}
\end{array}
\right. 
\end{eqnarray}

%%%%%%%%%%%%%%%%%%%%%%%%%%%%%%%%%%%%%%%%%%%%%%%%%%%%
\subsection{Random noise method}
\label{sec:rnm}

\begin{figure}[t]
\begin{center}
\includegraphics[width=2.4in]{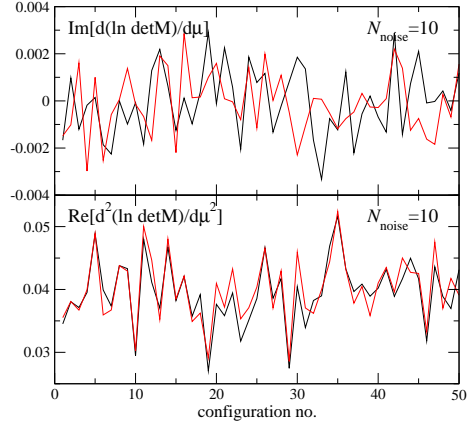}
\vskip -0.2cm
\caption{Time history of ${\cal D}_1 \times (N_f N_s^3 N_t)^{-1}$ (top) and 
${\cal D}_2 \times (N_f N_s^3 N_t)^{-1}$ 
(bottom) obtained by different noise sets at 
$T/T_{pc}=0.925, m_{\rm PS}/m_{\rm V}=0.8$.}
\label{fig:timehis}
\end{center}
\vskip -0.3cm
\end{figure} 

\begin{figure}[t]
\begin{center}
\includegraphics[width=2.4in]{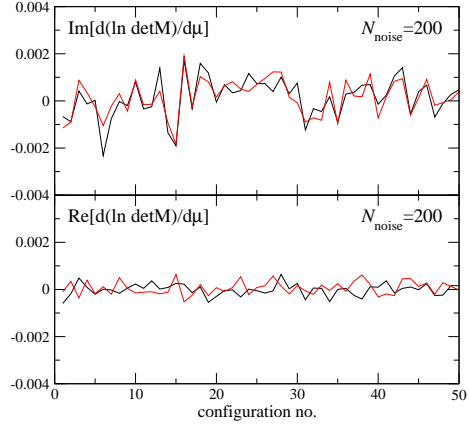}
\vskip -0.2cm
\caption{Time history of the imaginary part (top) and real part (bottom) 
of ${\cal D}_1 \times (N_f N_s^3 N_t)^{-1}$ obtained by different noise sets at 
$T/T_{pc}=0.925, m_{\rm PS}/m_{\rm V}=0.8$.}
\label{fig:timehis2}
\end{center}
\vskip -0.3cm
\end{figure} 

We apply a random noise method to evaluate the traces in Eq.~(\ref{eq:dermu}).
As we will see later, this method is effective when off-diagonal elements of the matrix are small. 
Therefore, the method works well for traces over spatial indices:
Because the inverse of the quark matrix $M^{-1}(x,y)$ decreases as 
a function of $|x-y|$, the off-diagonal elements in the spatial coordinate will be smaller than the diagonal ones. 
The random noise method will work well to suppress these small contaminations of off-diagonal elements. 
On the other hand, the off-diagonal elements in the color and spinor indices at the same spatial point are not suppressed by $|x-y|$, and will have the same magnitude as the diagonal elements. 
Because a staggered-type quark does not have the spinor index at 
a spatial point, the number of off-diagonal elements is only 6 in 
the $3 \times 3$ matrix, the contamination of off-diagonal elements 
may be not so serious. 
However, for Wilson-type quarks, because the number of color-spinor index is 
$3 \times 4$, 
the number of the off-diagonal elements in the quark matrix is 11 times 
larger than the diagonal one,
so that the color-spinor index should be treated more carefully with Wilson-type quarks. 
In this study, we apply the random noise method for the spatial coordinates only, repeating the calculation for each of the color and spinor indices. 

We generate noise vectors 
$\left( \eta_{i, \alpha} \right)_{x, \beta} \equiv 
\eta(i,x) \, \delta_{\alpha, \beta}$, which satisfy
\begin{eqnarray}
\frac{1}{N_{\rm noise}} \sum_{i=1}^{N_{\rm noise}} \eta(i,x) \eta^*(i,y) \approx \delta_{x,y}
\label{eq:noisedelta}
\end{eqnarray}
for large $N_{\rm noise}$. 
We adopt U(1) random numbers as $\eta$, 
which are complex random numbers with $|\eta|=1$ and are generated from 
uniform random numbers $\theta \in [0, 2 \pi)$ with $\eta=e^{i \theta}$. 
For each color-spinor index ($\alpha = 1, \cdots, 12$), we generate $N_{\rm noise}$ noise vectors
($i=1 \sim N_{\rm noise}$). 
Then
$\lim_{N_{\rm noise} \to \infty} (1/N_{\rm noise}) \sum_{i=1}^{N_{\rm noise}} \sum_{\alpha=1}^{12} 
\left( \eta_{i, \alpha} \right)_{x, \beta}
\left( \eta_{i, \alpha}^* \right)_{y, \gamma}
=\delta_{x,y} \delta_{\beta, \gamma}$, hence
\begin{eqnarray}
{\rm tr} \left( \frac{\partial^{n_1} M}{\partial \mu^{n_1}} M^{-1}
\frac{\partial^{n_2} M}{\partial \mu^{n_2}} \cdots M^{-1} \right)
& \approx & \frac{1}{N_{\rm noise}} \sum_{i=1}^{N_{\rm noise}} \sum_{\alpha=1}^{12}
\eta_{i, \alpha}^{\dagger} 
\frac{\partial^{n_1} M}{\partial \mu^{n_1}} X_{i, \alpha}, 
\hspace{1cm} (n=1, 2, \cdots),
\label{eq:noise}
\end{eqnarray}
where 
$X_{i, \alpha}= M^{-1} (\partial^{n_2} M / \partial \mu^{n_2}) 
\cdots M^{-1} \eta_{i,\alpha}$.
To obtain $X$, we solve equations $MX_n= Y_n$ recursively with 
$Y_1=\eta$, $Y_2=(\partial^{n} M / \partial \mu^{n}) M^{-1} \eta 
=(\partial^{n} M / \partial \mu^{n}) X_1$, etc.
%The computational time is drastically reduced if we can estimate the trace with a number of noise vectors $N_{\rm noise}$ which is much smaller than $N_{site}$. 

Because $N_{\rm noise}^{-1} \sum_i \eta (i,x) \eta^* (i,y)$ is 
$O(\sqrt{1/N_{\rm noise}}\,)$ for $x \neq y$, 
errors due to finite $N_{\rm noise}$ decrease as $O(\sqrt{1/N_{\rm noise}}\,)$. 
However, these errors are produced from all off-diagonal elements of the matrix in Eq.~(\ref{eq:noise}), hence these are proportional to the magnitude and number of the off-diagonal elements. 
Therefore, when the off-diagonal elements are not smaller than the diagonal elements, a number of noise vectors are needed to remove the error.
This is the reason why we do not use the random noise method for the color-spinor index.

For a product of traces, the random noise vectors for each trace must be independent.
We compute such product by subtracting the contribution of the same noise vector from the naive product of two noise averages for each trace.
This effectively increases the number of noises to $N_{\rm noise} (N_{\rm noise} - 1)$ for the products and thus suppresses their errors due to the noise method. 

We then average over configurations to evaluate the expectation values in Eq.~(\ref{eq:AB}). 
In addition to the errors due to the noise method, the statistical fluctuation of configurations contributes to the final error.
To check the relative amount of the errors from the noise method, we calculate 
the operators ${\cal D}_n$ ($n=1$-4) using two independent sets of 
noise vectors with $N_{\rm noise}=10$ on the same configurations.
Figure \ref{fig:timehis} shows the time history of the imaginary part 
of ${\cal D}_1$ and the real part of ${\cal D}_2$ computed 
using these two sets of noise vectors.
The operator ${\cal D}_n$ is real for even $n$ and purely imaginary 
for odd $n$ \cite{BS02}. Therefore, the average of ${\cal D}_1$ is zero 
because the expectation value is always real at $\mu_q=0$.
We find that two results of ${\cal D}_2$ obtained by different 
noise sets are consistent with each other on each configuration, 
while two results of ${\cal D}_1$ are sensibly different. 
This means that, in the evaluation of ${\cal D}_1$ with $N_{\rm noise}=10$, the error from the noise method is larger than the error from the statistical fluctuation of configurations. 
We can reduce the error from the noise method by increasing $N_{\rm noise}$. 
We plot the time history with $N_{\rm noise}=200$ in Fig.~\ref{fig:timehis2}. 
Two results of ${\rm Im} [{\cal D}_1]$ using different noise sets are almost consistent, i.e., the error in ${\cal D}_1$ is now dominated by the statistical fluctuation of configurations with this $N_{\rm noise}$.

The required number of noise vectors depends on each operator.
%A large value of $N_{\rm noise}$ is, however, demanding. 
Here, we note that, in the evaluation of $c_4$ and $c_4^I$ through 
Eq.~(\ref{eq:AB}), the errors due to the error of ${\cal D}_1$ is dominant. 
In order to efficiently reduce the total errors of $c_4$ and $c_4^I$, we adopt large $N_{\rm noise}$ only for ${\cal D}_1$, keeping $N_{\rm noise}$ for other operators small.
The values of $N_{\rm noise}$ we adopt are summarized in Table \ref{tab:c2c4}.
We choose $N_{\rm noise}=10$ for the calculations of 
the operators in Eq.~(\ref{eq:AB}) except for the operators 
${\rm tr}[(\partial^n M/\partial \mu^n) M^{-1}]$, where $n=1-4$, for which 
we adopt $N_{\rm noise}=100$--400
(the first number in the column of $N_{\rm noise}$ in Table \ref{tab:c2c4}).

Finally, we take advantage of the knowledge that the odd derivatives are purely imaginary and the even derivatives are real. 
In the lower panel of Fig.~\ref{fig:timehis2}, we plot ${\rm Re} [{\cal D}_1]$ which should vanishes when $N_{\rm noise}$ is large enough.
We find that, unlike the case of ${\rm Im} [{\cal D}_1]$ shown in the upper panel of the same figure, ${\rm Re} [{\cal D}_1]$ data from two sets of random noises show no correlations in the time history even with small $N_{\rm noise}$. 
Therefore, to further reduce errors from the random noise method, we can put the real and imaginary parts of the odd and even derivatives to zero, respectively.

%%%%%%%%%%%%%%%%%%%%%%%%%%%%%%%%%%%%%%%%%%%%%%%%%%%%
\subsection{Quark number density, quark number susceptibility 
and isospin susceptibility}
\label{sec:qns}

\begin{figure}[t]
\begin{center}
\includegraphics[width=2.4in]{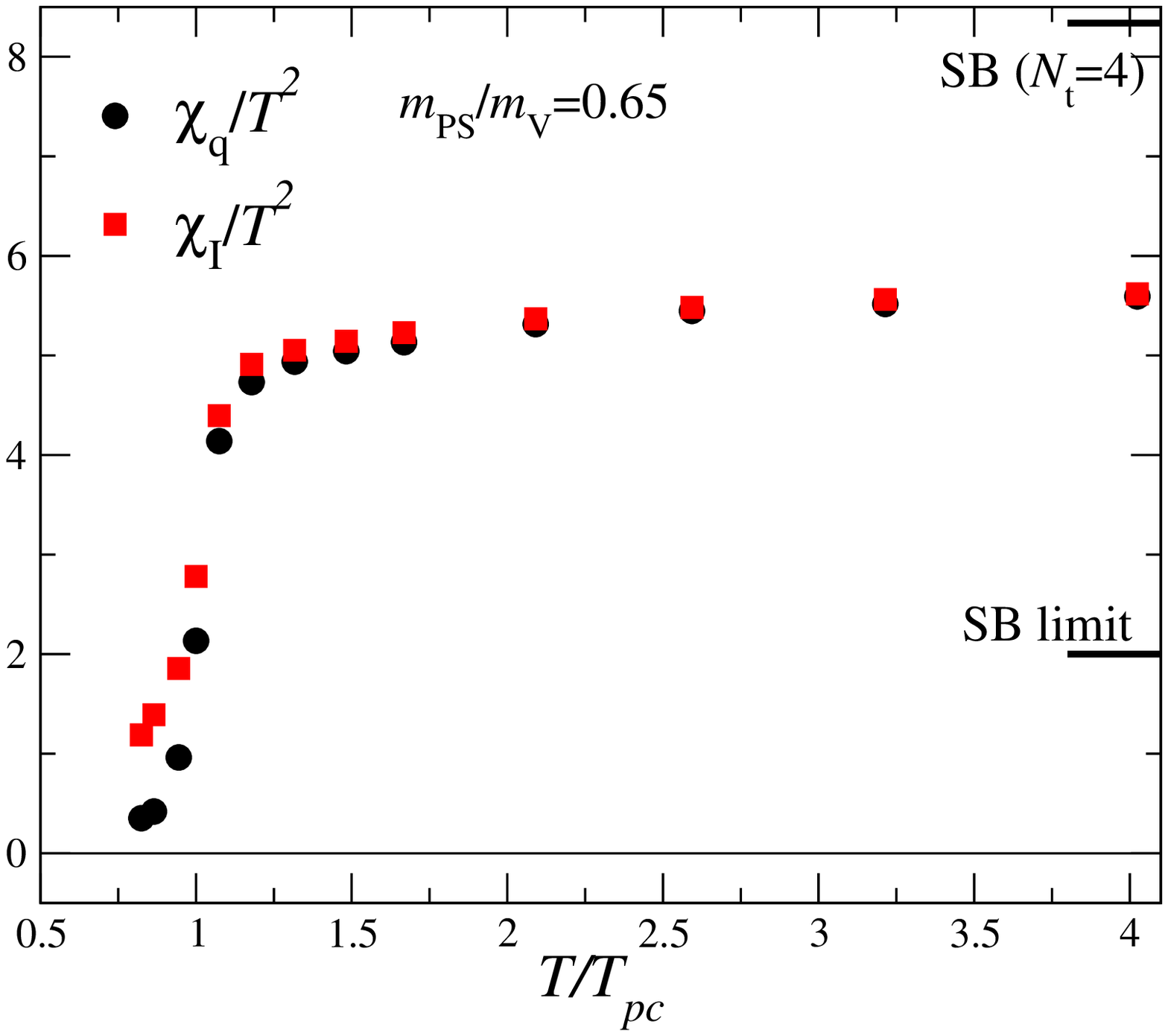}
\hskip 0.5cm
\includegraphics[width=2.4in]{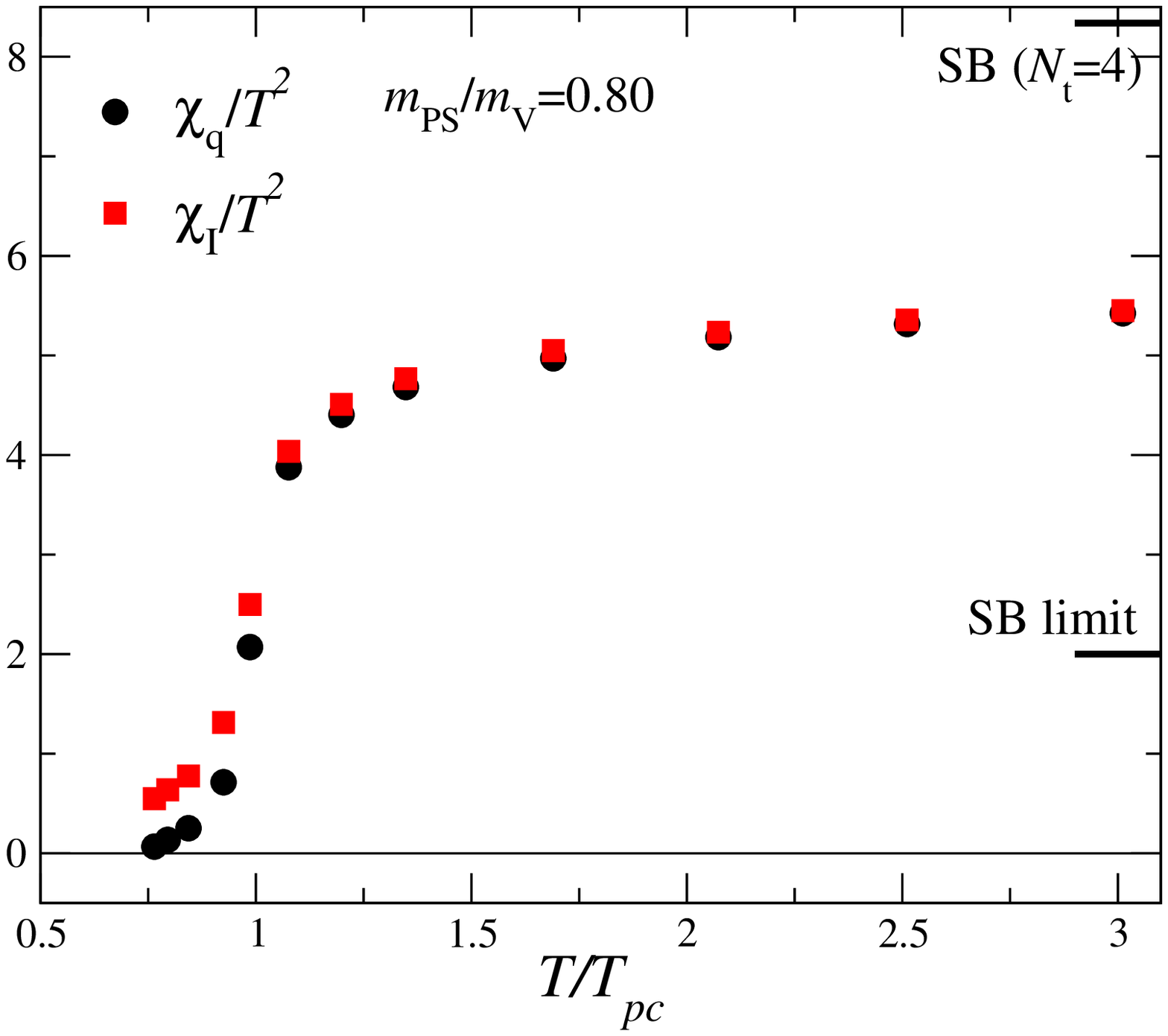}
\vskip -0.2cm
\caption{Quark number (black) and isospin (red) susceptibilities 
at $\mu_q=0$ for $m_{\rm PS}/m_{\rm V}=0.65$ (left) and $0.80$ (right).}
\label{fig:qns}
\end{center}
\vskip -0.3cm
\end{figure} 

\begin{table}[tbp]
 \begin{center}
 \caption{Results of the Taylor expansion coefficients
 for $m_{\rm PS}/m_{\rm V}=0.65$ and $0.80$.
 The first number in the column of $N_{\rm noise}$ is $N_{\rm noise}$ for the calculations of ${\rm tr}[(\partial^n M/\partial \mu^n) M^{-1}]$, and the second number for other traces. See text for details.}
 \label{tab:c2c4}
 {\renewcommand{\arraystretch}{1.2} \tabcolsep = 3mm
 \newcolumntype{a}{D{.}{.}{6}}
 \newcolumntype{d}{D{.}{.}{0}}
 \begin{tabular}{|aaaaad|}
 \hline
 \multicolumn{1}{|c}{$T/T_{pc}$} & 
 \multicolumn{1}{c}{$c_2 \times 2$} & 
 \multicolumn{1}{c}{$c_4 \times 4!$} &
 \multicolumn{1}{c}{$c_2^I \times 2$} & 
 \multicolumn{1}{c}{$c_4^I \times 4!$} &
 \multicolumn{1}{c|}{$N_{\rm noise}$} \\
 \hline
 \multicolumn{6}{|c|}{$m_{\rm PS}/m_{\rm V}=0.65$} \\
 \hline
0.82(3)  & 0.352(59)  & 6.3(108)  & 1.189(6)  & 1.41(49) & 400,10 \\
0.86(3)  & 0.420(71)  & 2.6(154)  & 1.392(6)  & 1.81(46) & 400,10 \\
0.94(3)  & 0.963(64)  & 10.5(103) & 1.857(10) & 2.88(64) & 400,10 \\
1.00(4)  & 2.134(53)  & 24.4(107) & 2.780(21) & 7.83(111)& 200,10 \\
1.07(4)  & 4.140(27)  & 8.7(21)   & 4.396(16) & 5.58(34) & 200,10 \\
1.18(4)  & 4.732(21)  & 7.8(11)   & 4.910(8)  & 4.82(19) & 200,10 \\
1.32(5)  & 4.938(20)  & 7.1(14)   & 5.052(6)  & 4.65(13) & 100,10 \\
1.48(5)  & 5.042(17)  & 5.6(12)   & 5.143(6)  & 4.72(14) & 100,10 \\
1.67(6)  & 5.133(15)  & 4.0(11)   & 5.229(5)  & 4.67(13) & 100,10 \\
2.09(7)  & 5.314(11)  & 5.0(6)    & 5.368(4)  & 4.65(8)  & 100,10 \\
2.59(9)  & 5.447(13)  & 4.8(6)    & 5.482(4)  & 4.72(5)  & 100,10 \\
3.22(12) & 5.517(12)  & 6.4(7)    & 5.562(4)  & 5.05(8)  & 100,10 \\
4.02(15) & 5.593(12)  & 5.8(6)    & 5.618(4)  & 5.03(7)  & 100,10 \\
 \hline
 \multicolumn{6}{|c|}{$m_{\rm PS}/m_{\rm V}=0.80$} \\
 \hline
0.76(4)  & 0.066(34) & 3.8(51)  & 0.549(4)  & 0.37(19) & 400,10 \\
0.80(4)  & 0.134(33) & 1.9(39)  & 0.637(5)  & 0.35(23) & 400,10 \\
0.84(4)  & 0.251(35) & 0.0(37)  & 0.776(6)  & 0.80(27) & 400,10 \\
0.93(5)  & 0.713(40) & 2.0(48)  & 1.313(9)  & 1.94(34) & 400,10 \\
0.99(5)  & 2.071(34) & 17.4(47) & 2.498(17) & 5.13(53) & 400,10 \\
1.08(5)  & 3.877(19) & 8.0(10)  & 4.036(10) & 4.92(18) & 200,10 \\
1.20(6)  & 4.403(14) & 7.8(9)   & 4.508(7)  & 4.63(14) & 200,10 \\
1.35(7)  & 4.682(11) & 5.8(5)   & 4.767(5)  & 4.50(7)  & 200,10 \\
1.69(8)  & 4.970(10) & 5.9(4)   & 5.048(5)  & 4.62(7)  & 200,10 \\
2.07(10) & 5.184(9)  & 5.8(3)   & 5.234(5)  & 4.71(5)  & 200,10 \\
2.51(13) & 5.315(8)  & 5.9(3)   & 5.357(4)  & 4.72(4)  & 200,10 \\
3.01(15) & 5.424(9)  & 6.0(3)   & 5.451(4)  & 4.83(3)  & 200,10 \\
\hline
 \end{tabular}}
 \end{center}
\end{table}

\begin{figure}[t]
\begin{center}
\includegraphics[width=2.4in]{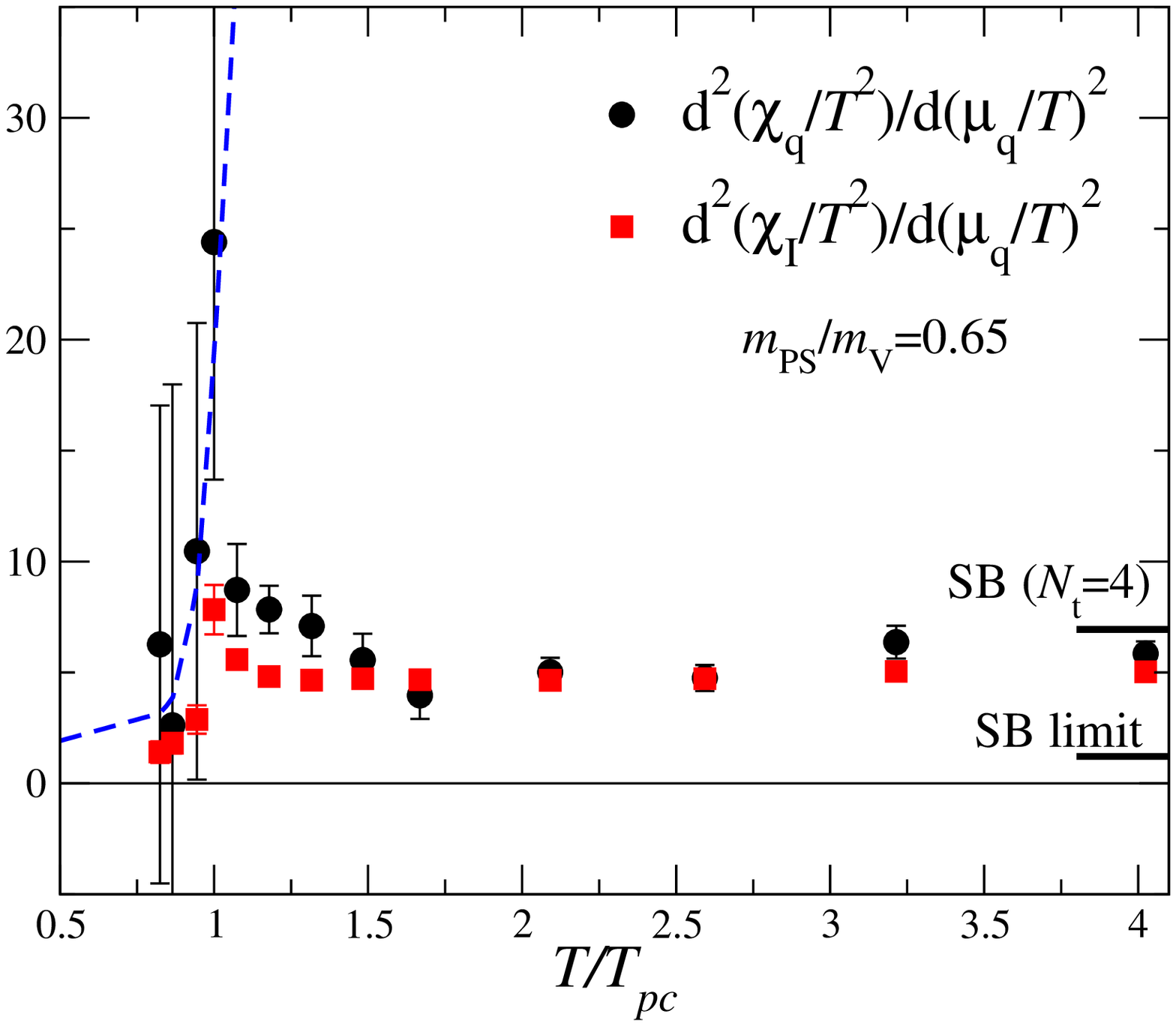}
\hskip 0.5cm
\includegraphics[width=2.4in]{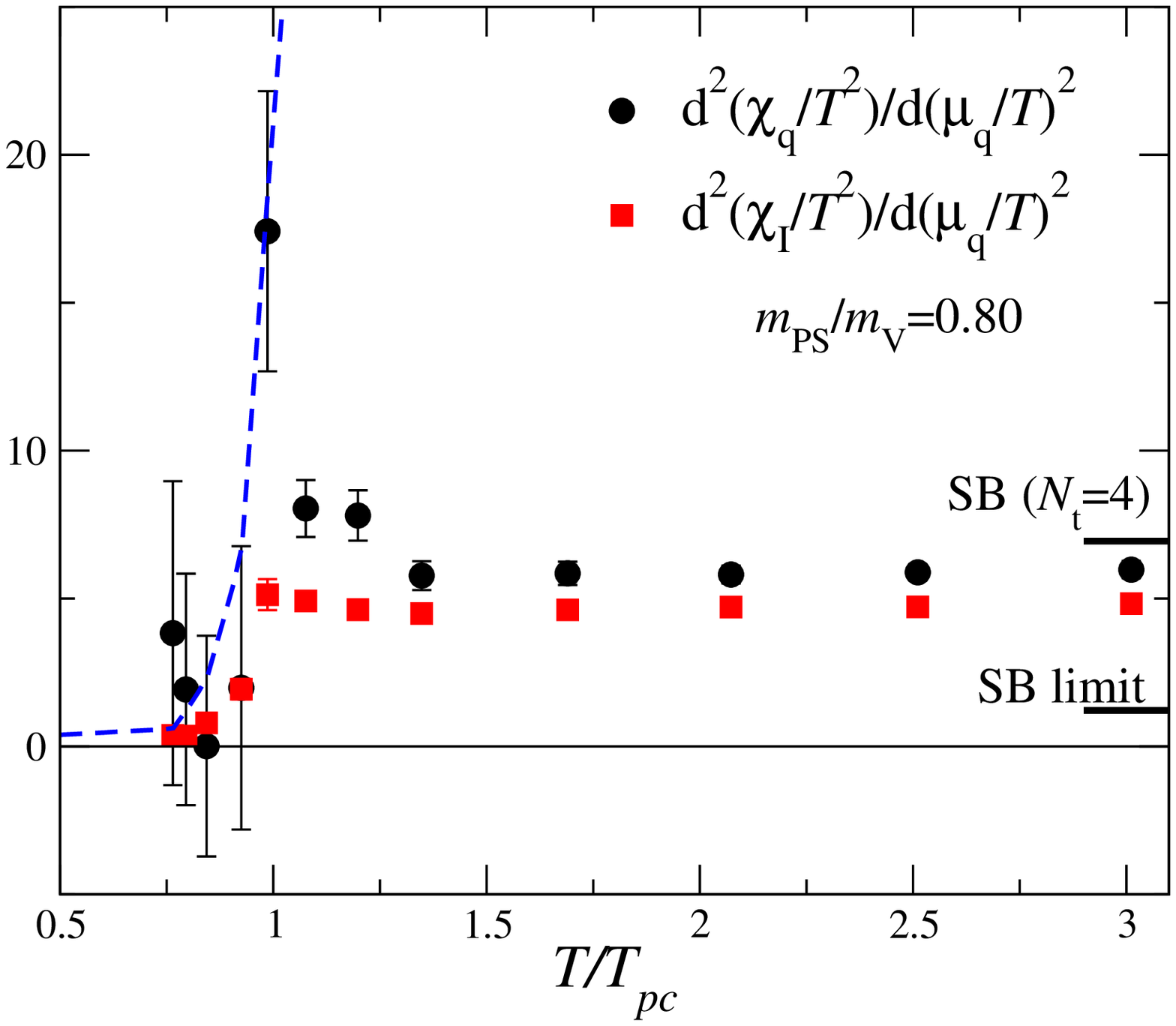}
\vskip -0.2cm
\caption{The second derivatives of quark number (black) and 
isospin (red) susceptibilities at $\mu_q=0$ for 
$m_{\rm PS}/m_{\rm V}=0.65$ (left) and $0.80$ (right).
The dashed line is a prediction from the hadron resonance gas model: 
$\partial^2 \chi_q/ \partial \mu_q^2 \approx 9 \chi_q/T^2$.}
\label{fig:d2qns}
\end{center}
\vskip -0.3cm
\end{figure} 

We perform a series of simulations along LCP's for two quark masses corresponding to $m_{\rm PS}/m_{\rm V}=0.65$ and $0.80$
to calculate the expansion coefficients $c_2, c_4, c_2^I$ and $c_4^I$ defined in Eq.~(\ref{eq:cncnI}).
The results are summarized in Table \ref{tab:c2c4}.

The results for $\chi_q/T^2$ and $\chi_I/T^2 $ at $\mu_q=0$ are plotted in Fig.~\ref{fig:qns}.
The circle and square symbols are for $\chi_q/T^2$ and $\chi_I/T^2$, respectively. 
The short lines in the right end denote the values in the free quark-gluon gas (Stefan-Boltzmann) limit, both for $N_t=4$ and in the continuum (cf. Appendix \ref{ap:free}).
%Note that the lattice discretization error in the equation of state is known to be large for $N_t=4$ with our action.
%For a more quantitative study, we need to increase $N_t$ decreasing $m_{\rm PS}/m_{\rm V}$.

At $\mu_q=0$, $\chi_q/T^2=2c_2$ and $\chi_I/T^2=2c_2^I$.
Because ${\cal D}_1$ is a pure imaginary number, ${\cal D}_1^2$ is negative in Eq.~(\ref{eq:AB}) and thus $\chi_I/T^2$ will be larger than $\chi_q/T^2$, 
while the difference should vanishes in the high temperature limit according to Eq.~(\ref{eq:psb}) for the free quark gluon gas.
In the low temperature phase, $\chi_q/T^2$ and $\chi_I/T^2$ correspond 
to the fluctuations of baryon and isospin numbers, respectively. 
Since the fluctuation of isospin number is mainly caused by pions, 
the fluctuation should be larger than that of the baryon number. 
Moreover, because the pion mass is more sensitive to the quark mass than baryon 
masses, $\chi_I/T^2$ will show more sensitivity to the quark mass 
than $\chi_q/T^2$. 

As seen from Fig.~\ref{fig:qns}, both $\chi_q/T^2$ and $\chi_I/T^2$ 
increase sharply at $T_{pc}$, in accordance with an expectation 
that the fluctuations in the quark-gluon plasma phase are much larger 
than those in the hadronic phase.
We find that $\chi_I/T^2$ is larger than $\chi_q/T^2$ at low temperatures 
and the difference vanishes in the high temperature region.
Also, the isospin susceptibility increases as $m_{\rm PS}/m_{\rm V}$ decreases 
at low temperatures, while $\chi_q/T^2$ does not change very much.
These results agree qualitatively with previous results obtained with  
staggered-type quarks \cite{BS03,BS05,milc07,rbcb08,Gottlieb}. 

The quark number and isospin susceptibilities are expected 
to show quite different behaviors near the critical point at finite density.  
When the quark mass is nonzero, iso-triplet mesons are massive and thus are irrelevant 
to the critical behavior. 
Therefore, the iso-triplet susceptibility $\chi_I$ will not show singularity.
On the other hand, if there is a critical point in the $(T, \mu_q)$ plane, 
scalar sectors, $\bar{\psi} \psi$ and $\bar{\psi} \gamma_0 \psi$, 
may become massless at the critical point. 
We then expect divergence in the fluctuations of the chiral condensate and quark number 
towards the critical point. 

%Therefore, to investigate the critical point, 
%it is useful to compare $\chi_q$ and $\chi_I$ at finite density. 
%We calculate the second derivatives of these susceptibilities. 
%These are the leading contributions to the $\mu_q$-dependence of them. 
Figure \ref{fig:d2qns} shows our results for 
$\left. \partial^2 (\chi_q/T^2)/ \partial (\mu_q/T)^2 \right|_{\mu_q=0}= 24 c_4$ (circles) 
and $\left. \partial^2 (\chi_I/T^2)/ \partial (\mu_q/T)^2\right|_{\mu_q=0} = 24 c_4^I$ 
(squares). 
We also plot $9 \chi_q/T^2$ as a dashed line in this figure to compare with the prediction from the hadron resonance gas model in Eq.~(\ref{eq:chiHRG}), 
i.e. $\partial^2 \chi_q/ \partial \mu_q^2 \approx 9 \chi_q/T^2$.
These results are consistent within the error at $T < T_{pc}$ .

Although the statistical errors are not quite small yet, the two susceptibilities show quite different behaviors near $T_{pc}$. 
$\partial^2 (\chi_q/T^2)/ \partial (\mu_q/T)^2$ near $T_{pc}$ is more 
than three times larger than that at high temperatures, 
suggesting the large enhancement in the quark number fluctuations as the density is increased. 
Moreover, the peak height is larger for smaller $m_{\rm PS}/m_{\rm V}$. 
On the other hand, no such sharp peak appears for $\partial^2 (\chi_I/T^2)/ \partial (\mu_q/T)^2$,
in accordance with the expectation that $\chi_I$ is analytic at the critical point. 
These observations suggest the existence of the critical point. 
Similar results were obtained by p4-improved staggered fermions \cite{BS03,BS05,milc07,rbcb08}.  

\begin{figure}[t]
\begin{center}
\includegraphics[width=2.4in]{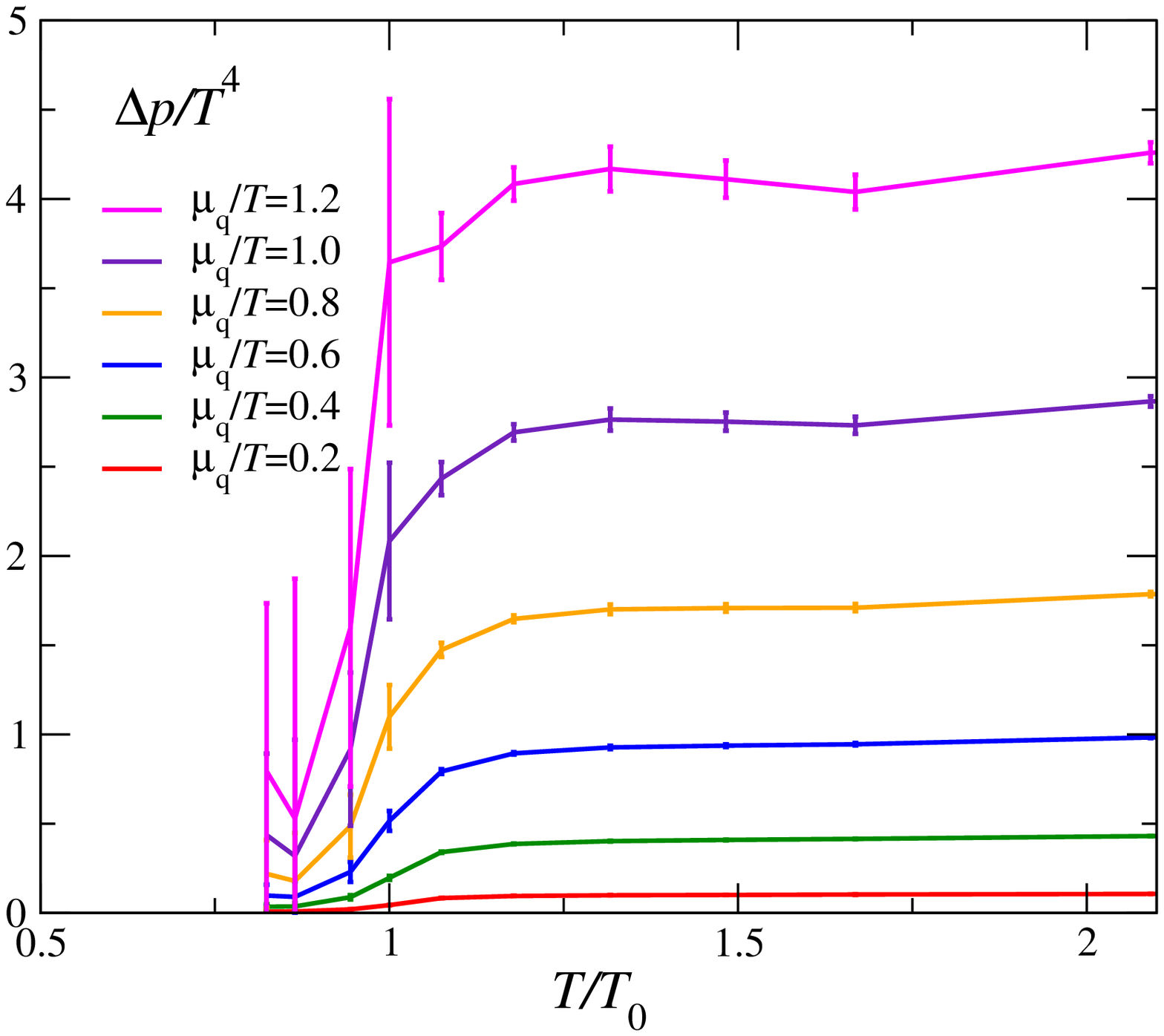}
\hskip 0.5cm
\includegraphics[width=2.4in]{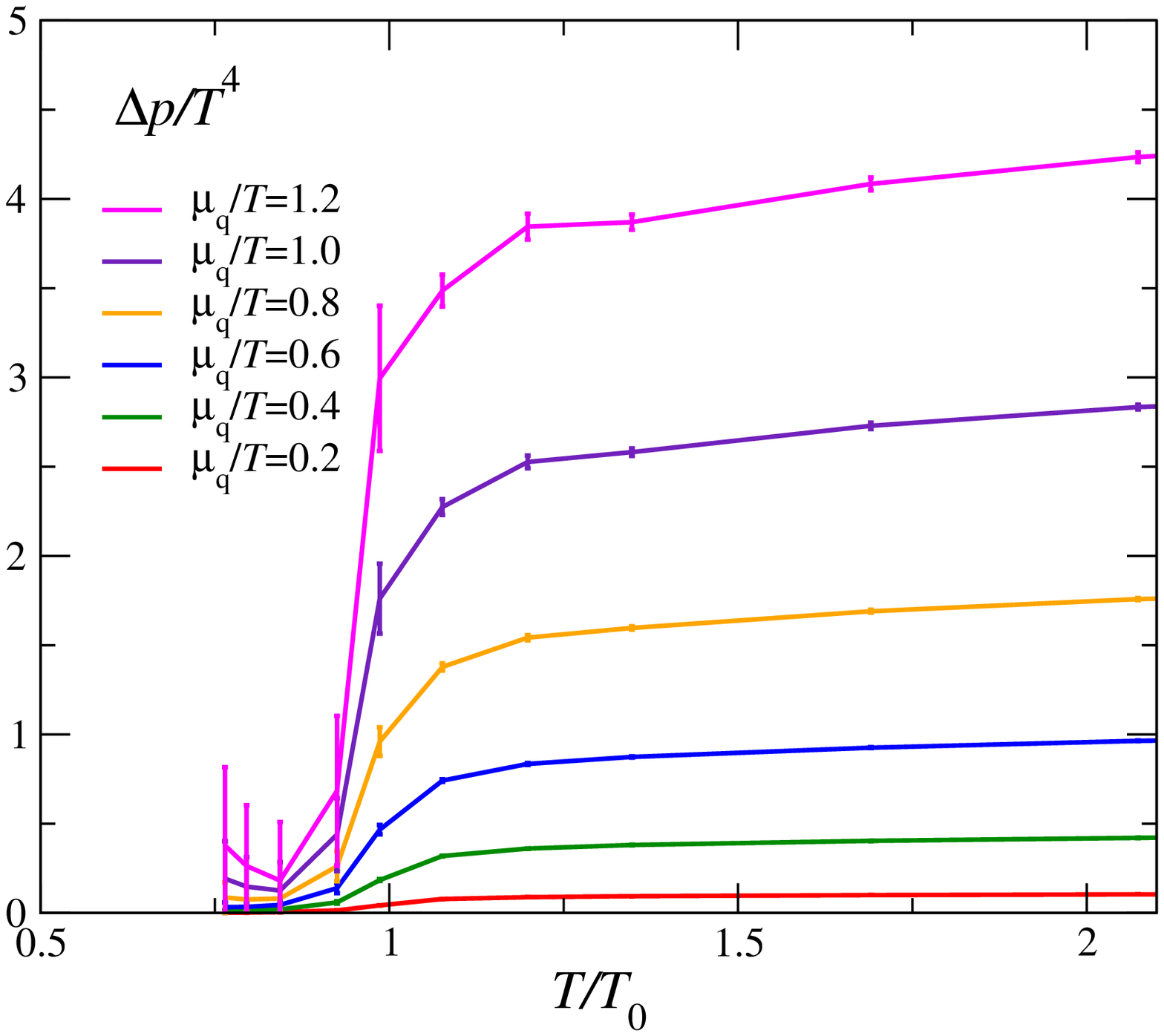}
\vskip -0.2cm
\caption{$T$-dependence of the $\mu_q$-dependent contribution to the pressure, $\Delta p/T^4 \equiv p(\mu_q)/T^4 - p(0)/T^4$,  
at $m_{\rm PS}/m_{\rm V}=0.65$ (left) and $0.80$ (right).
$T_0$ is $T_{pc}$ at $\mu_{q}=0$.}
\label{fig:prsmu}
\end{center}
\vskip -0.3cm
\end{figure}

\begin{figure}[t]
\begin{center}
\includegraphics[width=2.4in]{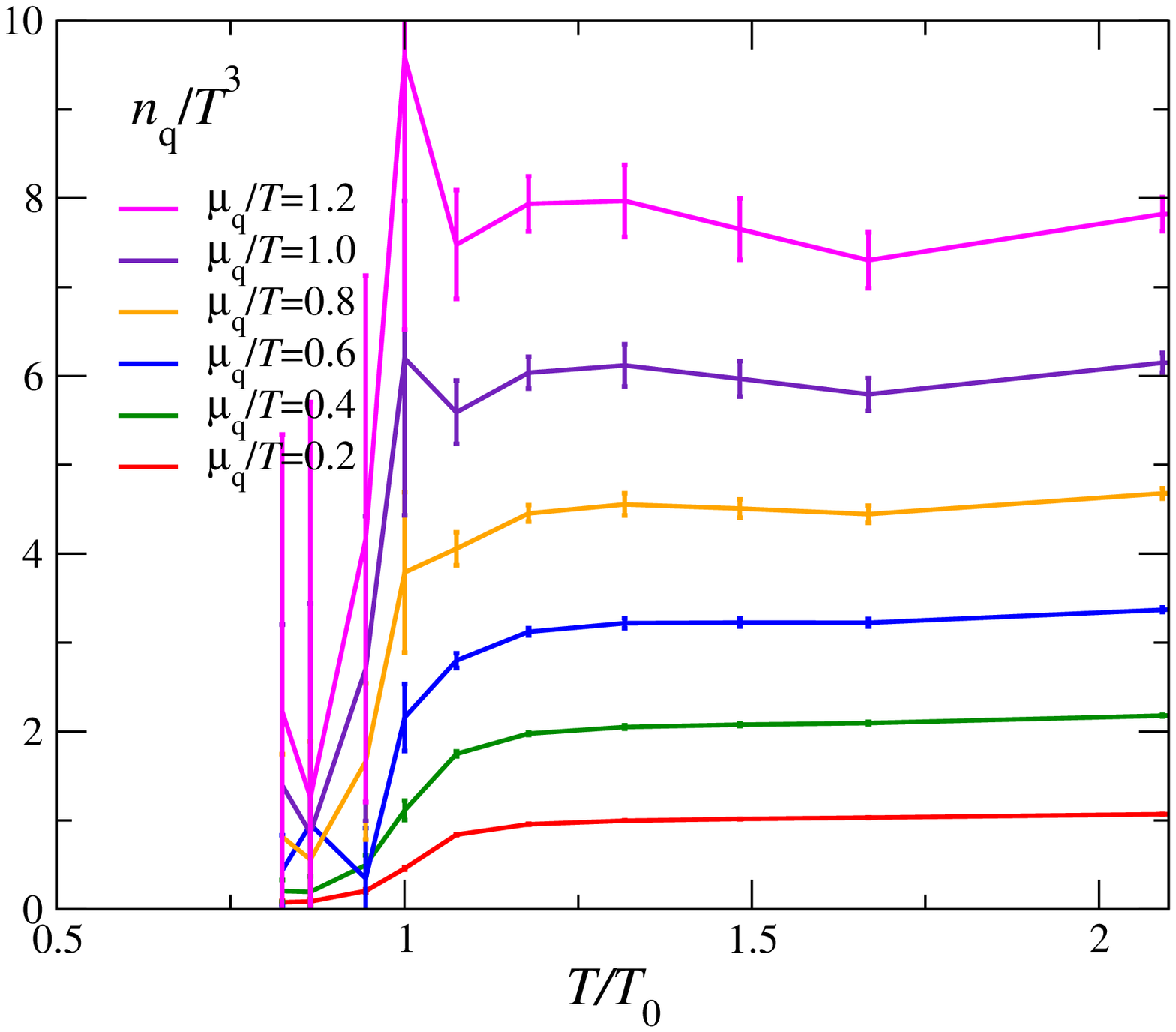}
\hskip 0.5cm
\includegraphics[width=2.4in]{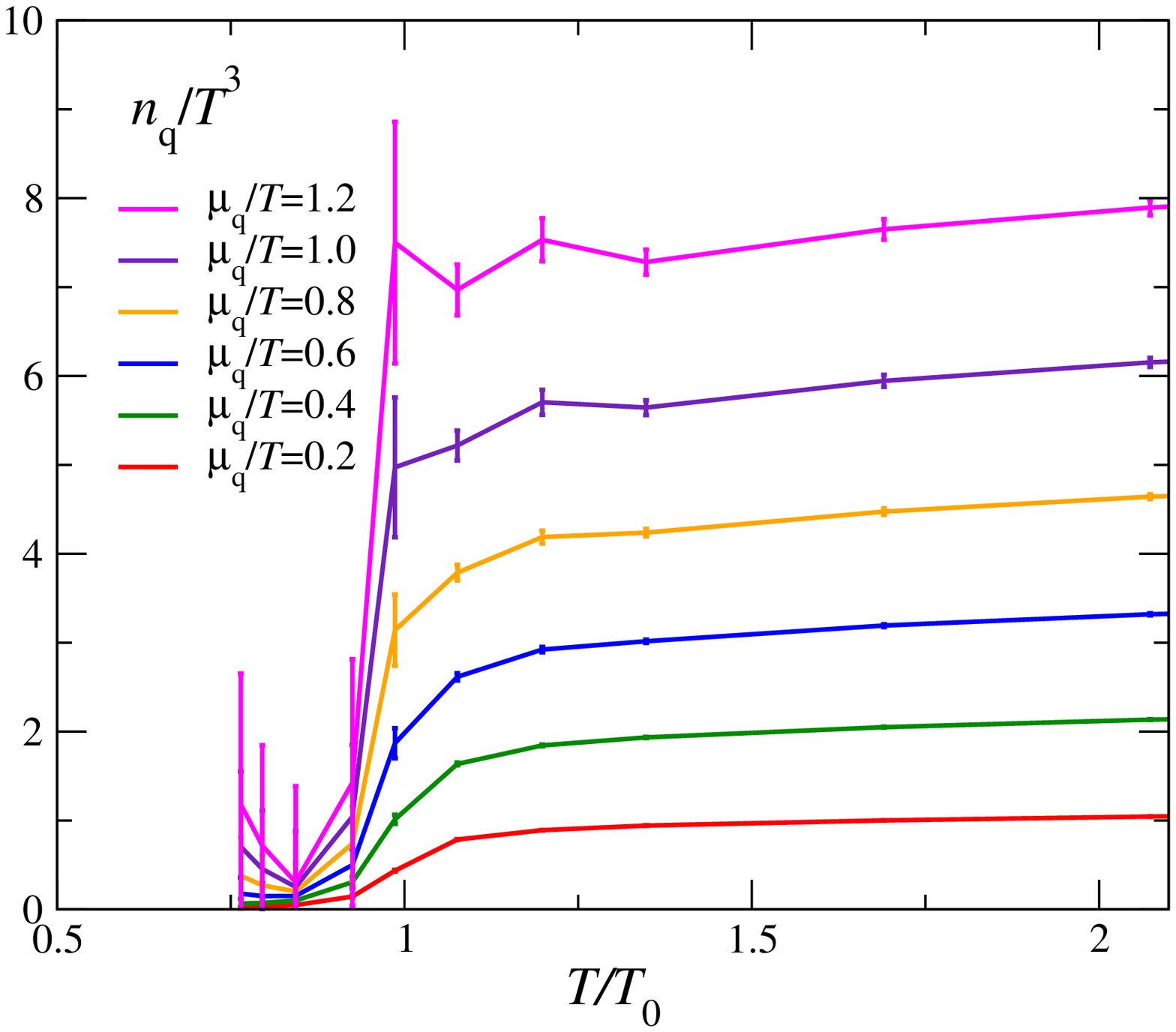}
\vskip -0.2cm
\caption{Quark number density at finite $\mu_q$ 
for $m_{\rm PS}/m_{\rm V}=0.65$ (left) and $0.80$ (right). $T_0$ is $T_{pc}$ at $\mu_{q}=0$.}
\label{fig:qndmu}
\end{center}
\vskip -0.3cm
\end{figure}

\begin{figure}[t]
\begin{center}
\includegraphics[width=2.4in]{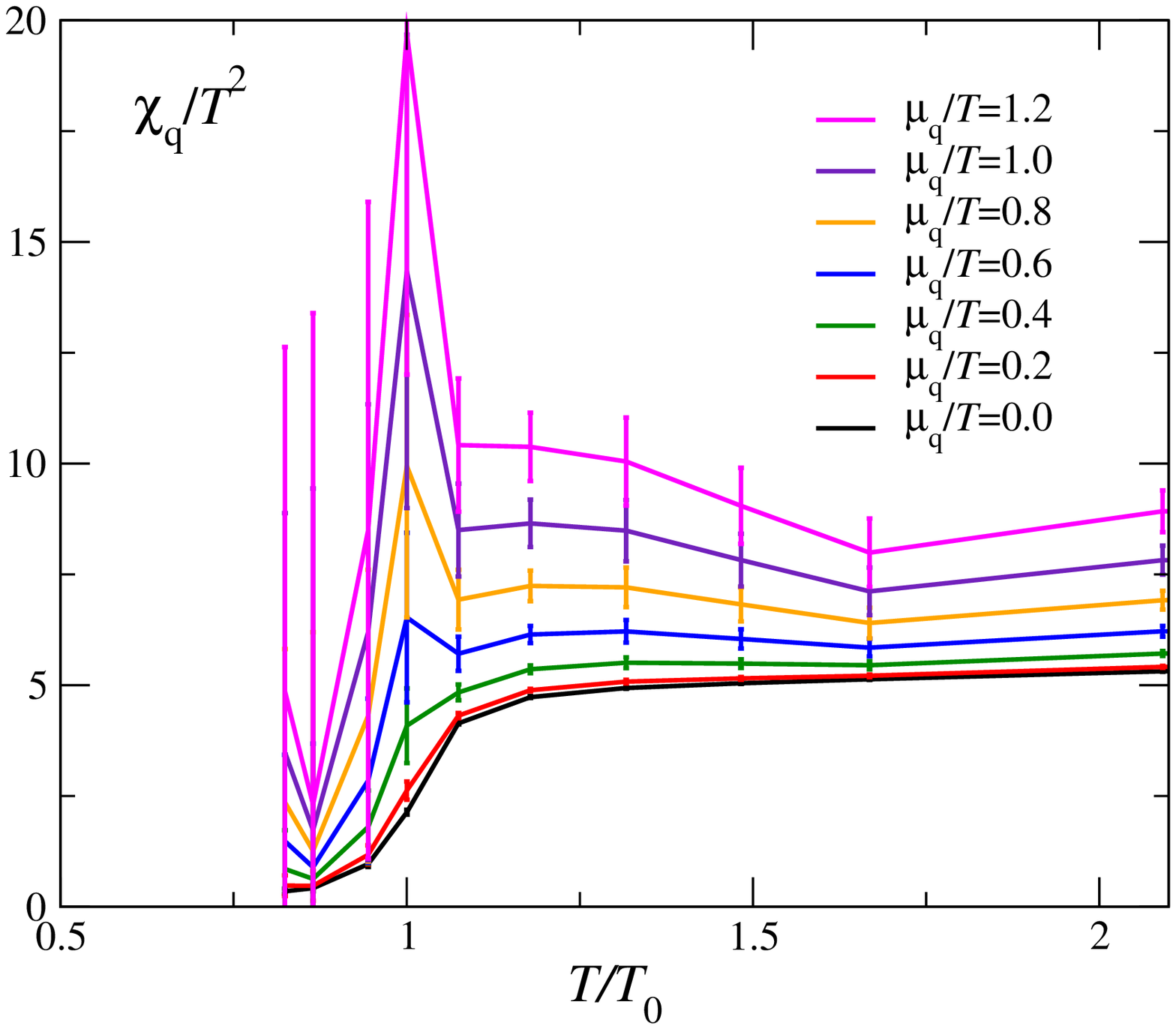}
\hskip 0.5cm
\includegraphics[width=2.4in]{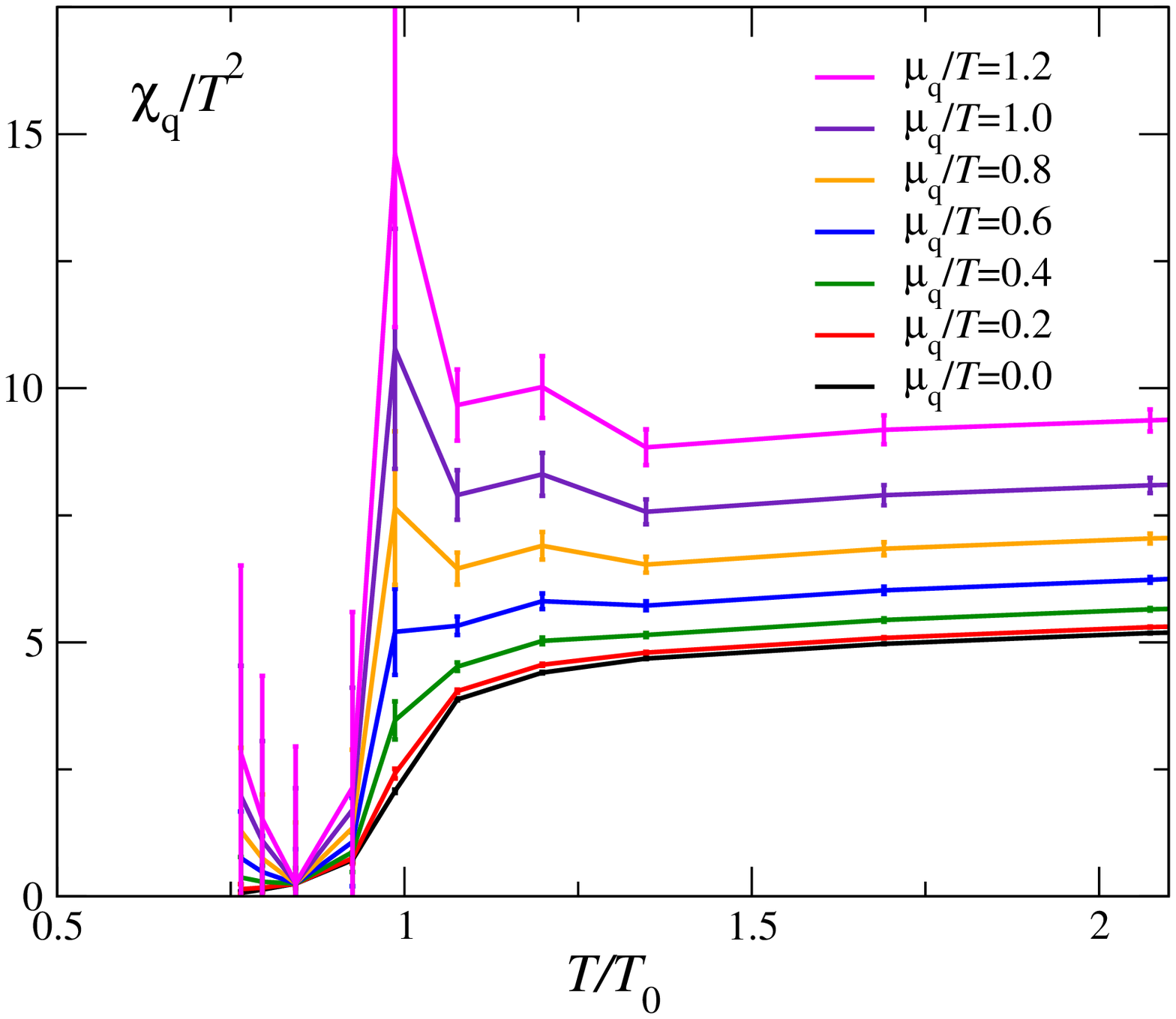}
\vskip -0.2cm
\caption{Quark number susceptibility 
at finite $\mu_q$ for $m_{\rm PS}/m_{\rm V}=0.65$ (left) and $0.80$ (right).}
\label{fig:qnsmu}
\end{center}
\vskip -0.3cm
\end{figure}

\begin{figure}[t]
\begin{center}
\includegraphics[width=2.4in]{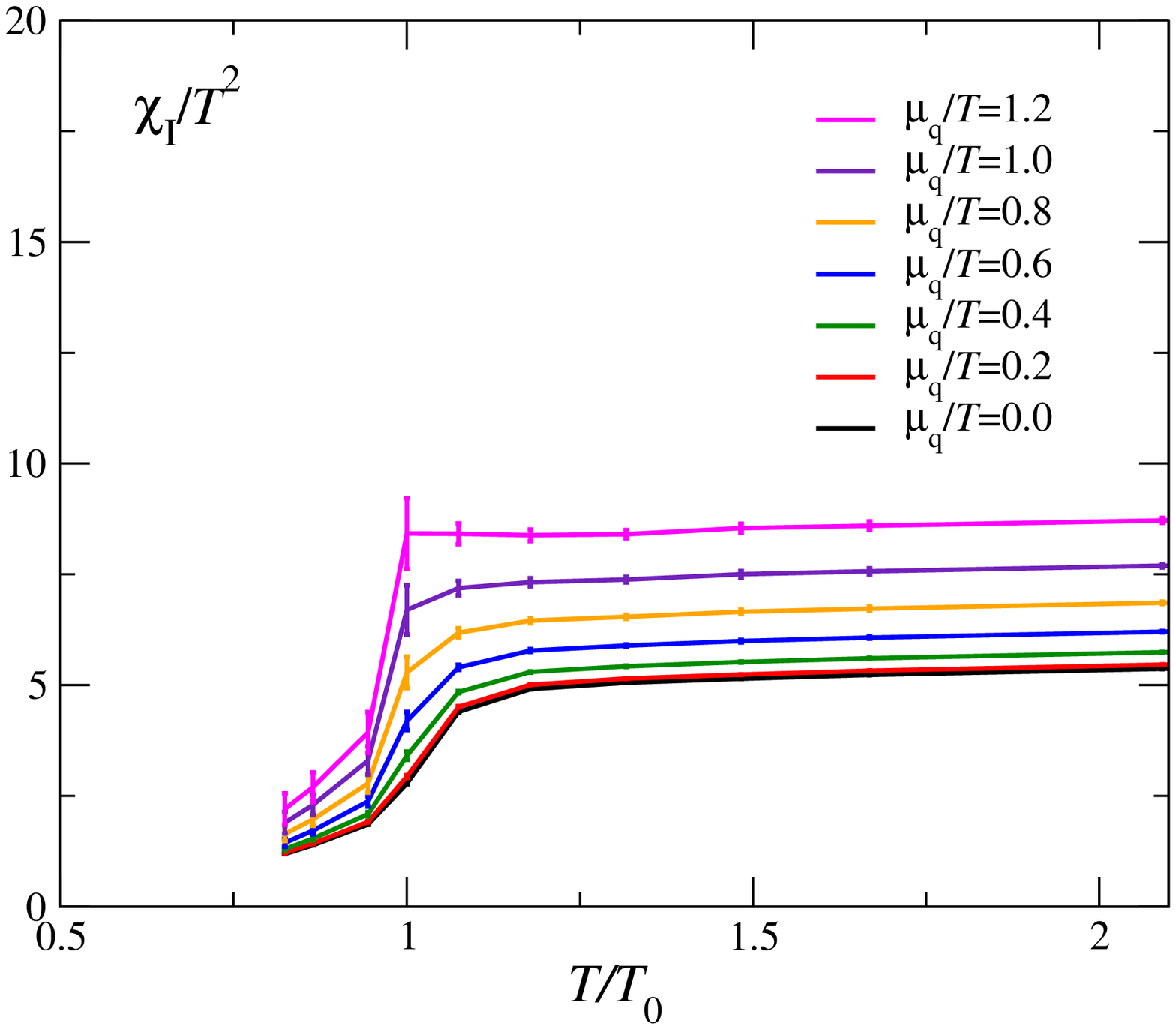}
\hskip 0.5cm
\includegraphics[width=2.4in]{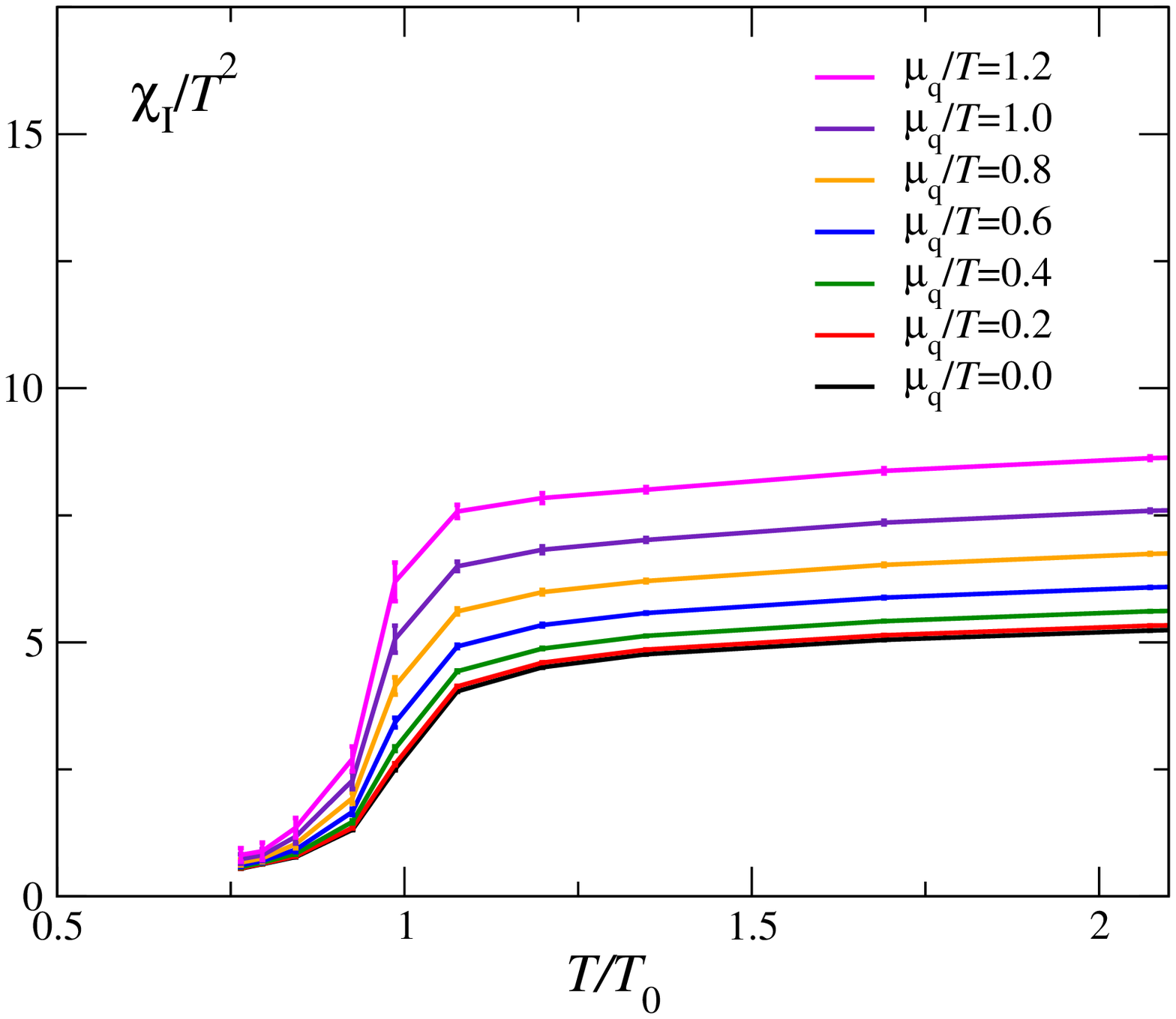}
\vskip -0.2cm
\caption{Isospin susceptibility 
at finite $\mu_q$ for $m_{\rm PS}/m_{\rm V}=0.65$ (left) and $0.80$ (right).}
\label{fig:issmu}
\end{center}
\vskip -0.3cm
\end{figure}

Finally, we evaluate the equation of state at finite $\mu_q$ combining  
the results of derivatives. %$c_2$, $c_4$, $c_2^I$ and $c_4^I$. 
Figure \ref{fig:prsmu} shows the $\mu_q$-dependent contribution of 
the pressure, $\Delta p/T^4 \equiv p(\mu_q)/T^4 - p(0)/T^4 
= c_2 (\mu_q/T)^2 + c_4 (\mu_q/T)^4$, at $m_{\rm PS}/m_{\rm V}=0.65$ (left) and $0.80$ (right). 
The truncation error is $O(\mu_q^6)$.
$T_{0}$ is $T_{pc}$ at $\mu_q=0$.
The finite density correction for $p/T^4$ becomes the same size as 
$p/T^4$ at $\mu_q=0$ around $\mu_q/T \sim O(1)$, and the correction 
$\Delta p/T^4$ increases rapidly around $T_{pc}$ in comparison with 
the behavior of $p/T^4$ at $\mu_q=0$. 
This suggests that the pressure changes more sharply as $\mu_q$ is increased.
The quark number density, 
$n_q/T^3 
\equiv (n_u+n_d)/T^3
%\equiv \partial (p/T^4)/ \partial (\mu_q/T) 
= 2 c_2 (\mu_q/T) + 4 c_4 (\mu_q/T)^3 +O(\mu_q^5)$, 
is shown in Fig.~\ref{fig:qndmu}. 
The quark number susceptibility and isospin susceptibility are shown 
in Fig.~\ref{fig:qnsmu} and Fig.~\ref{fig:issmu}, respectively. 
As discussed above, we find large quark number fluctuations near $T_{pc}$ when $\mu_q$ is increased. 
On the other hand, such an enhancement around $T_{pc}$ is not visible in the isospin fluctuations.
These results are consistent with the observations with staggered-type quarks 
and suggest a critical point at finite $\mu_q$.
%However, because the statistical errors in $n_q/T^3$ and $\chi_q/T^2$ are still large, 
%further studies are needed for more definite discussions with this approach.

%%%%%%%%%%%%%%%%%%%%%%%%%%%%%%%%%%%%%%%%%%%%%%%%%%%%
\section{Equation of state from the Gaussian approximation}
\label{sec:eosrew}

In the previous section, we have studied the equation of state at finite density by computing the Taylor expansion coefficients $c_n$ up to the fourth order, based on the calculation of 
$ {\cal D}_n = N_f [\partial^n \ln \det M/\partial \mu^n]$ 
for $n \leq 4$.
We found, however, that the statistical errors in $n_q/T^3$ and $\chi_q/T^2$ are not small.
Furthermore, the statistical errors will be larger when we include higher order terms, $c_6$, $c_8$ etc. 

In this connection, we recall that, in a previous study with staggered-type 
quarks \cite{eji07}, a hybrid method of the reweighting technique and 
Taylor expansion \cite{BS02}, combined with a Gaussian approximation 
for the complex phase distribution of quark determinant, 
was efficient to suppress statistical fluctuations at finite densities.
We call the method simply the Gaussian approximation.
In this section, we apply the Gaussian approximation to the calculation of EOS with improved Wilson quarks.

In the evaluation of higher order Taylor coefficients $c_n$ with $n>4$, the calculation of ${\cal D}_n$ at large $n$ is quite demanding. 
However, the free quark-gluon gas leads to ${\cal D}_n = 0$ for $n>4$ in the continuum limit.
Therefore, we may approximately evaluate higher order coefficients by keeping ${\cal D}_n$ for $n\leq 4$ only.
The approximation should work at least at high temperatures. 
Therefore, we consider the following approximate grand canonical potential, 
\begin{eqnarray}
-\frac{\omega(T,\mu_q)}{T^4} 
&=& \frac{1}{VT^3} \ln \left[ \int 
{\cal D}U \left( \det M(\mu) \right)^{N_{\rm f}} e^{-S_g} \right]
= \frac{1}{VT^3} \ln {\cal Z}(T,0) + \frac{1}{VT^3} \ln \left\langle 
\left( \frac{\det M(\mu)}{\det M(0)} 
\right)^{N_{\rm f}} \right\rangle_{(\mu=0)} \nonumber  \\
& \approx &
\frac{1}{VT^3} \ln {\cal Z}(T,0) 
+ \frac{1}{VT^3} \ln \left\langle \exp \left[ 
\sum_{n=1}^{N_{\rm max}} \frac{1}{n!} {\cal D}_n \, \mu^n 
\right] \right\rangle_{(\mu=0)}, 
\label{eq:pranpotap} 
\end{eqnarray}
where $\mu \equiv \mu_q a=\mu_q/(TN_t)$ and $N_{\rm max}=4$.
Here, $\langle \cdots \rangle_{(\mu =0)}$ is the average over 
configurations at $\mu =0$.
This approximate grand canonical potential is equal to the exact potential up to $O(\mu^{N_{\rm max}})$, and most of higher order contributions are contained except for terms including ${\cal D}_n$ for $n> N_{\rm max}$.
In this context, the method would be better than the truncated Taylor expansion method discussed in the previous section.

%%%%%%%%%%%%%%%%%%%%%%%%%%%%%%%%%%%%%%%%%%%%%%%%%%%%
\subsection{Gaussian approximation for the $\theta$ distribution}
%Problem of complex determinant (sign problem)} 
\label{sec:signpro}

\begin{figure}[t]
\begin{center}
\includegraphics[width=2.4in]{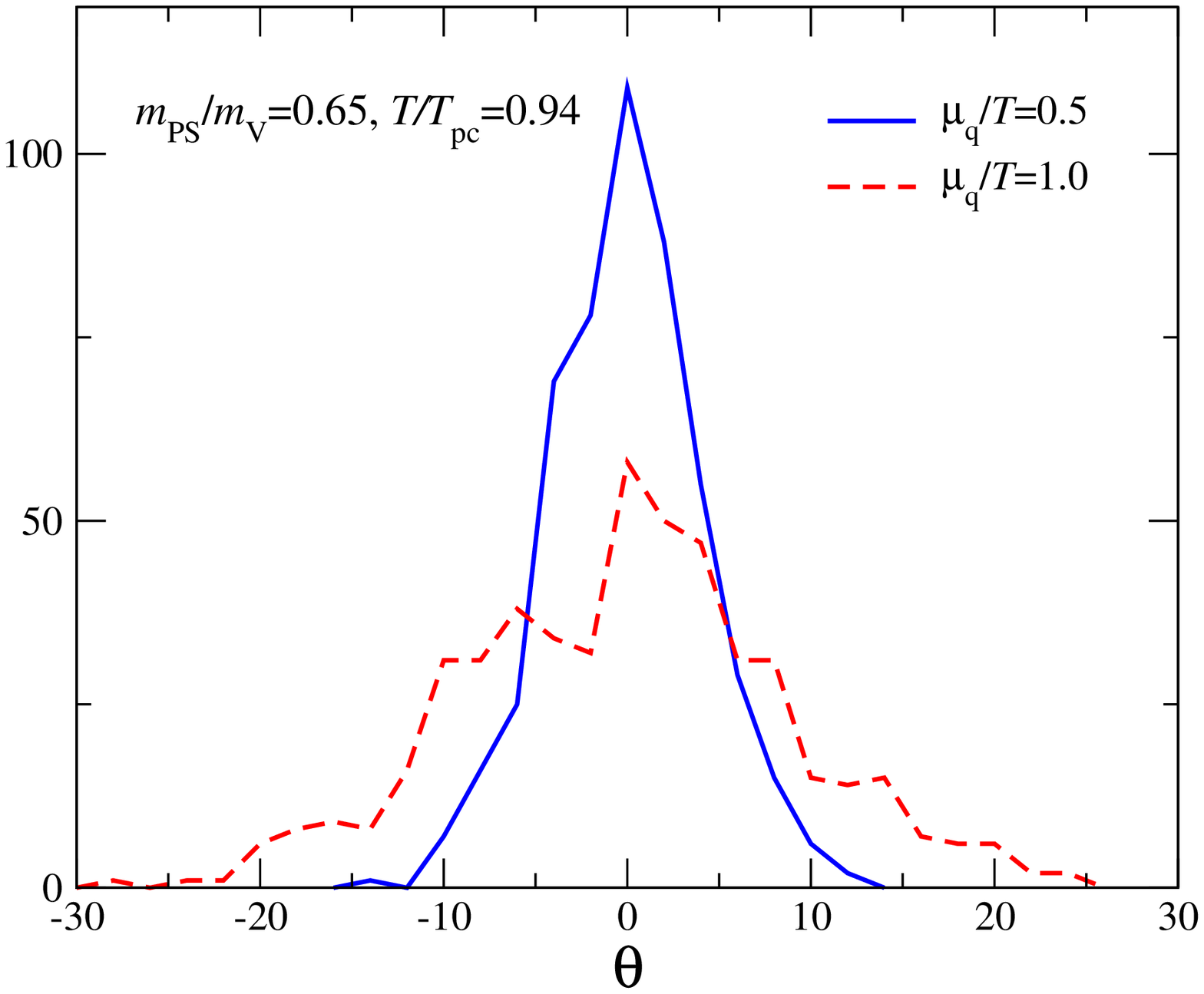}
\hskip 0.5cm
\includegraphics[width=2.4in]{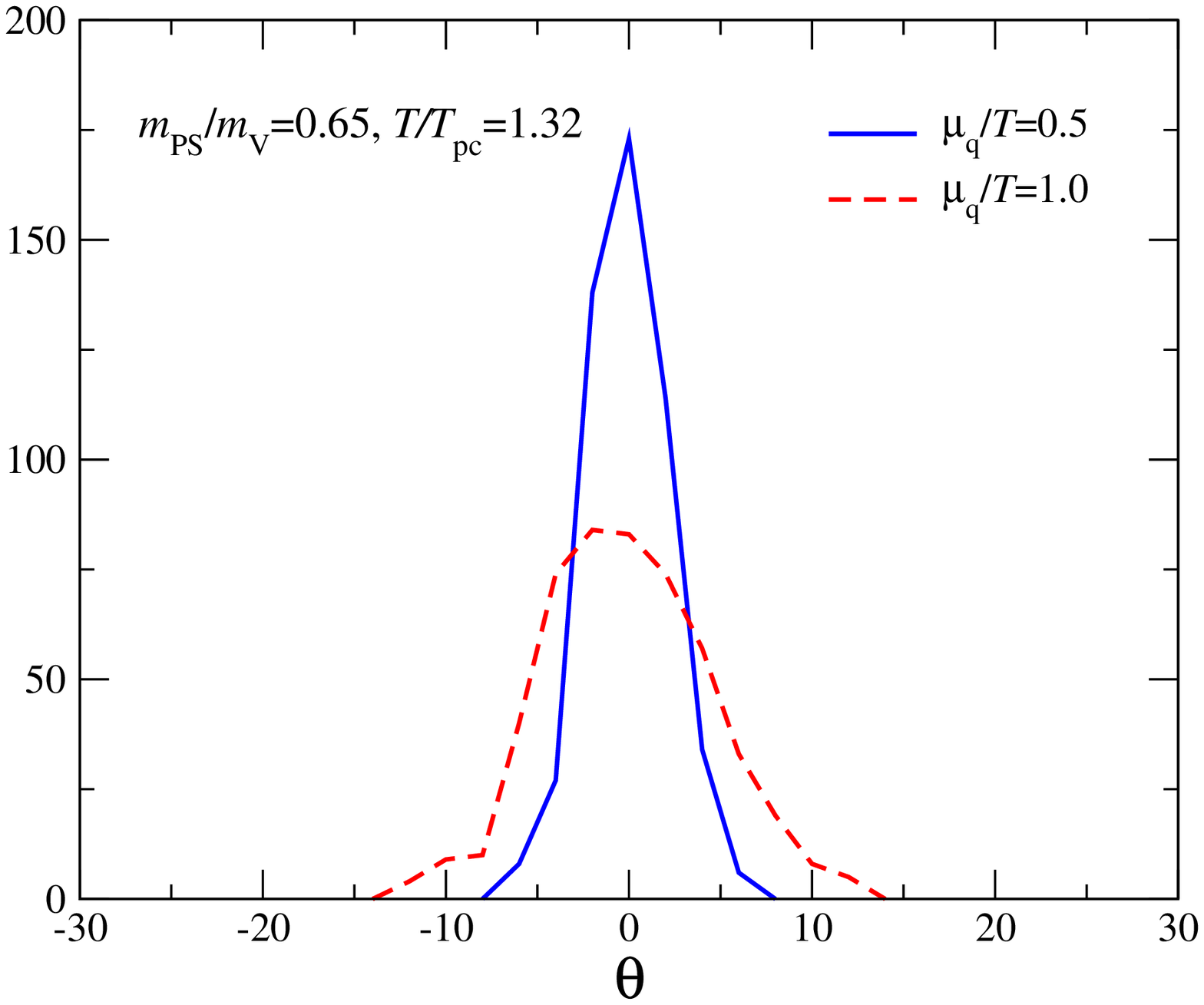}
\vskip 0.1cm
\includegraphics[width=2.4in]{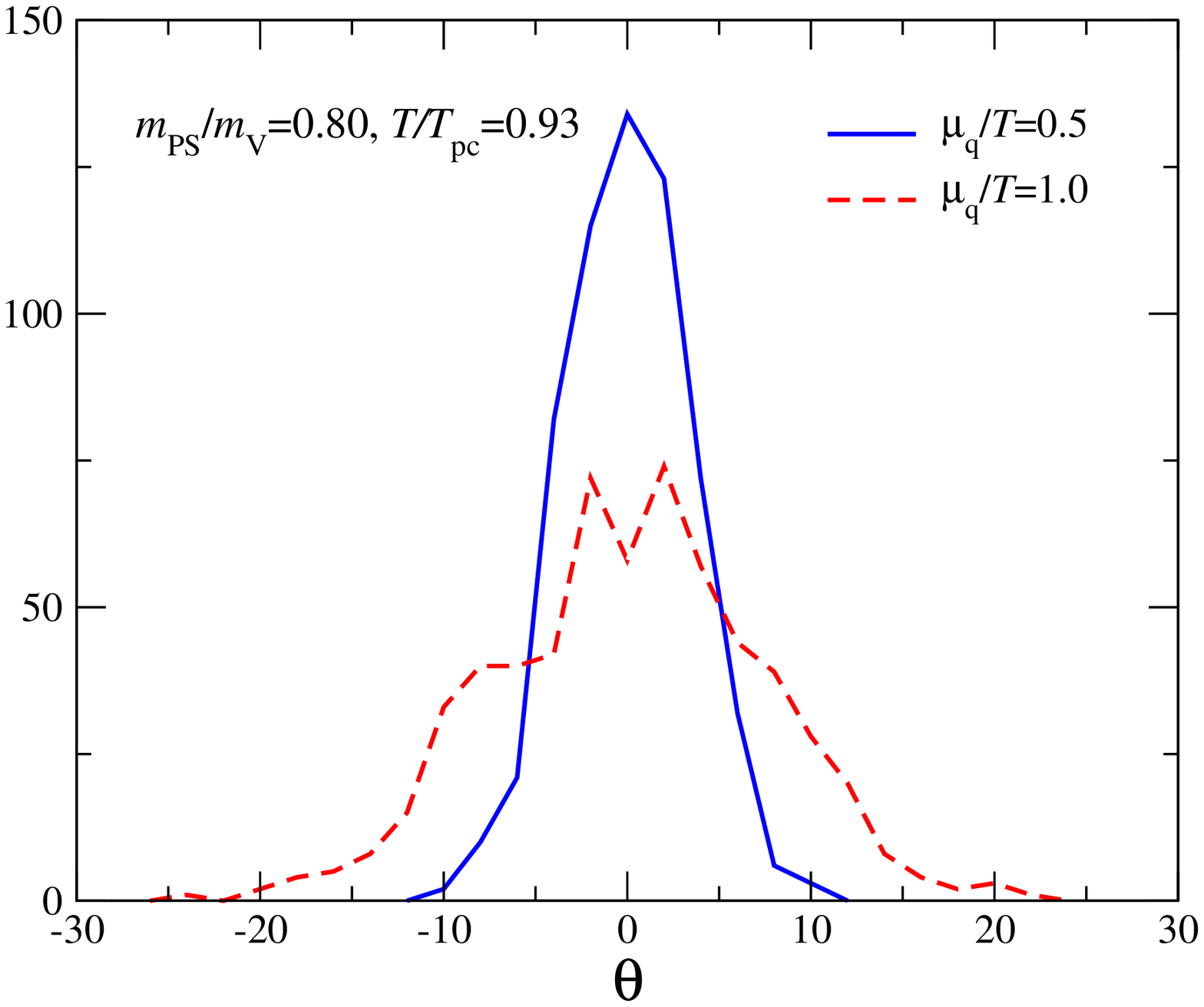}
\hskip 0.5cm
\includegraphics[width=2.4in]{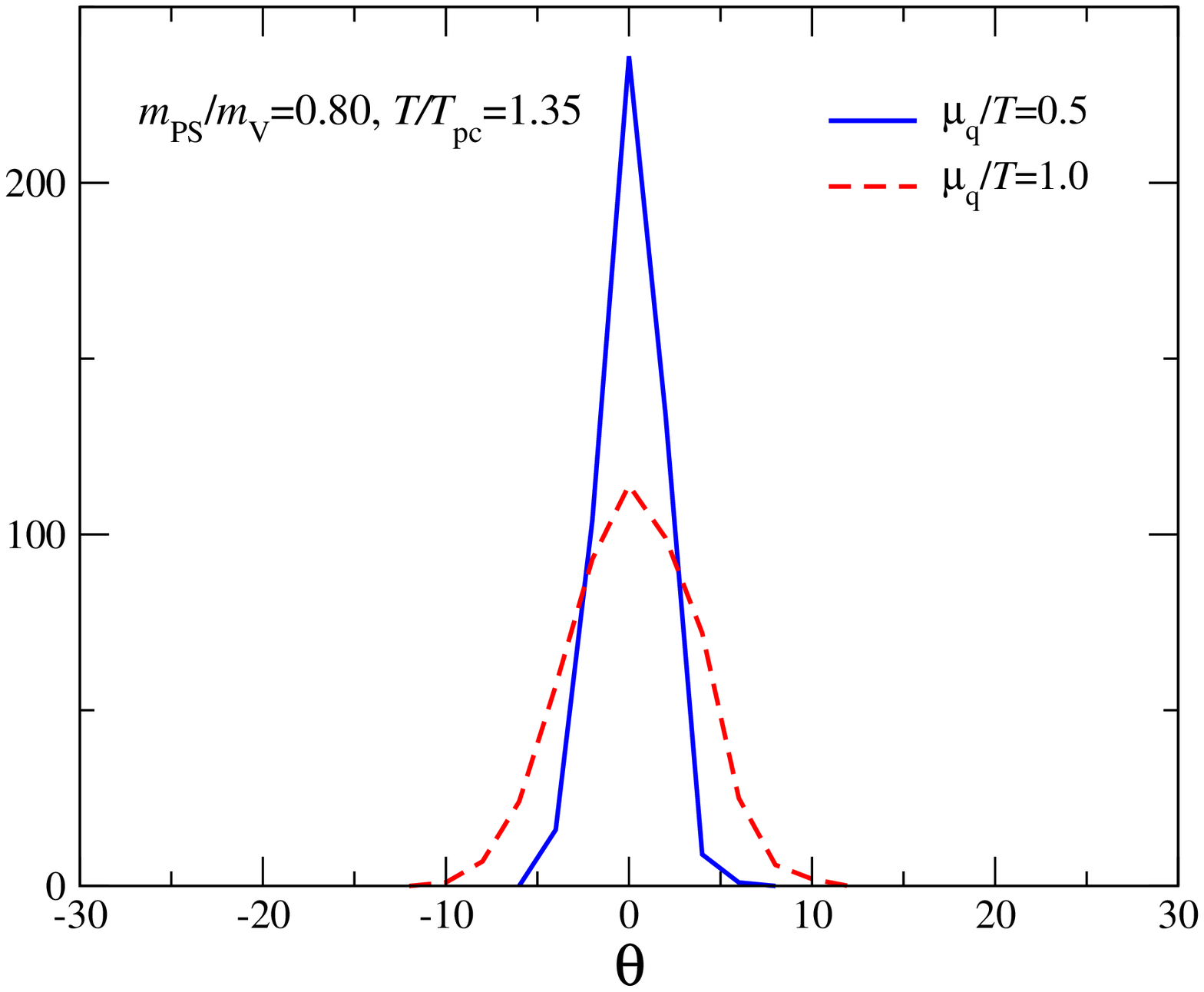}
\vskip -0.2cm
\caption{The histogram of $\theta$ for simulations 
at $(m_{\rm PS}/m_{\rm V}, T/T_{pc})=(0.65, 0.94)$ (top left), 
$(0.65, 1.32)$ (top right), $(0.80, 0.93)$ (bottom left) 
and $(0.80, 1.35)$ (bottom right).}
\label{fig:phhis}
\end{center}
\vskip -0.3cm
\end{figure} 

\begin{figure}[t]
\begin{center}
\includegraphics[width=2.4in]{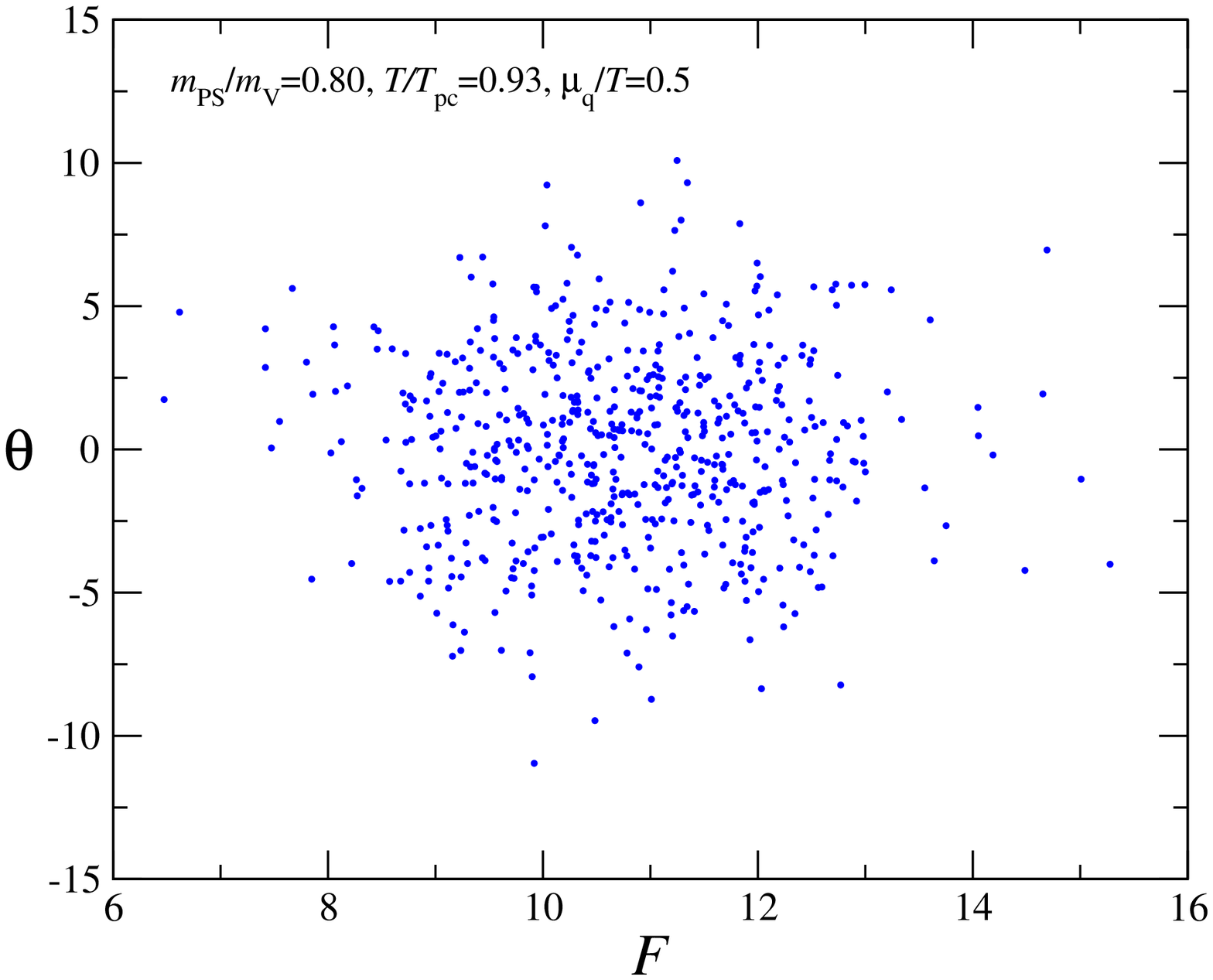}
\hskip 0.5cm
\includegraphics[width=2.4in]{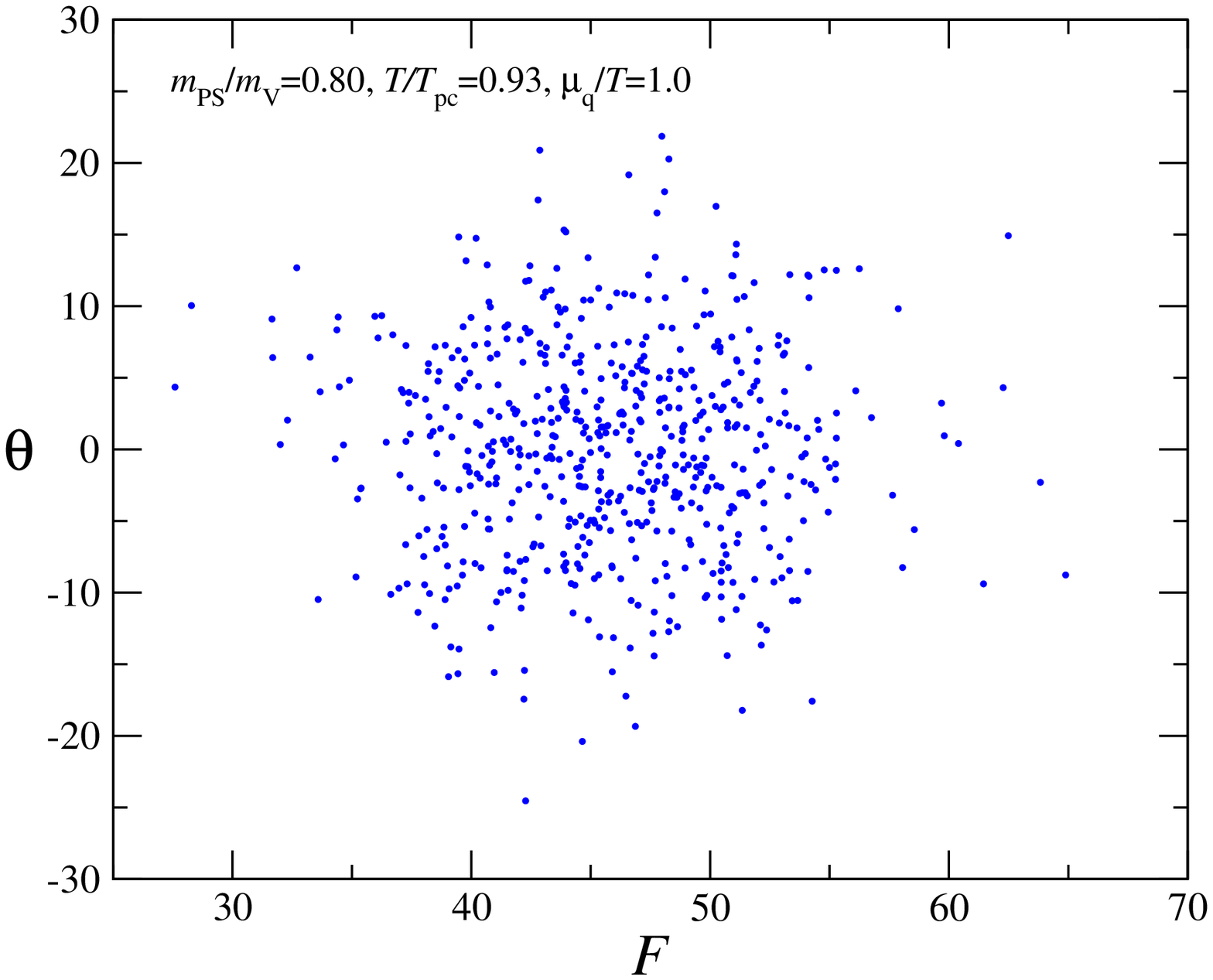}
\vskip -0.2cm
\caption{The distribution in the $(F, \theta)$ plane for 
$\mu_q/T=0.5$ (left) and $1.0$ (right) at $(m_{\rm PS}/m_{\rm V}, T/T_{pc})=(0.80, 0.93)$.}
\label{fig:thdis}
\end{center}
\vskip -0.3cm
\end{figure} 

%%\begin{figure}[t]
%%\begin{center}
%%\includegraphics[width=2.4in]{figureX/f2_080125.eps}
%%\vskip -0.2cm
%%\caption{$\langle (F - \langle F \rangle)^2 \rangle$ at 
%%$\mu_q/T=0.5$ and $1.0$.}
%%\label{fig:f2}
%%\end{center}
%%\vskip -0.3cm
%%%\setcounter{figure}{1}
%%\end{figure} 

\begin{figure}[t]
\begin{center}
\includegraphics[width=2.4in]{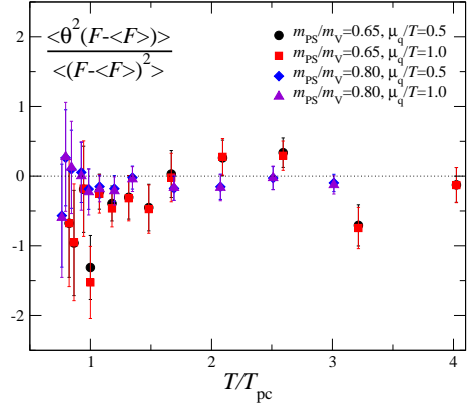}
\vskip -0.2cm
\caption{$\langle \theta^2 (F- \langle F \rangle) \rangle / 
\langle (F - \langle F \rangle)^2 \rangle$ for 
$\mu_q/T=0.5$ and $1.0$.}
\label{fig:thcor}
\end{center}
\vskip -0.3cm
\end{figure} 

We calculate the grand canonical potential (\ref{eq:pranpotap}) following the method of Ref.~\cite{eji07}.
We first rewrite the grand canonical partition function as follows. 
\begin{eqnarray}
{\cal Z}(T,\mu_q) 
= {\cal Z}(T,0) \left\langle \left( \frac{\det M(\mu)}{\det M(0)} 
\right)^{N_{\rm f}} \right\rangle_{(\mu_q =0)} 
\equiv {\cal Z}(T,0) \left\langle e^{F(\mu)} e^{i \theta(\mu)} \right\rangle_{(\mu_q =0)},
\label{eq:partition} 
\end{eqnarray}
where $F(\mu)$ and $\theta (\mu)$ are the real and imaginary parts of
$ N_{\rm f} \ln (\det M(\mu)/\det M(0))$, respectively, and
they can be calculated by the Taylor 
expansion in $\mu$.
Since odd (even) derivatives of $\ln (\det M(\mu)/\det M(0))$ are purely imaginary (real), we have 
\begin{eqnarray}
F (\mu) 
& \equiv & N_{\rm f} {\rm Re} \left[ \ln \left( \frac{\det M(\mu)}{\det M(0)} \right) \right]
\nonumber\\
& = & N_{\rm f} \sum_{n=1}^{\infty} \frac{1}{(2n)!} {\rm Re}
\left[ \frac{\partial^{2n} (\ln \det M)}{\partial \mu^{2n}} 
\right]_{(\mu=0)} \mu^{2n} 
= \sum_{n=1}^{\infty} \frac{1}{(2n)!} {\rm Re} {\cal D}_{2n} \mu^{2n} .
\label{eq:teabs} 
\end{eqnarray}
In this paper, we study terms up to $\mu^4$.
For the complex phase $\theta$, we have
\begin{eqnarray}
\theta (\mu) 
& = & N_{\rm f} {\rm Im} \,[\ln \det M(\mu)]
\nonumber\\
&=& N_{\rm f} \sum_{n=0}^{\infty} \frac{1}{(2n+1)!} {\rm Im}
\left[ \frac{\partial^{2n+1} (\ln \det M(\mu))}{\partial \mu^{2n+1}} 
\right]_{(\mu=0)} \mu^{2n+1} 
= \sum_{n=0}^{\infty} \frac{1}{(2n+1)!} {\rm Im} {\cal D}_{2n+1} \mu^{2n+1} ,
\label{eq:tatheta}
\end{eqnarray}
We note that $\ln \det M(\mu)$ is not uniquely defined for complex $\det M(\mu)$.
On the other hand, the $\mu$ derivatives of $\ln \det M(\mu)$ are unique.
We regard the Taylor expansion in (\ref{eq:tatheta}) as our definition of $\theta$.
Note that the $\theta$ thus defined is {\em NOT} restricted to be in the range $-\pi$ to $\pi$, and the maximum value of $|\theta|$ is infinite in the large volume limit. 
The principal value of $N_{\rm f} \ln \det M(\mu)$ is recovered by identifying $\theta+2n\pi$ with $\theta$ in the range $-\pi$ to $\pi$.

Histograms of $\theta$ are shown in Fig.~\ref{fig:phhis} for $\mu_q/T=0.5$ and $1.0$ 
at $(m_{\rm PS}/m_{\rm V}, T/T_{pc})=(0.65, 0.94)$ (top left), $(0.65, 1.32)$ (top right), 
$(0.80, 0.93)$ (bottom left) and $(0.80, 1.35)$ (bottom right). 
We find that the fluctuations in $\theta$ become larger as $\mu_q$ increases. 
Note that the width of the distribution is larger than $2 \pi$ at $T<T_{pc}$.
A large fluctuation in $\theta$ makes the calculation of $\ln {\cal Z}(T, \mu_q)$ difficult due to a rapid change of the factor $e^{i \theta}$. This is the origin of the sign problem.
On the other hand, these figures suggest that the distribution of $\theta$ defined in this way is almost of Gaussian. 
In Sec.\ref{sec:gaussphase}, we discuss that the Gaussian approximation corresponds to the leading order approximation of the cumulant expansion and confirm the validity of the Gaussian approximation.
This is a key observation to avoid the sign problem:
In a previous study with staggered quarks, using the fact that the $\theta$-distribution is well described by a Gaussian form, the $\theta$-averaging has been carried out. 
The resulting errors for observables turn out to be smaller than those with the naive averaging,
and thus the method may enable us to perform a reliable evaluation at a wider range of $\mu_q$ \cite{eji07}.

To implement this assumption, we define the distribution function $w(F, \theta)$ as
\begin{eqnarray}
w(\bar{F}, \bar{\theta}) \equiv 
\int {\cal D}U \; \delta (\bar{F}- F(\mu)) \; 
\delta (\bar{\theta}- \theta(\mu)) \; 
[\det M(0)]^{N_{\rm f}} e^{-S_g} = 
Z(T,0)\langle \delta (\bar{F}- F(\mu)) \; 
\delta (\bar{\theta} - \theta(\mu)) \rangle_{(\mu =0)}
\end{eqnarray}
where $\theta (\mu)$ and $F(\mu)$ are defined in Eq.~(\ref{eq:tatheta}) and Eq.~(\ref{eq:teabs}).
Note that $w(F,\theta)$ depend implicitly on $\mu$.
Figure \ref{fig:thdis} shows a typical distribution of $(F, \theta)$ at $(m_{\rm PS}/m_{\rm V}, T/T_{pc})=(0.80, 0.93)$.  
The Gaussian $\theta$-distribution means that 
\begin{eqnarray}
w(F, \theta) \approx \sqrt{\frac{a_2(F)}{\pi}} 
w_0(F) e^{-a_2(F) \theta^2} .
\end{eqnarray}
With this form,  it is easy to carry out the $\theta$-integration as follows.
\begin{eqnarray}
{\cal Z}(T,\mu) & = &
\int dF \int d \theta \ w(F, \theta) \ e^F e^{i \theta} 
% \nonumber \\ 
\; \approx \; \int dF \int d \theta \ e^F w_0(F)
\sqrt{\frac{a_2(F)}{\pi}} 
e^{i \theta} e^{-a_2(F) \theta^2 } 
\nonumber \\
& = & \int dF \ e^F w_0(F) e^{ -1/(4a_2(F))}
\; = \; 
%\int {\cal D}U \ e^F e^{ -1/(4a_2)} \left( \det M(0) \right)^{N_{\rm f}}e^{-S_g} \nonumber \\
%&=& 
{\cal Z}(T,0) \left\langle e^{F(\mu)} 
e^{ -1/(4a_2(F(\mu))} \right\rangle_{(\mu=0)}. 
\label{eq:parap}
\end{eqnarray}
In the last line we use the fact that
\begin{eqnarray}
w_0(\bar{F}) = \int {\cal D}U \; \delta (\bar{F}- F(\mu)) \; 
  [\det M(0)]^{N_{\rm f}} e^{-S_g} 
= Z(T,0) \langle \delta (\bar{F}- F(\mu))\rangle_{(\mu =0)}
\end{eqnarray}
holds within this assumption.
%This $\theta$-integration can be safely carried out for the width up to $O(2\pi)$.
Note that the problematic factor $e^{i\theta}$ in Eq.~(\ref{eq:partition}) is now replaced by a positive definite factor $e^{-1/(4a_2)}$.
Thus the statistical error of Eq.~(\ref{eq:parap}) is always smaller than its expectation value, i.e. there is no sign problem. 

Of course, one may replace the Gaussian distribution function $w(F, \theta)$ with 
a periodic distribution function given by 
\begin{eqnarray}
\lim_{N \to \infty} \frac{1}{2N+1} \sum_{n=-N}^{N} w(F, \theta +2 \pi n).
\end{eqnarray}
However, the integral of $e^{i \theta}$ does not change simply because 
$\int e^{i \theta} w(F, \theta +2 \pi n) d \theta$ gives the same answer as
$\int e^{i \theta} w(F, \theta) d \theta$. 
Hence, the absence of the periodicity of $2\pi$ in $w(F, \theta)$ 
is not a problem for the integral of $e^{i \theta}$.

The validity of this method can be discussed more precisely based on the Taylor expansion of the partition function at least in the low density region. 
In Appendix \ref{ap:gaussian} we compare the 
derivatives of $\ln {\cal Z}$ in the Gaussian approximation 
with the exact calculations up to $O(\mu_q^4)$. We find that 
the Gaussian approximation does not affect up to $O(\mu_q^2)$. 
At the fourth order in $\mu_q$,  
$\left\langle {\cal D}_1^4 \right\rangle$ 
of Eq.~(\ref{eq:dermu}) is replaced by 
$3\left\langle {\cal D}_1^2 \right\rangle^2$ in the Gaussian case. 
%Since ${\cal D}_1$ is proportional to the total quark number density at 
%$\mu_q=0$, which is expected to be non-singular for two-flavor QCD with finite quark mass, the histogram of ${\cal D}_1$ should also be of Gaussian in the sufficiently large volume.
%This implies $\left\langle {\cal D}_1^4 \right\rangle= 3\left\langle {\cal D}_1^2 \right\rangle^2$, so that the Gaussian approximation Eq.~(\ref{eq:parap}) is valid. See Appendix \ref{ap:gaussian} for more details.
In Ref.~\cite{eji08}, the effects caused by deviations from the Gaussian distribution in $w(F, \theta)$ are estimated 
assuming $w(F, \theta) \sim \exp[ -a_2 \theta^2 -a_4 \theta^4 ]$. 
It turned out that the additional term $a_4$ 
does not affect the terms up to $\mu_q^4$ as far as $a_4/a_2 \leq O(1)$. 

Now the problem is reduced to a determination of the coefficient $a_2(F)$: 
\begin{eqnarray}
\frac{1}{2a_2(\bar{F})} =
\left\langle \theta^2 \right\rangle_{\bar{F}} \equiv 
\frac{\left\langle \theta^2(\mu) \delta (\bar{F}- F(\mu))
\right\rangle_{(\mu=0)} }{ \left\langle \delta (\bar{F}- F(\mu))
\right\rangle_{(\mu=0)} }=
\frac{ \int {\cal D}U  \ \theta^2(\mu) \ 
\delta (\bar{F}- F(\mu)) (\det M(0))^{N_{\rm f}} e^{-S_g} }{
\int {\cal D}U \ \delta (\bar{F}- F(\mu))
(\det M(0))^{N_{\rm f}} e^{-S_g}}.
\label{eq:a2}
\end{eqnarray}
The distribution shown in Fig.~\ref{fig:thdis} suggests that the $F$-dependence in $\langle \theta^2 \rangle_{F}$ is mild.
Unfortunately, the limitation of the statistics makes a precise evaluation of $\langle \theta^2 \rangle_{F}$ for each thin slices of $F$ difficult.
However, when we restrict ourselves to calculate the equation of state up to $O(\mu_q^4)$, we only need to evaluate the first derivative of $\langle \theta^2 \rangle_F$ in terms of $F$:
Because ${\cal D}_1$ and ${\cal D}_2$ represent the leading $\mu_q$-dependence of $\theta$ and $F$, respectively, consulting Eq.~(\ref{eq:AB}), we note that the $F$-dependence of $\langle \theta^2 \rangle_{F}$ affects only in the $\langle {\cal D}_2 {\cal D}_1^2 \rangle$ term for the $O(\mu_q^4)$ coefficients $c_4$ and $c_4^I$. 
(See Appendix \ref{ap:gaussian} too.)
This quantity, i.e. the $O(\mu_q^4)$ contribution of $\langle F \theta^2 \rangle$, corresponds to the first derivative of $\langle \theta^2 \rangle_F$ because
\begin{eqnarray} 
\langle \theta^2(\mu) (F(\mu)- \langle F \rangle ) \rangle_{(\mu=0)} 
&=& \int \left[ \langle \theta^2 \rangle_{\langle F \rangle} 
(F- \langle F \rangle ) 
+ \left[\frac{d \langle \theta^2 \rangle_F}{dF}\right]_{\langle F \rangle} 
(F- \langle F \rangle )^2  + \cdots \right ] \frac{w_0(F)}{{\cal Z}(T,0)}dF \nonumber \\
&\approx& \left[\frac{d \langle \theta^2 \rangle_F}{dF}\right]_{\langle F \rangle} 
\langle (F- \langle F \rangle )^2 \rangle_{(\mu=0)},
\end{eqnarray} 
when the $F$-dependence in $\langle \theta^2 \rangle_F$ is mild. 
Using this relation,
we then estimate the first derivative of $\langle \theta^2 \rangle_F$ with respect to $F$ as
\begin{eqnarray} 
\left[\frac{d \langle \theta^2 \rangle_F}{dF}\right]_{\langle F \rangle} 
\approx \frac{\langle \theta^2 (F- \langle F \rangle ) \rangle }{ 
\langle (F- \langle F \rangle )^2 \rangle} ,
\end{eqnarray} 
which is shown in Fig.~\ref{fig:thcor}.
We find that $[d \langle \theta^2 \rangle_F/dF]_{\langle F \rangle} $ is actually smaller than statistical errors, so that $\langle \theta^2\rangle_F \simeq \langle \theta^2\rangle_{\langle F\rangle} $ is a good approximation.
This point is also suggested in chiral perturbation theory \cite{Lomb09}.
To include the small $F$ dependence of $a_2(F)$,
we assume a simple ansatz function: 
\begin{eqnarray}
\frac{1}{2a_2(F)}&=&\langle \theta^2\rangle_F= f(F)=\exp[x_1 + x_2 F] ,
\end{eqnarray}
where we take into account the fact that $\theta^2$ is positive for all $F$.
The two parameters are sufficient for the exact calculation up to $O(\mu^4)$. 
We thus determine fit parameters $x_1$ and $x_2$, by minimizing $\chi^2 \equiv \sum_i [ \theta^2_i -f(F_i) ]^2$, where the summation is taken over configurations.

Finally, we integrate over $F$.
The factor $e^F$ in Eq.~(\ref{eq:parap}) is a potential danger in the integration because it can easily shift the central contribution for the average to a statistically poor region of $F$.
This will be the case when $\mu_q$ is not small ($\langle F \rangle$ is not small).
At small $\mu$, this problem can be removed in part by a reweighting in the $\beta$-direction of the coupling parameter space such that the fluctuation in $e^{F(\mu)}$ is compensated by that in the gauge action. This is possible
since $F$ is strongly correlated with $P=-S_g/(6 N_{\rm site} \beta)$, where the gauge action $S_g$ is defined in Eq.~(\ref{eq:action}), and 
$N_{\rm site} = N_s^3 \times N_t$.
By reweighting, the expectation value of an operator ${\cal O}$ at $\beta$ is 
evaluated from a simulation at $\beta_0$ as
\begin{eqnarray}
\left\langle {\cal O} \right\rangle_{(\beta, \mu=0)}
=\frac{ \left\langle {\cal O}(P) e^{6 N_{\rm site} (\beta-\beta_0)P} 
\right\rangle_{\beta_0} }{ \left\langle 
e^{6 N_{\rm site} (\beta-\beta_0)P} \right\rangle_{\beta_0}}.
\end{eqnarray}
To calculate $\langle e^{F(\mu)} e^{-1/(4a_2(F))} \rangle$, we adjust $\beta$ such that the value of $e^F e^{-1/(4a_2)} e^{6 N_{\rm site} (\beta- \beta_0) P}$ is stabilized during the Monte Carlo steps.
%the center of the distribution of 
%$e^F e^{-1/(4a_2)} e^{6 N_{\rm site} (\beta- \beta_0) P}$ remains the same. 
In practice, since $e^{F(\mu)} e^{-1/(4a_2(F))}=1$ at $\mu_q=0$, we start with $\beta=\beta_0$ at $\mu_q=0$ and find $\beta$ for finite $\mu_q$ at which the fluctuation of $e^F e^{-1/(4a_2)} e^{6 N_{\rm site} (\beta- \beta_0) P} \equiv X$,
\begin{eqnarray}
\left. \left\langle \left( X - 
\langle X \rangle_{(\mu=0)} \right)^2 
\right\rangle_{(\mu=0)} \right/
\left\langle X \right\rangle^2_{(\mu =0)} ,
%&&
%\left. \left\langle \left(e^{F(\mu)} e^{-1/(4a_2(F))} 
%e^{6 N_{\rm site} (\beta -\beta_0) P} - 
%\left\langle e^{F(\mu)} e^{-1/(4a_2(F))} 
%e^{6 N_{\rm site} (\beta -\beta_0) P} \right\rangle_{(\mu=0)} \right)^2 
%\right\rangle_{(\mu=0)} \right/
%\nonumber \\
%&&  \left\langle e^{F(\mu)} e^{-1/(4a_2(F))} 
%e^{6 N_{\rm site} (\beta -\beta_0) P} \right\rangle^2_{(\mu =0)} ,
\end{eqnarray}
is minimized.
Since $F$ becomes larger for larger $P$, $\beta < \beta_0$. 
The resulting shift in $\beta$ is translated to the temperature scale using a cubic spline interpolation of the temperature data. 
Because we do not shift the hopping parameter, a shift in $\beta$ leads to a slight deviation from the original line of constant physics (LCP). 
In our study, however, the shifts in $\beta$ turn out to be smaller than $0.03$. 
Since these shifts are negligible in Fig.~\ref{fig:simpara}, we disregard the resulting small deviation from the LCP, and simply translate the shifts in $\beta$ to shifts in $T$ for the final plots.

To conclude we summarize the final formulae:
\begin{eqnarray}
\frac{ {\cal Z}(T,\mu)}{{\cal Z}(T,0)} &=&
\frac{ \langle e^{F(\mu)}  e^{- \langle \theta^2 \rangle_F /2} e^{6 N_{\rm site} (\beta -\beta_0) P} \rangle_{\beta_0}}
{ \langle  e^{6 N_{\rm site} (\beta -\beta_0) P} \rangle_{\beta_0}},
\quad
\langle \theta^2 \rangle_F = \exp (x_1+x_2 F) .
\label{eq:gaps}
\end{eqnarray}

%%%%%%%%%%%%%%%%%%%%%%%%%%%%%%%%%%%%%%%%%%%%%%%%%%%%
\subsection{Gaussian approximation as the lowest order approximation of cumulant expansion} 
\label{sec:gaussphase}

The only difference between the Gaussian approximation (\ref{eq:gaps}) and its exact 
formula is the replacement of $\langle \exp(i \theta) \rangle_F$ by 
$\exp[ -\langle \theta^2 \rangle_F /2 ]$.
The meaning of the replacement can be understood in the context of the cumulant expansion,
\begin{eqnarray}
\langle \exp{i \theta} \rangle_F = 
\exp \left[i \left\langle \theta \right\rangle_c
- \frac{1}{2} \langle \theta^2 \rangle_c
- \frac{i}{3!} \left\langle \theta^3 \right\rangle_c
+ \frac{1}{4!} \langle \theta^4 \rangle_c 
+ \frac{i}{5!} \langle \theta^5 \rangle_c 
- \frac{1}{6!} \langle \theta^6 \rangle_c + \cdots \right],
\label{eq:cum}
\end{eqnarray}
where $\langle \theta^n \rangle_c$ is the $n^{\rm th}$ order cumulant, e.g.
\begin{eqnarray}
\left\langle \theta^2 \right\rangle_c 
&=& \left\langle \theta^2 \right\rangle_F , 
\hspace{3mm}
\left\langle \theta^4 \right\rangle_c 
= \left\langle \theta^4 \right\rangle_F
-3 \left\langle \theta^2 \right\rangle_F^2, 
\hspace{3mm}
\left\langle \theta^6 \right\rangle_c 
= \left\langle \theta^6 \right\rangle_F
-15 \left\langle \theta^4 \right\rangle_F 
\left\langle \theta^2 \right\rangle_F 
+30 \left\langle \theta^2 \right\rangle_F^3 .
\end{eqnarray}
Note that $\langle \theta^n \rangle_c =0$ for odd $n$
due to the symmetry under $\theta \rightarrow -\theta$. 
Because only the odd-order cumulants are the source of the complex phase in 
$\langle \exp(i \theta) \rangle_F$, the value of 
$\langle \exp(i \theta) \rangle_F$ is guaranteed to be real and positive 
from this symmetry if the cumulant expansion converges. 
There is thus no source of the sign problem once we eliminate the odd terms.
%The sign problem is thus solved by eliminating the odd terms.

When the distribution of $\theta$ is of Gaussian, 
the $O(\theta^n)$ terms vanish for $n >2$ in Eq.~(\ref{eq:cum}). 
Hence, the Gaussian approximation is equivalent to the approximation 
that the higher order cumulants are neglected except for the first 
nonzero term.
If one wants to improve the Gaussian approximation, it is achieved 
by adding higher order terms.

Moreover, the cumulant expansion can be regarded as a power expansion 
in terms of $\mu_q$ because $\theta \sim O(\mu_q)$. 
Therefore, if we take into account the cumulants up to the $n^{\rm th}$ order, 
the truncation error does not affect the Taylor expansion up to $O(\mu_q^n)$.
The Gaussian approximation corresponds to the leading non-trivial 
order approximation of the Taylor expansion in $\mu_q$.

On the other hand, a careful discussion about the infinite volume 
$(V)$ limit is required.
Because the operator $\theta$ is roughly proportional to $V$, 
the $n^{\rm th}$ order cumulant $\langle \theta^n \rangle_c$ may 
increase as $O(V^n)$ naively.
In such a case, the cumulant expansion does not converge at large $V$.
%To suppress the higher order cumulants, the distribution of $\theta$ 
%is required to be of Gaussian approximately. 
However, the following argument suggests that the convergence property of the cumulant expansion is independent of the volume when the correlation length of the system is finite. 
Note that, since no critical point is expected to exist in two-flavor QCD 
at $m_q > 0$ and $\mu_q=0$, the correlation length between quarks is finite.

The expansion coefficients of $\theta$ in Eq.~(\ref{eq:tatheta}) are 
given by combinations of traces of products of $M^{-1}$, 
$\partial^n M/ \partial (\mu_q/T)^n$ and so on.
For example, ${\cal D}_1$ is given by the trace of 
$N_f [M^{-1} (\partial M/ \partial (\mu_q/T))]$ 
and the diagonal element of this matrix is the local quark number density
operator $(\sim \bar{\psi} \gamma_0 \psi(x))$ at $\mu_q=0$. 
If the correlation length of the local number density operator is much shorter than the system size, we may decompose ${\cal D}_1$ into independent contributions from spatially separated regions. 
The same discussion can be applied to higher order coefficients ${\cal D}_n$ too. 

In this case, one can write the phase as $\theta=\sum_x \theta_x$, 
where $\theta_x$ is the contribution from a spatial region labeled by $x$ 
and these contributions are independent.
The average of $\exp(i\theta)$ is thus 
\begin{eqnarray}
\left\langle e^{i\theta} \right\rangle 
\approx \prod_x \left\langle e^{i\theta_x} \right\rangle 
= \exp \left( \sum_x \sum_n \frac{i^n}{n!} 
\left\langle \theta_x^n \right\rangle_c \right).
\end{eqnarray}
This equation suggests that all cumulants $\langle \theta^n \rangle_c 
\approx \sum_x \left\langle \theta_x^n \right\rangle_c$ 
increase in proportion to the volume as the volume increases. 
Therefore, while the width of the distribution, i.e. the phase fluctuation, 
increases in proportion to the volume, 
the ratios of the cumulants are independent of the volume.
The higher order terms in the cumulant expansion are well under control in the large volume limit.

Because $\theta$ is $O(\mu_q)$ and $\langle \theta^n \rangle_c$ is $O(\mu_q^n)$, 
the Gaussian approximation is valid at small $\mu_q$ and 
the higher order cumulants will become visible at large $\mu_q$. 
The application range of the Gaussian approximation in terms of $\mu_q$ must 
be checked for each analysis by calculating the ratio of cumulants. 
However, the volume-dependence of the ratios of cumulants suggests that the application 
range does not change once the system size becomes larger than the correlation length.
This means that the qualification of the Gaussian distribution on a small lattice is enough to verify the Gaussian approximation.

\begin{figure}[t]
\begin{center}
\includegraphics[width=2.4in]{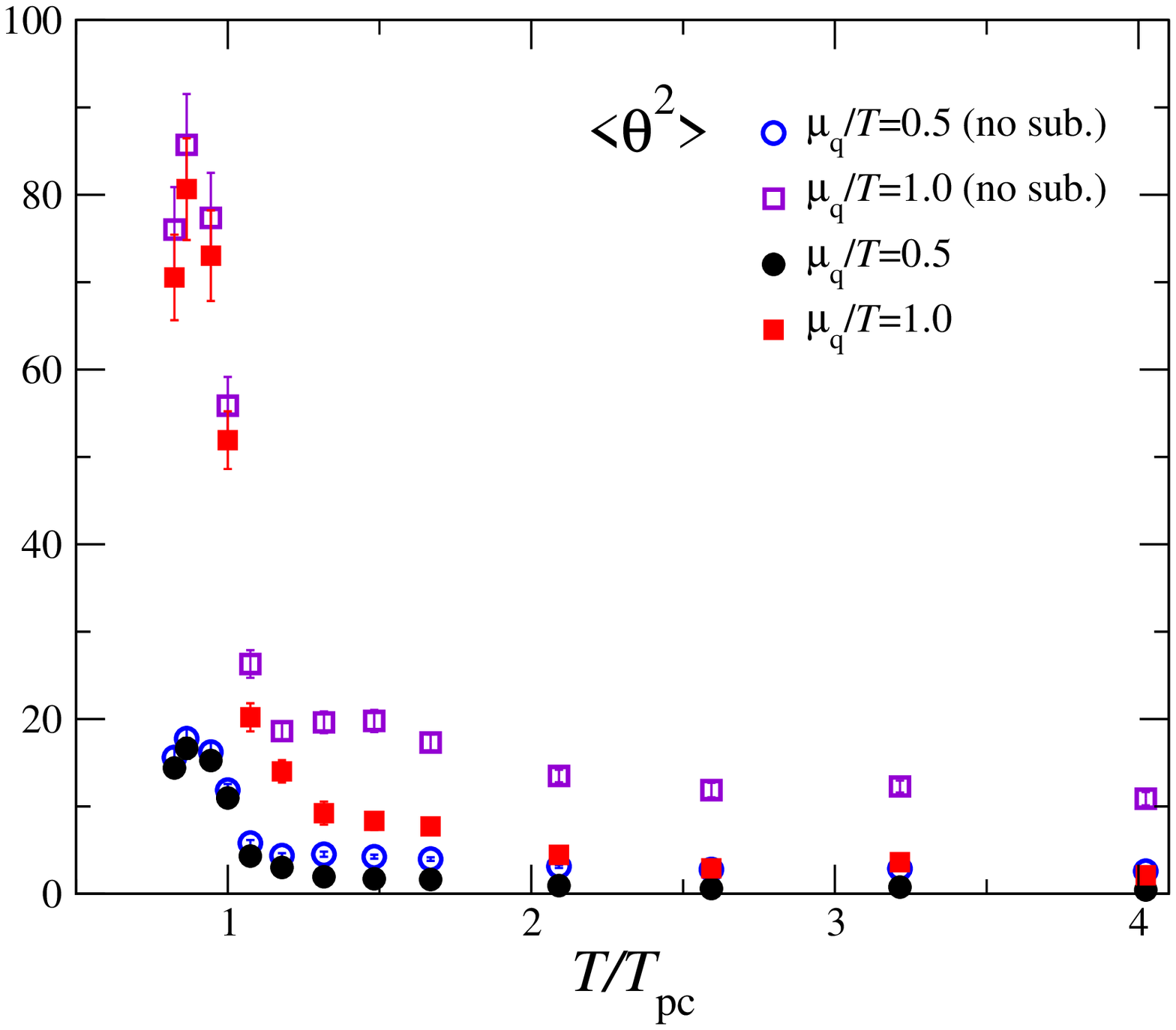}
\hskip 0.5cm
\includegraphics[width=2.4in]{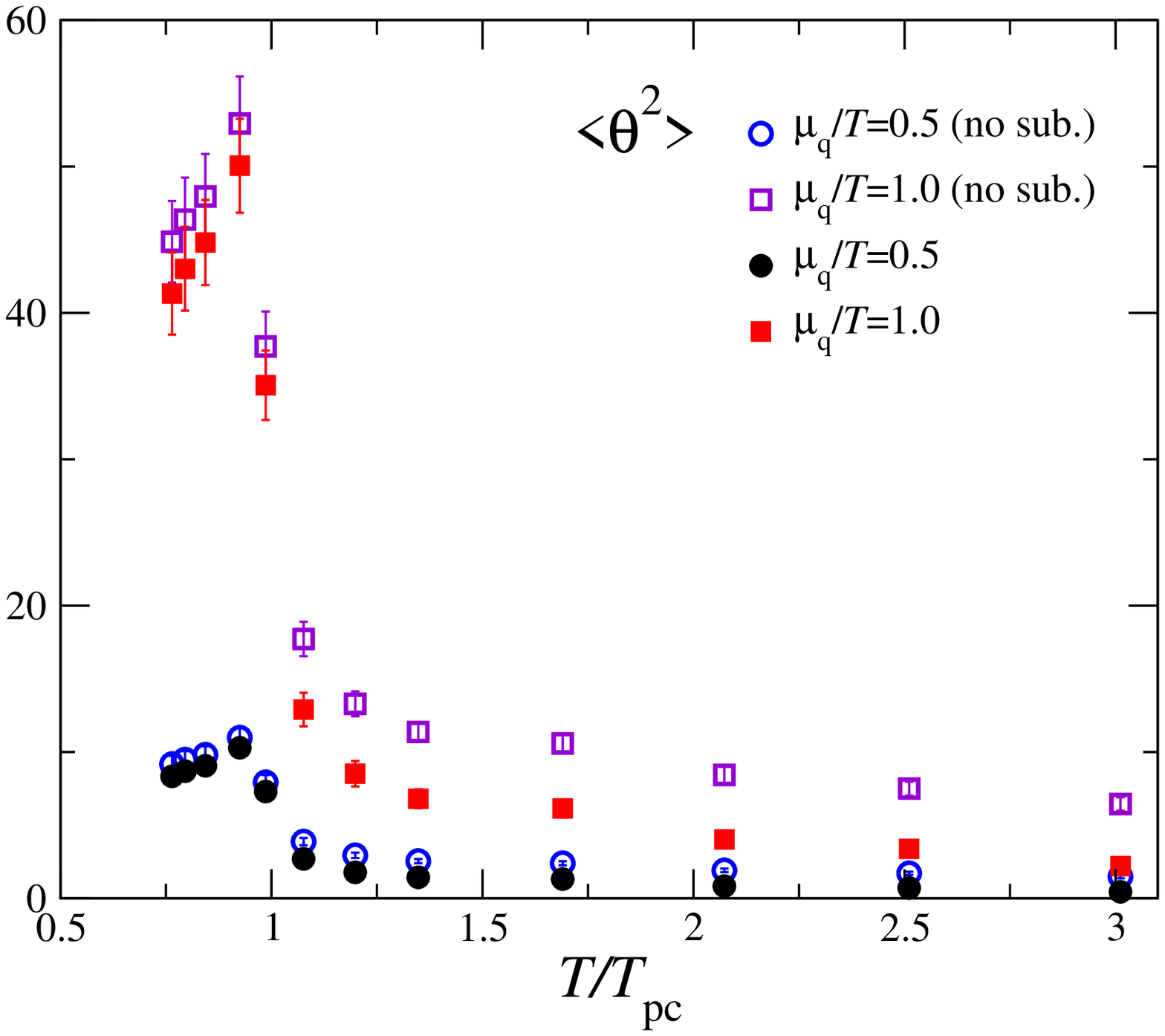}
\vskip -0.2cm
\caption{The expectation value of $\theta^2$ for each temperature 
with $m_{\rm PS}/m_{\rm V}=0.65$ (left) and $0.80$ (right).}
\label{fig:th2}
\end{center}
\vskip -0.3cm
\end{figure} 

\begin{figure}[t]
\begin{center}
\includegraphics[width=2.4in]{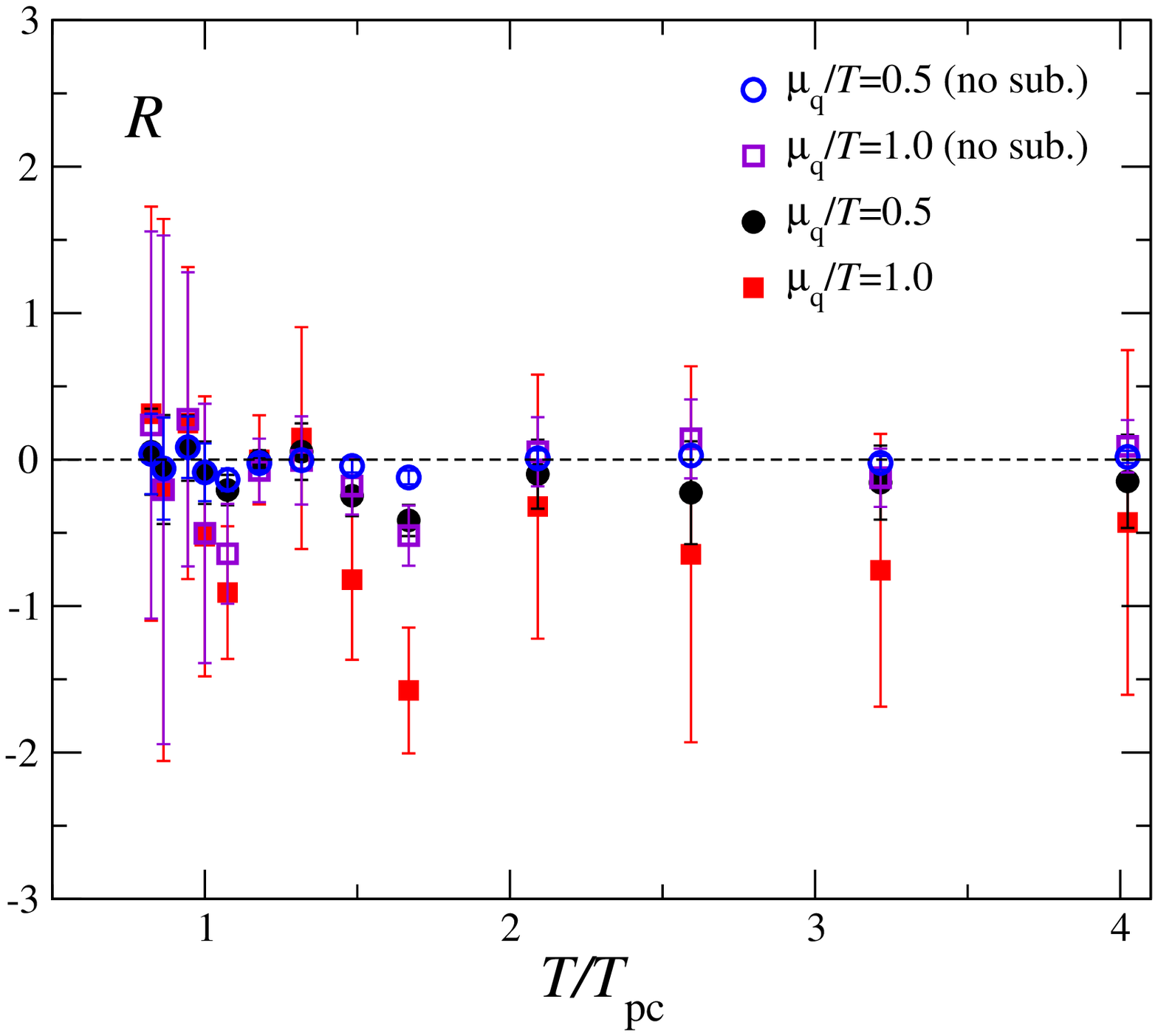}
\hskip 0.5cm
\includegraphics[width=2.4in]{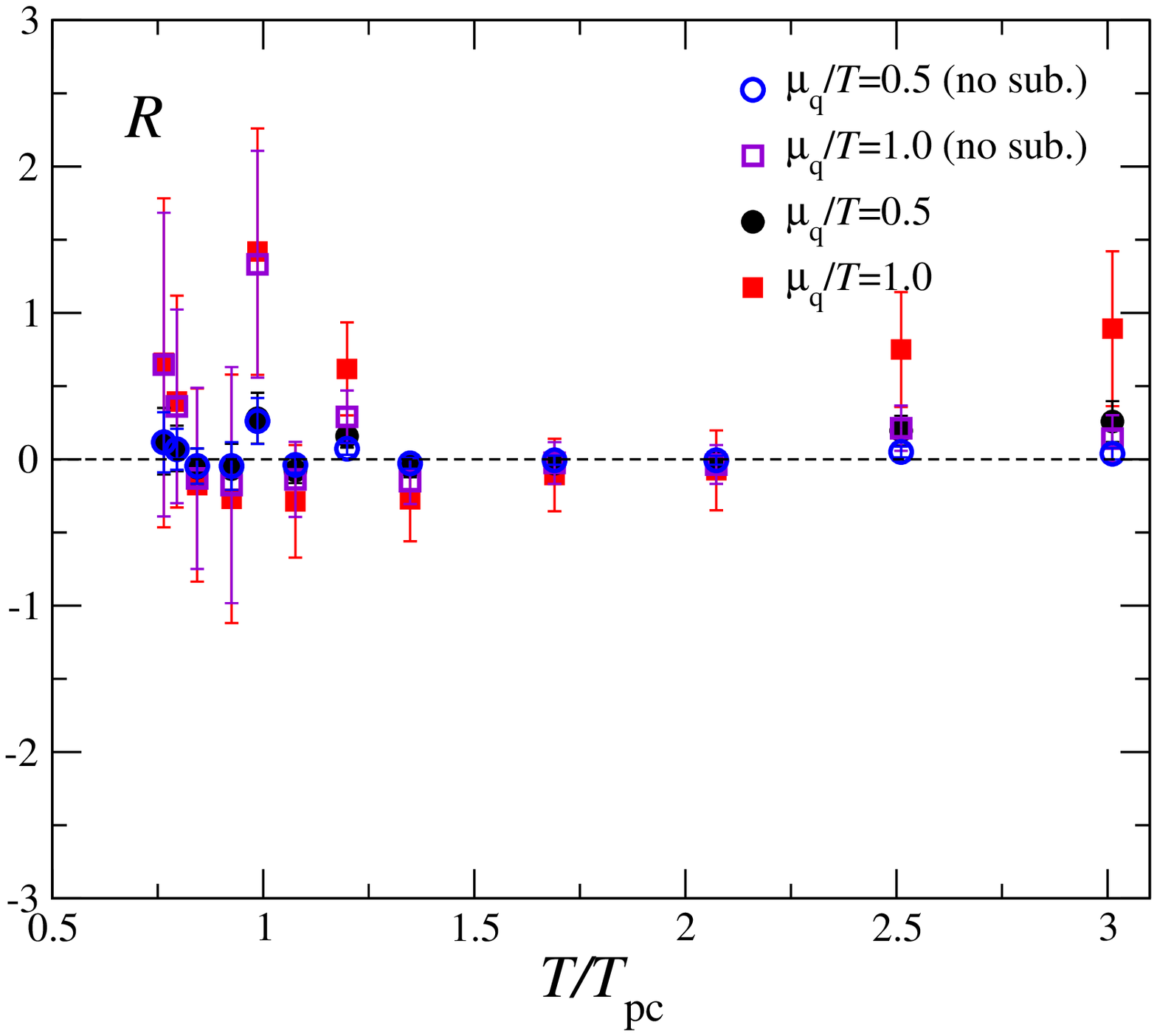}
\vskip -0.2cm
\caption{The relative magnitude of the fourth order cumulant contribution 
to the leading order contribution as a function of the temperature 
for $m_{\rm PS}/m_{\rm V}=0.65$ (left) and $0.80$ (right).}
\label{fig:c4th}
\end{center}
\vskip -0.3cm
\end{figure} 

We study the validity of the Gaussian approximation %for the $\theta$ distribution
by examining the relative magnitude of the fourth order cumulant contribution to the leading order contribution in Eq.(\ref{eq:cum}):
\begin{eqnarray}
R \equiv 
\left(\frac{1}{4!} \langle \theta^4 \rangle_c \right)
\left/
\left(\frac{1}{2} \langle \theta^2 \rangle_c \right) \right.
=
\frac{\langle \theta^4 \rangle_c}{12\langle \theta^2 \rangle_c}
\label{eq:c4theta}
\end{eqnarray}
The Gaussian approximation is valid if $R \ll O(1)$ is satisfied.
In this paper, we will check whether $R$ is consistent with zero,
which is a less stringent condition when the statistical error is large.

Here, we note a caveat in the evaluation of 
$\left\langle \theta^2 \right\rangle$ from the histogram. 
Because we calculate $\theta$ using the random noise method, 
the fluctuation of $\theta$ contains a contribution due to the finite 
number of noise vectors $(N_{\rm noise})$. 
This makes the width of the $\theta$ histogram wider than that of 
the true distribution. 
True width is given by $\sqrt{\langle \theta^2 \rangle }$ in the limit of 
large $N_{\rm noise}$. 
To reduce the errors in $\langle \theta^2 \rangle$ due to finite 
$N_{\rm noise}$, we adopt the subtraction method discussed in 
Sec.~\ref{sec:rnm} for the calculation of products of traces. 
The expectation value of $\theta^2$ is summarized in Fig.~\ref{fig:th2}. 
Filled symbols in Fig.~\ref{fig:th2} are the results of the subtraction method. 
We have checked that the $N_{\rm noise}$-dependence in 
$\langle \theta^2 \rangle$ is negligible with our choices of $N_{\rm noise}$.
We find that $\langle \theta^2 \rangle$ becomes 
larger than $O(\pi^2)$ from $\mu_q/T \sim 0.5$ in the low temperature phase 
while, in the high temperature phase, the complex phase fluctuations 
decrease as $T$ increases, in accordance with our expectation that 
the quark determinant is real in the high temperature limit. 
On the other hand, the width of the histogram shown in Fig.~\ref{fig:phhis} 
corresponds to $\sqrt{\langle \theta^2 \rangle}$ obtained by the naive 
calculation without subtraction, 
which is plotted with open symbols in Fig.~\ref{fig:th2}.
The difference between the results by the subtraction and naive methods 
decreases as $N_{\rm noise}$ increases but is almost the same size 
for all temperatures, and the error due to 
finite $N_{\rm noise}$ is larger than the expectation value of 
$\langle \theta^2 \rangle$ at high temperature.
Therefore, the subtraction is indispensable for 
a calculation of the width of the $\theta$ distribution.

We summarize the results for $R$ in Fig.~\ref{fig:c4th}. 
The circle and square symbols are the results for $\mu_q/T$ = 0.5 and 1.0, 
respectively. Filled symbols are the results of the subtraction method, 
while open symbols are the results of naive calculations without 
the subtraction. 
Although errors become gradually larger as $\mu_q/T$ increases and 
are as large as $O(1)$ for $\mu_q/T=1.0$, the central values 
of $R$ are consistent with zero for all temperatures and $\mu_q/T$ \footnote{
Because the complex phase vanishes in the high temperature limit, 
$\langle \theta^2 \rangle$ becomes smaller as $T$ increases. 
The small $\langle \theta^2 \rangle$ causes the large statistical 
error of $R$ at large $T$ for the subtraction method. 
Where $\langle\theta^2\rangle$ is small, however, the correction due to 
the phase fluctuation itself is small, and thus a deviation from 
the Gaussian approximation does not affect the results.}. 
However, we need higher statistics to identify the 
actual magnitude of $R$ and to check the validity 
range of the Gaussian approximation in terms of $\mu_q/T$, which is left 
for future investigations.

%%%%%%%%%%%%%%%%%%%%%%%%%%%%%%%%%%%%%%%%%%%%%%%%%%%%
\subsection{Results for the equation of state and quark number susceptibility} 
\label{sec:eosmurew}

\begin{figure}[t]
\begin{center}
\includegraphics[width=2.4in]{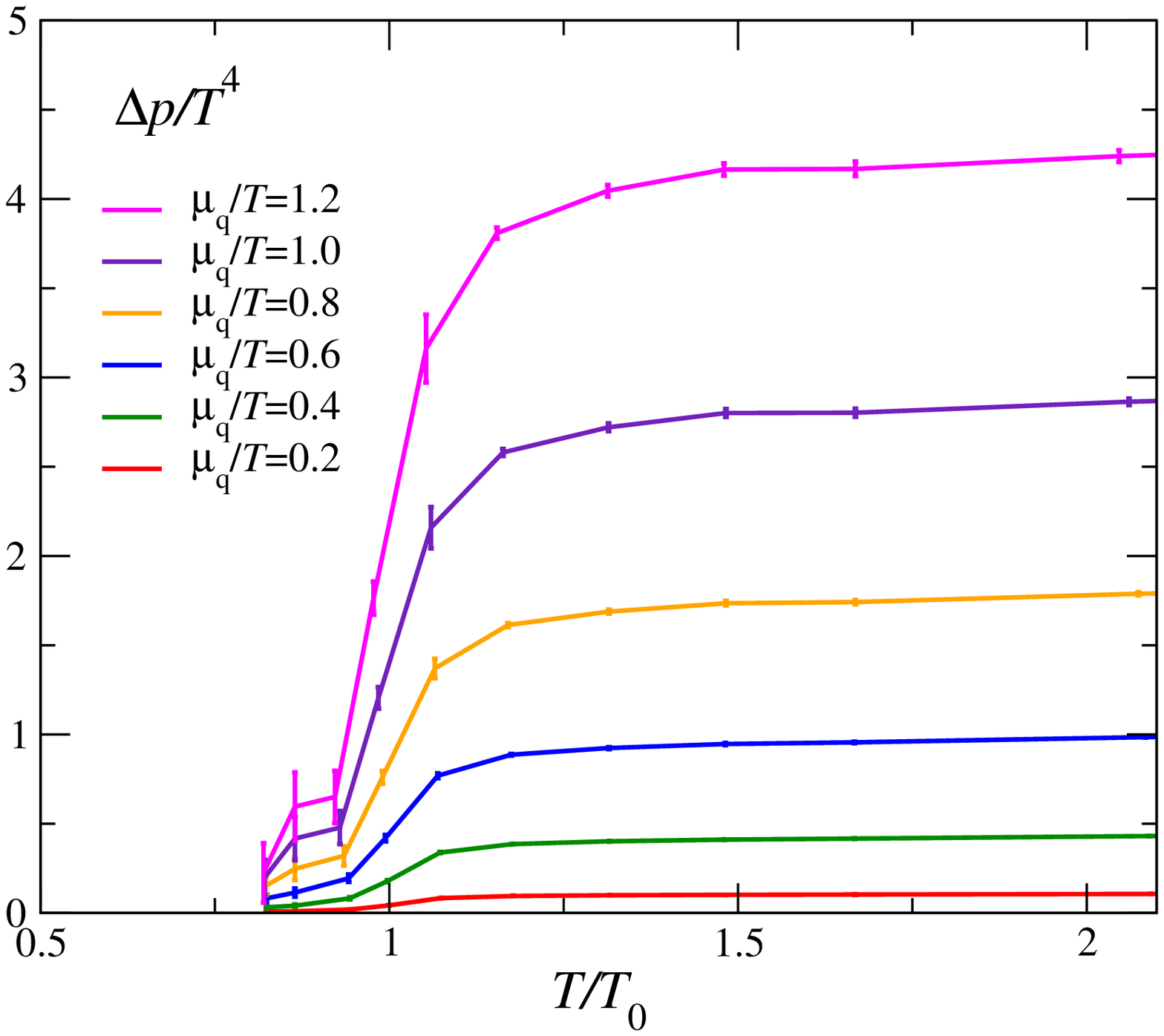}
\hskip 0.5cm
\includegraphics[width=2.4in]{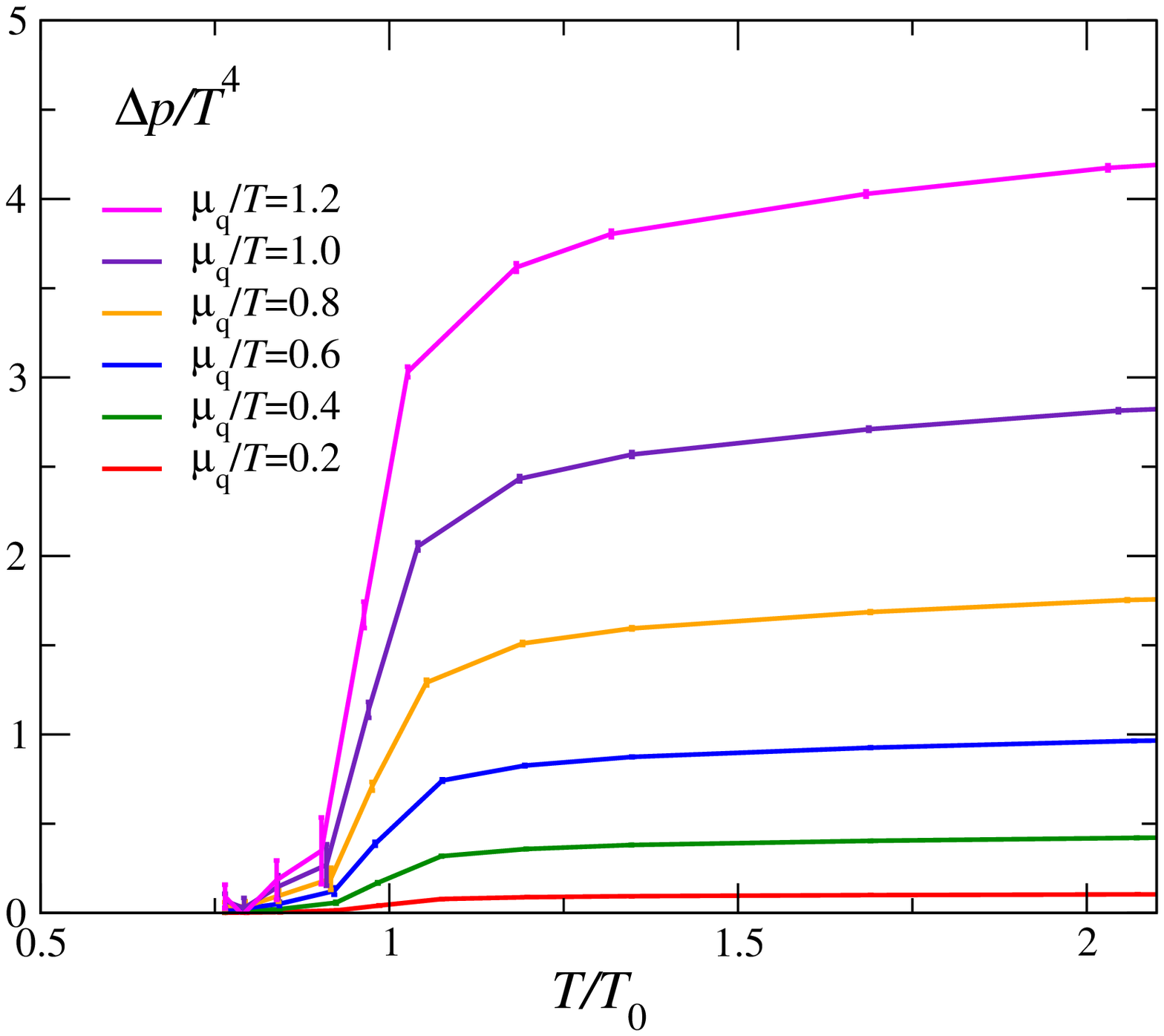}
\vskip -0.2cm
\caption{$\mu_q$-dependent contribution to pressure as a function of 
$T/T_0$ for each $\mu_q/T$ with $m_{\rm PS}/m_{\rm V}=0.65$ (left) and $0.80$ (right), 
where $T_0$ is $T_{pc}$ at $\mu_{q}=0$.}
\label{fig:prsrew}
\end{center}
\vskip -0.3cm
\end{figure} 

\begin{figure}[t]
\begin{center}
\includegraphics[width=2.4in]{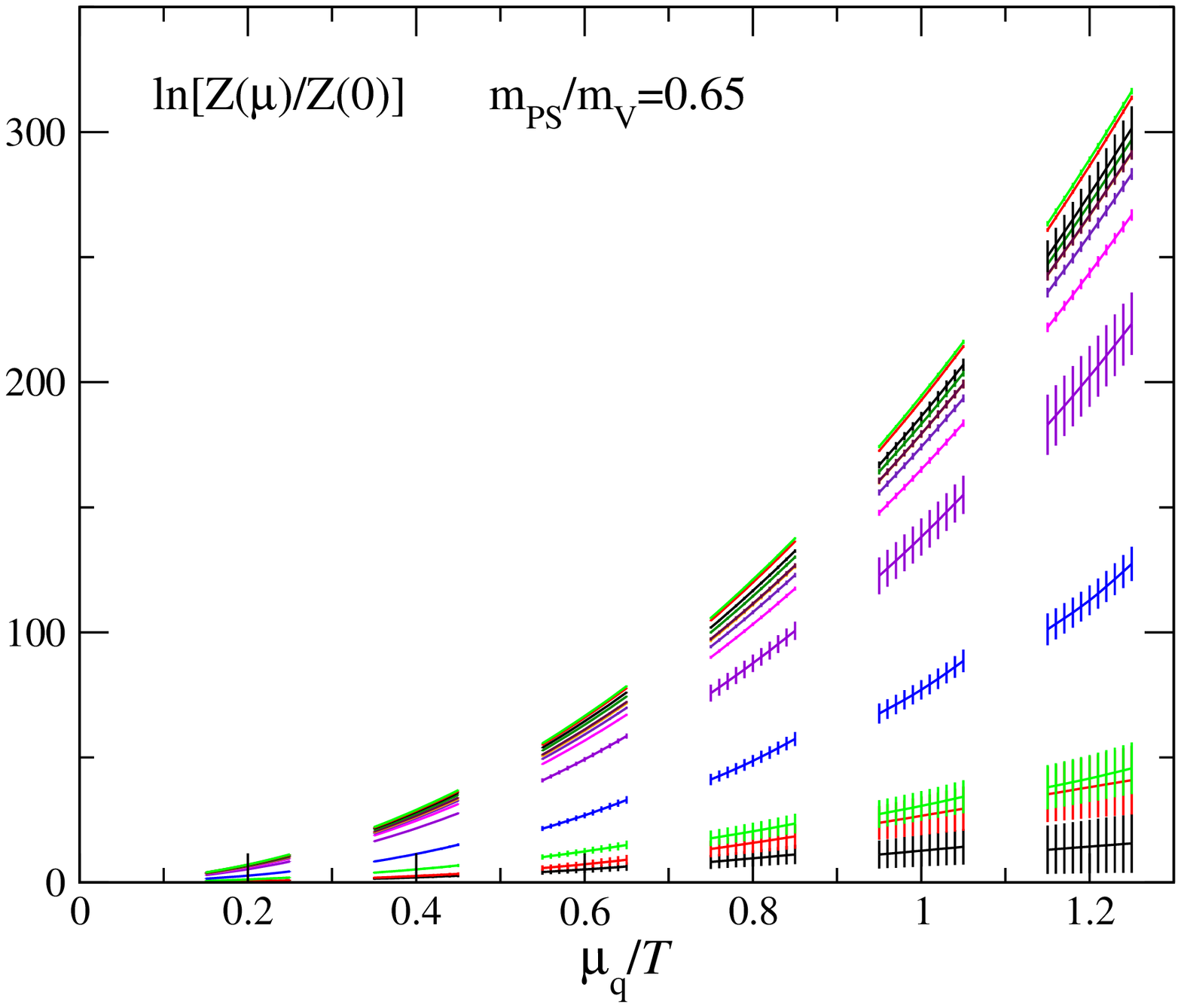}
\hskip 0.5cm
\includegraphics[width=2.4in]{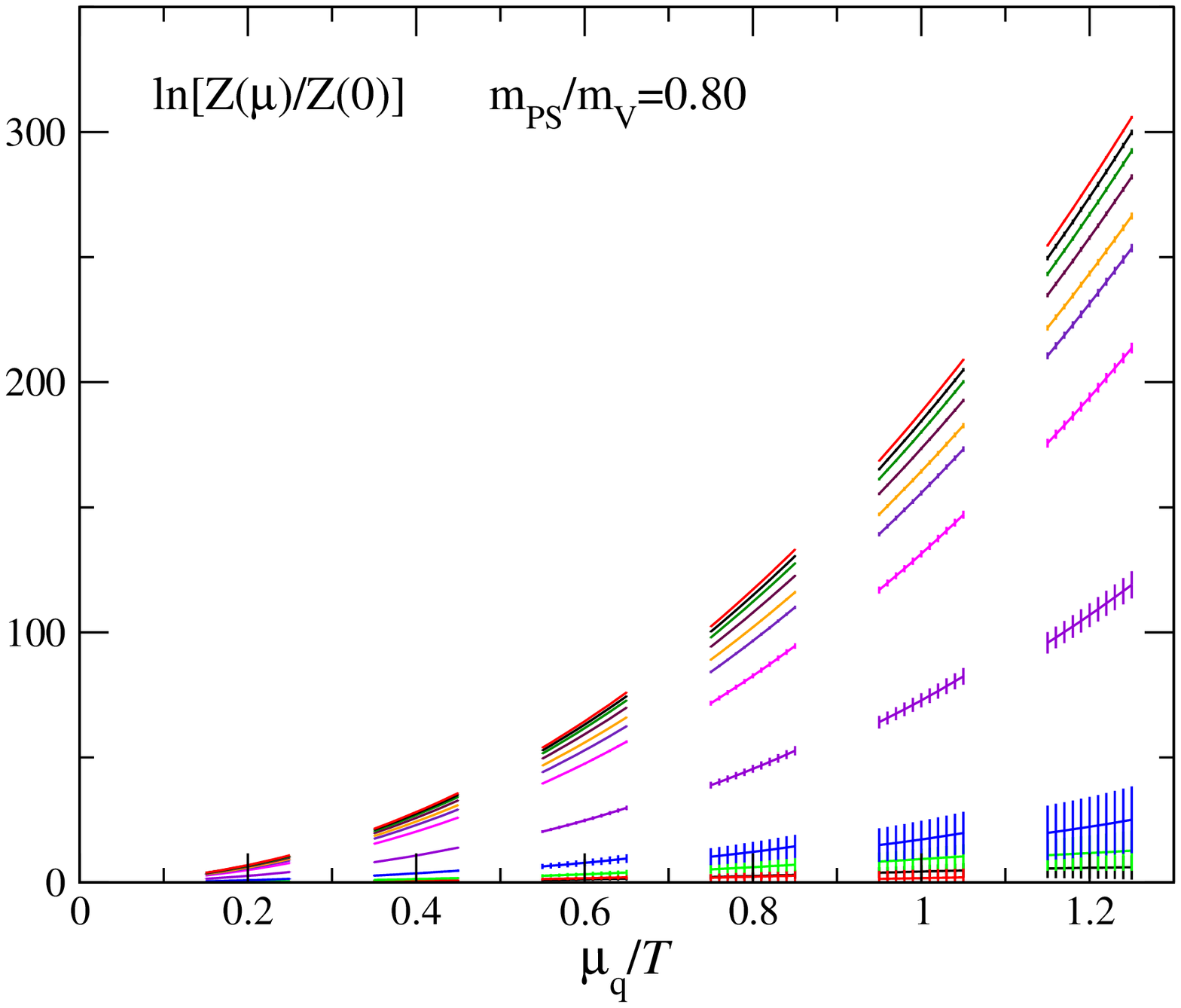}
\vskip -0.2cm
\caption{$\mu_q$-dependence of $\ln [{\cal Z}(\beta, \mu_q)/ {\cal Z}(\beta,0)]$
for each temperature. 
The values of $\ln [{\cal Z}(\beta, \mu_q)/ {\cal Z}(\beta,0)]$ 
increases as $T/T_0$ increases for each $\mu_q/T$.
The left and right figures are the results at $m_{\rm PS}/m_{\rm V}=0.65$ 
and $0.80$, respectively.}
\label{fig:lnr}
\end{center}
\vskip -0.3cm
\end{figure} 

\begin{figure}[t]
\begin{center}
\includegraphics[width=2.4in]{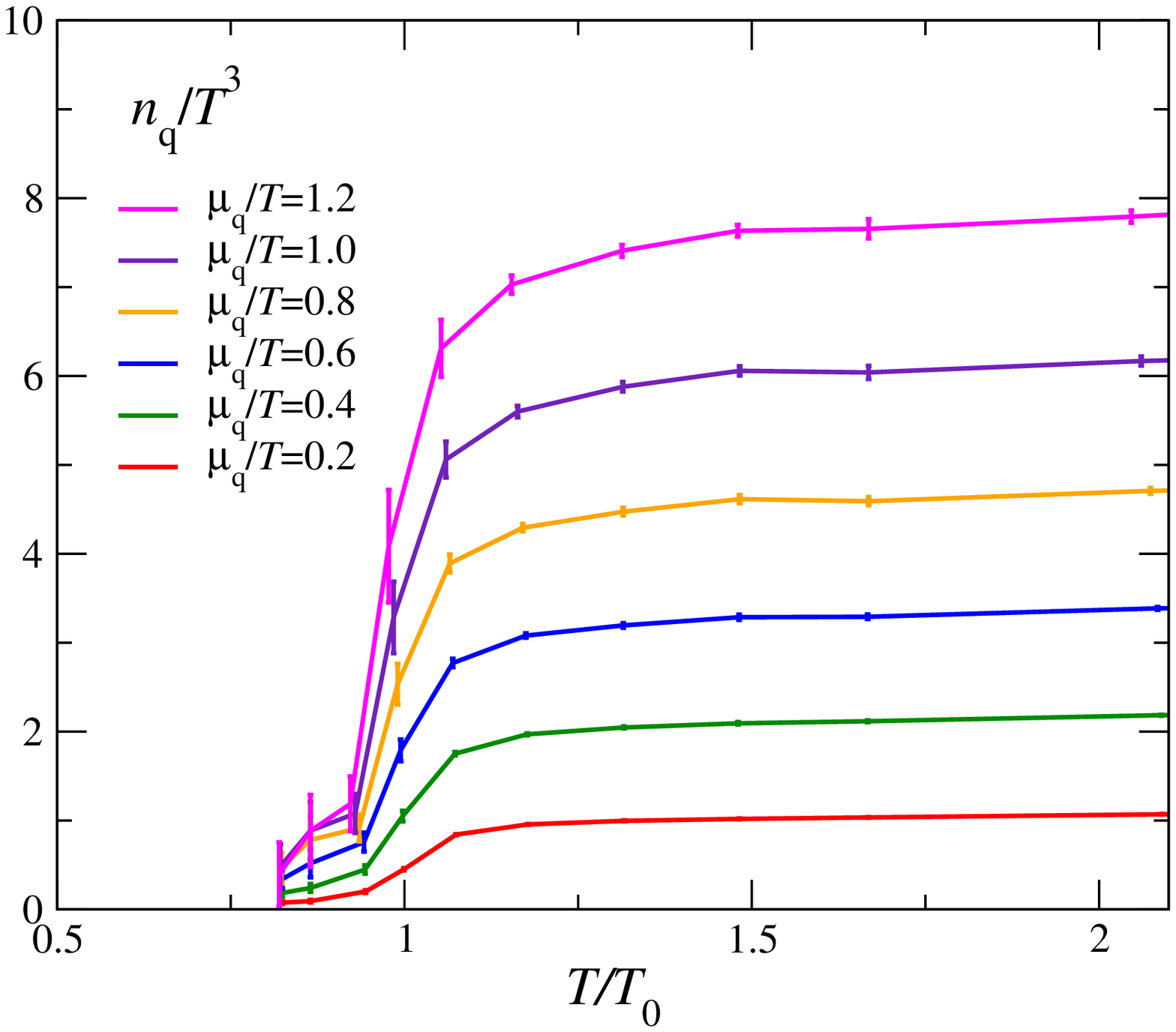}
\hskip 0.5cm
\includegraphics[width=2.4in]{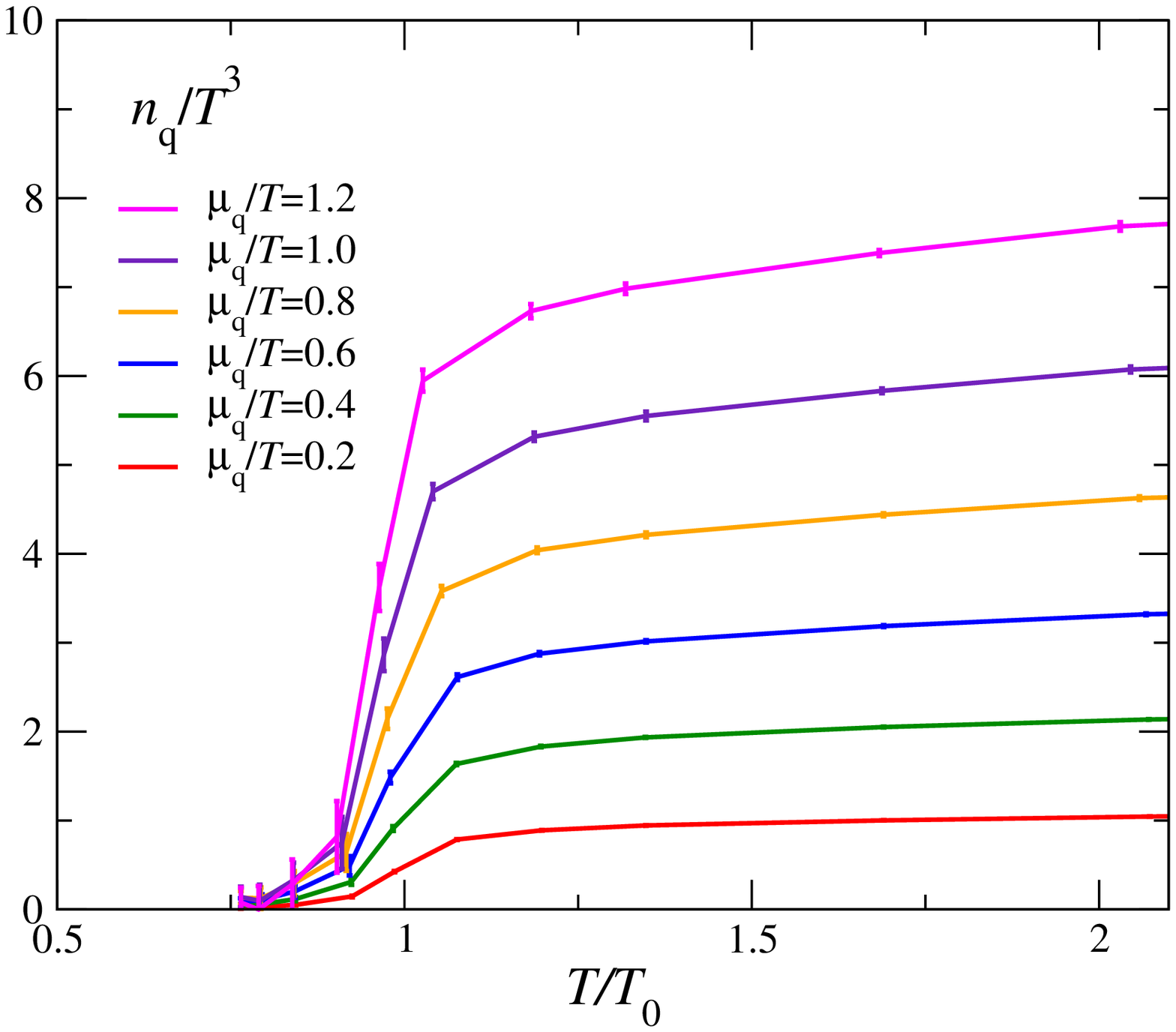}
\vskip -0.2cm
\caption{Quark number density for each $\mu_q/T$ 
at $m_{\rm PS}/m_{\rm V}=0.65$ (left) and $0.80$ (right).
$T_0$ is $T_{pc}$ at $\mu_{q}=0$.}
\label{fig:qndrew}
\end{center}
\vskip -0.3cm
\end{figure} 

\begin{figure}[t]
\begin{center}
\includegraphics[width=2.4in]{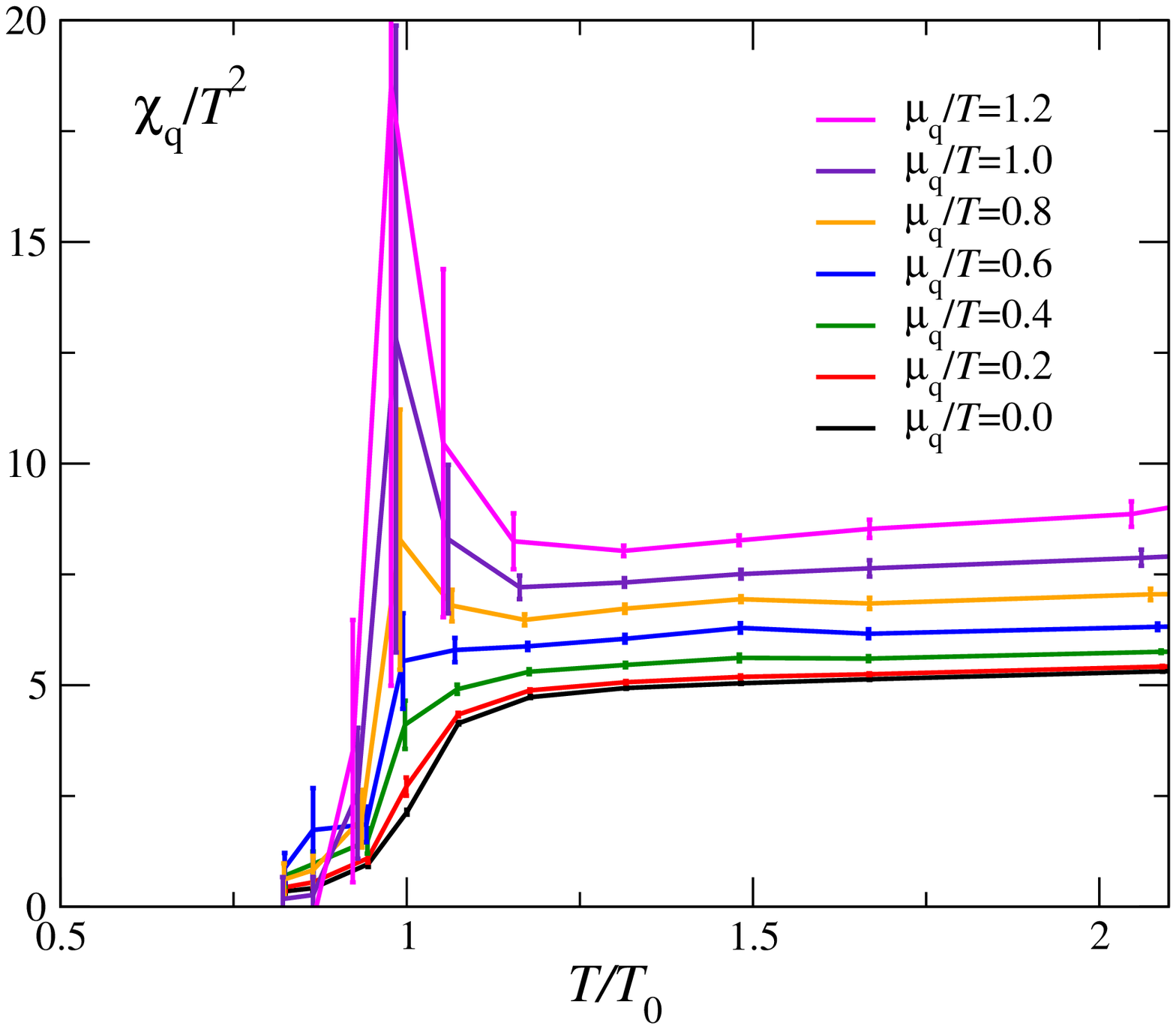}
\hskip 0.5cm
\includegraphics[width=2.4in]{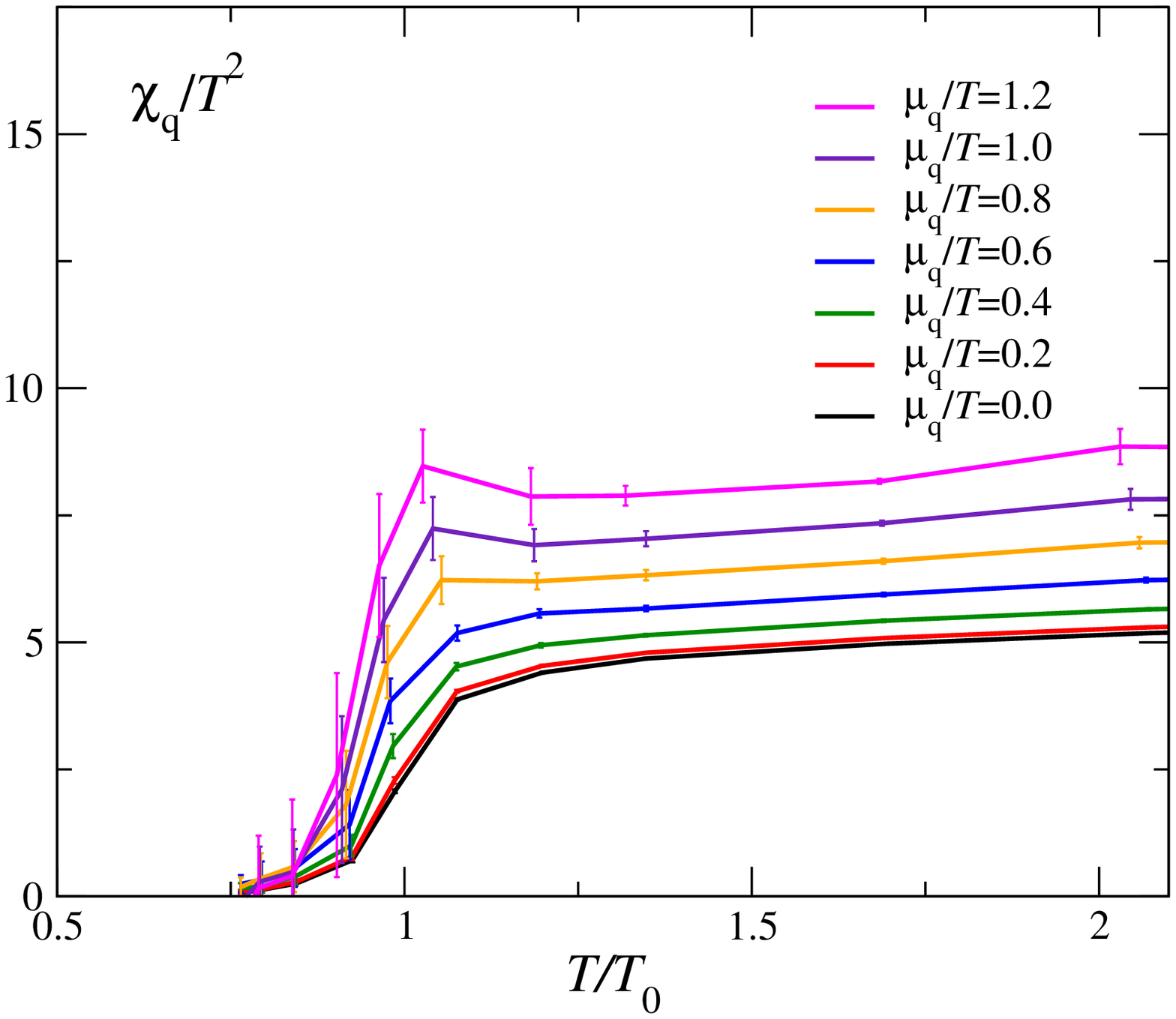}
\vskip -0.2cm
\caption{Quark number susceptibility for each $\mu_q/T$ 
at $m_{\rm PS}/m_{\rm V}=0.65$ (left) and $0.80$ (right).}
%$T_0$ is $T_{pc}$ at $\mu_{q}=0$.}
\label{fig:qnsrew}
\end{center}
\vskip -0.3cm
\end{figure} 

In Fig.~\ref{fig:prsrew}, we show the results for the $\mu_q$-dependent contribution to the pressure, $\Delta p/T^4 = p(\mu_q)/T^4 - p(0)/T^4$, obtained by the Gaussian approximation.
Comparing with Fig.~\ref{fig:prsmu}, improvement towards larger $\mu_q$ is clearly seen.

We calculate the quark number density $n_q$ and its susceptibility $\chi_q$ by the following numerical differentiations:
\begin{eqnarray}
\frac{n_{q}}{T^3} 
= \frac{N_t^3}{N_s^3} 
\frac{\partial (\ln {\cal Z})}{\partial (\mu_q/T)} , \hspace{5mm}
\frac{\chi_{q}}{T^2} 
= \frac{N_t^3}{N_s^3} 
\frac{\partial^2 (\ln {\cal Z})}{\partial (\mu_q/T)^2} .
\end{eqnarray}
Results of $\ln [{\cal Z}(T, \mu_q)/{\cal Z}(T, 0)]$  
around representative points $\tilde\mu_q/T = 0.2$, 0.4, $\cdots$, 1.2 
are shown in Fig.~\ref{fig:lnr} 
% for the ranges $\tilde\mu_q/T-0.05 \le \mu_q/T \le  \tilde\mu_q/T+0.05$ 
where $\beta$ is optimized at each $\tilde\mu_q/T$.
The value of $\ln [{\cal Z}(T, \mu_q)/{\cal Z}(T, 0)]$ increases 
as $T/T_0$ increases for each $\mu_q/T$, where $T_0$ is $T_{pc}$ at $\mu_q/T=0$. 
In Fig.~\ref{fig:lnr}, results at the optimized values of $T/T_0 (\beta)$ for simulations listed in Table~\ref{tab:parameter} are shown.
We then fit the data in the range $\tilde\mu_q/T-0.05 \le \mu_q/T \le  \tilde\mu_q/T+0.05$ 
by a quadratic function of $\mu_q/T$, 
\begin{eqnarray}
\frac{N_t^3}{N_s^3} 
\ln \left[ \frac{{\cal Z}(T,\mu_q)}{{\cal Z}(T,0)} \right]  
= \frac{n_q(\tilde\mu_q)}{T^3} \frac{\mu_q}{T} 
+ \frac{\chi_q(\tilde\mu_q)}{2T^2} \left( \frac{\mu_q}{T} \right)^2 +C(\tilde\mu_q) ,
\end{eqnarray}
with the fit parameters $n_q(\tilde\mu_q), \chi_q(\tilde\mu_q)$ and $C(\tilde\mu_q)$, 
for each values of $\tilde\mu_q/T$ and $T/T_0$.

The results of $n_q(\mu_q)$ and $\chi_q(\mu_q)$ are plotted in Figs.~\ref{fig:qndrew} 
and \ref{fig:qnsrew}. 
As is similar to the case of $p/T^4$, the statistical errors in these figures are much smaller than the results 
given in Sec.~\ref{sec:eostay}. 
Moreover, although simulations at different temperature are independent, the temperature dependence in these figures is smooth and natural. 
The reduced statistical fluctuations over the results of Sect. \ref{sec:qns} are mainly due to the Gaussian method for the $\theta$-averaging and the $\beta$-reweighting for the $F$-averaging.

At $m_{\rm PS}/m_{\rm V}=0.65$, we find a sharp peak 
in $\chi_q/T^2$ near $T_{pc}$. 
The peak becomes higher as $\mu_q$ increases.
These observations are consistent with the findings in Sec.~\ref{sec:eostay}, and suggests a critical point at finite $\mu_q$.
On the other hand, the peak is much milder at $m_{\rm PS}/m_{\rm V}=0.80$. 
This may be explained in part  by the expectation that the critical point locates at larger $\mu_q$ because the quark mass is larger than that for $m_{\rm PS}/m_{\rm V} = 0.65$.
% The mild enhancement shown in Fig.\ref{fig:qnsrew}(right) may  suggest that, at this quark mass, the critical point does not locate inside the applicability range in $\mu_q$ with our ${\cal O}(\mu_q^4)$ calculation.
Further studies with increased statistics around $T_0$ are needed for a more definite conclusion. A scaling analysis increasing the volume is also important.

%%%%%%%%%%%%%%%%%%%%%%%%%%%%%%%%%%%%%%%%%%%%%%%%%%%%%%%%%%%%
\section{Heavy-quark free energy and Debye screening mass at finite temperature and density}
\label{sec:hqfe}

In this section, we investigate
the heavy-quark free energies between static quark ($Q$) and antiquark ($\bar{Q}$), and between $Q$ and $Q$. 
These free energies are important inputs in phenomenologial studies of color-singlet quarkoniums such as charmoniums and bottomoniums in QGP \cite{Matsui:1986dk,Satz:2008zc} and of color non-singlet quark-quark states in QGP \cite{Shuryak:2004tx}. 
Lattice simulations for $Q\bar{Q}$ and $QQ$ free energies in different color channels
at $\mu_{q} = 0$ have been performed in $N_f=2$ QCD with the staggered fermion
\cite{Kaczmarek:2005ui,Doring:2007uh} and with the Wilson fermion \cite{Bornyakov:2004ii,Maezawa:2007fc}. In these works, Coulomb gauge fixing is
employed to define the Polyakov-loop correlations in different color channels.
Furthermore, the $Q \bar{Q}$ free energy at finite $\mu_{q}$ has been studied with the staggered fermion by the reweighting method in the $\mu$-$\beta$ parameter plane \cite{Fodor:2005qy}
and by the Taylor expansion method \cite{Doring}. Screening masses at finite $\mu_{q}$ have been also studied in dimensionally reduced effective field theory at high temperature \cite{Hart00}.

Here, we extend our previous study with two flavors of improved Wilson quarks at $\mu_{q}=0$ \cite{Maezawa:2007fc} to finite $\mu_{q}$ using the Taylor expansion method. Under Coulomb gauge fixing,
we calculate the expansion coefficients of the heavy-quark free energies up to the 2nd order with respect to $\mu_{q}/T$ for color-singlet $Q \bar{Q}$ channel, color-octet $Q \bar{Q}$ channel, color-sextet $QQ$ channel and color-antitriplet $QQ$ channel. The effective running coupling and Debye screening mass are also extracted
by fitting the screened Coulomb form expanded as a power series of $\mu_{q}/T$,
and compare with a prediction of the thermal perturbation theory. 

%%%%%%%%%%%%%%%%%%%%%%%%%%%%%%%%%%%%%%%%%%%%%%%%%%%%
\subsection{Taylor expansion of heavy-quark free energy}

The expectation value of an observable ${\cal O}$ for $\mu_u=\mu_d = \mu_q$ is defined as
\begin{eqnarray}
\langle {\cal O} \rangle_{\mu_q}= \frac{1}{{\cal Z}(T,\mu_q)} \int {\cal D} U \, {\cal O} \, \left[ \det M (\mu) \right]^{N_f} e^{-S_g} ,
\label{eq:EV}
\end{eqnarray}
where $\mu=\mu_{q} a$.
For ${\cal O}$ which does not depend on $\mu_q$ explicitly, $\langle {\cal O} \rangle_{\mu_q}$ can be expanded as a power series of $\mu=\mu_{q} a$ as follows \cite{Doring}: %
The quark determinant is expanded as,
\begin{eqnarray}
\left[ \det M(\mu) \right]^{N_f} &=&
\left[ \det M(0) \right]^{N_f} \left( 1 + M_1 \mu + M_2 \mu^2 + O(\mu^3) \right)
,
\label{eq:5_expansion_QD}
\end{eqnarray}
with the expansion coefficients
$
M_1 = {\cal D}_1
$,
$
M_2 = \frac{1}{2} \left({\cal D}_1^2 + {\cal D}_2 \right) $, etc.\
using ${\cal D}_n$ defined by (\ref{eq:calDn}). %
Then, using the fact that the system is symmetric under $\mu_q \rightarrow -\mu_q$, $\langle {\cal O} \rangle_{\mu_q}$ can be expanded as %
\begin{eqnarray}
\langle {\cal O} \rangle_{\mu_q}
&=&
\frac{\langle {\cal O} \rangle_0 + \langle {\cal O} M_1 \rangle_0\, \mu + \langle {\cal O} M_2 \rangle_0\, \mu^2}{1 + \langle M_2 \rangle_0 \mu^2 }
+ O(\mu^3)
\\ \nonumber
&=&
\langle {\cal O} \rangle_0 \left[ 1 + {\cal O}_1\, \mu + \left( - \langle M_2 \rangle_0 + {\cal O}_2 \right) \mu^2 + O(\mu^3) \right]
,
\end{eqnarray}
where $\langle {\cal O} \rangle_0 = \langle {\cal O} \rangle_{\mu_{q}=0}$ and ${\cal O}_i$ is defined by \begin{eqnarray}
{\cal O}_i &=& \frac{\langle {\cal O} M_i \rangle_0}{\langle {\cal O} \rangle_0}
\label{eq:On}
.
\end{eqnarray}

The heavy-quark free energies are defined by correlation functions between Polyakov loops,
$\Omega ( {\bf x} ) = \prod_{ \tau = 1}^{N_t} U_4 (\tau, {\bf x})$. At a fixed gauge, the $Q\bar{Q}$ correlation function can be decomposed into color singlet ({\bf 1}) and color octet ({\bf 8}) channels,
while the $QQ$ correlation function into color antitriplet (${\bf 3}^*$) and color sextet ({\bf 6}) channels
as follows \cite{Nadkarni1,Nadkarni2}:
\begin{eqnarray}
\Osi (r) &=&
\frac{1}{3} \tr \Omega^\dagger({\bf x}) \Omega ({\bf y}) \label{eq:1}
, \\
\Oad (r) &=&
\frac{1}{8} \tr \Omega^\dagger({\bf x}) \tr \Omega ({\bf y}) - \frac{1}{24} \tr \Omega^\dagger({\bf x}) \Omega ({\bf y}) \label{eq:8}
, \\
\Osy (r) &=&
\frac{1}{12} \tr \Omega({\bf x}) \tr \Omega ({\bf y}) + \frac{1}{12} \tr \Omega({\bf x}) \Omega ({\bf y}) , \\
\Oan (r) &=&
\frac{1}{6} \tr \Omega({\bf x}) \tr \Omega ({\bf y}) - \frac{1}{6} \tr \Omega({\bf x}) \Omega ({\bf y}) ,
\end{eqnarray}
where $r=| {\bf x} - {\bf y}|$.
The free energy ${\cal F}^R$ for color channel $R$ ($R= {\bf 1}$, ${\bf 8}$, ${\bf 6}$, ${\bf 3^*}$) is defined as \begin{eqnarray}
e^{- {\cal F}^R (r,T,\mu_{q}) /T} = \langle \Om \rangle_{\mu_{q}} \label{eq:3_NFE}
.
\end{eqnarray}

Above $T_{pc}$, we introduce normalized free energies 
$(\Vsi, \Vad, \Vsy, \Van)$
by dividing the right-hand side of (\ref{eq:3_NFE}) by
   $\langle L \rangle_{\mu_q} \langle L \rangle_{\mu_q}^\ast$ for $Q \bar{Q}$ free energies and 
   $\langle L \rangle_{\mu_q}^2$ for $QQ$ free energies,
  where $L = \tr \Omega$.
$\Vm$ vanishes at $r\rightarrow\infty$.
The Taylor expansion
 of $\Vm$ with respect to $\mu_{q}/T$ is given by
\begin{eqnarray}
\Vm(r,T,\mu_{q}) = 
v^R_0 +
v^R_1 \left( \frac{\mu_{q}}{T} \right) +
v^R_2 \left( \frac{\mu_{q}}{T} \right)^2 +
O(\mu^3)
,
\label{eq:ENFE}
\end{eqnarray}
where
\begin{eqnarray}
% Q\bar{Q} \ {\bf channel} && \nonumber \\
% 0th
\frac{v^R_0 (r,T)}{T} &=& -\ln \left( 
\frac{ \langle \Om \rangle_{0} }{ \ell_{0}^2 } \right)
\label{eq:QQb0}
, \\
% 1st
\frac{v^R_1 (r,T)}{T} &=& 0
, \\
% 2nd
\frac{v^R_2 (r,T)}{T} &=& \frac{1}{N_t^2}
\left( \langle M_2 \rangle_0  - \Om_2 \right)
+ \frac{4 \ell_0 \ell_2
- \left( \ell_1^2 + {\ell^\ast_1}^2 \right) }
{2 \ell_0^2}
,
\end{eqnarray}
for color singlet and octet $Q \bar{Q}$ channels, and 
\begin{eqnarray}
%QQ \ {\bf channel} && \nonumber \\
% 0th
\frac{v^R_0 (r,T)}{T} &=& -\ln \left( 
\frac{ \langle \Om \rangle_{0} }{ \ell_{0}^2 } \right)
, \\
% 1st
\frac{v^R_1 (r,T)}{T} &=& 
- \frac{1}{N_t} \Om_1 
+ 2 \frac{\ell_1}{\ell_0}
,\\
% 2nd
\frac{v^R_2 (r,T)}{T} &=& \frac{1}{N_t^2}
\left( \langle M_2 \rangle_0 + \frac{1}{2} (\Om_1)^2 - \Om_2 \right)
+ 2 \frac{\ell_2}{\ell_0}
-\frac{\ell_1^2}{\ell_0^2} 
\label{eq:QQ2}
,
\end{eqnarray}
for color sextet and antitriplet $QQ$ channels.
Here $\Om_n = \langle \Om M_n \rangle_0/\langle \Om \rangle_0$,
 and the $\ell_n$ is an $n$-th order coefficient 
  of the Taylor expansion of the Polyakov loop:
\begin{eqnarray}
\langle L \rangle_{\mu_{q}} &=& 
\ell_0 +
\ell_1 \left( \frac{\mu_{q}}{T} \right) +
\ell_2 \left( \frac{\mu_{q}}{T} \right)^2 +
O(\mu^3)
.
\end{eqnarray}
Note that
 the color singlet and octet channels do not have the odd orders in
  the Taylor expansion since the free energies for both channels are
  symmetric under $\mu_{q} \rightarrow - \mu_{q}$, 
  i.e., the $Q\bar{Q}$ free energies are invariant under the charge conjugation.

%%%%%%%%%%%%%%%%%%%%%%%%%%%%%%%%%%%%%%%%%%%%%%%%%%%%%%%%%%%

\subsection{Results for expansion coefficients of normalized free energies}

Heavy quark free energies are calculated in the high temperature phase on the lines of constant physics at $m_{\rm PS}/m_{\rm V}=0.65$ and 0.80 (see Table \ref{tab:parameter}).
%The Taylor expansion coefficients of the quark determinant,  ${\cal D}_1$ and ${\cal D}_2$, are calculated using the random noise method discussed in Sec.~[{\bf IIIB ref.}].
Observables are measured every ten trajectories at each quark mass and temperature, and the statistical errors are estimated by a jackknife method with the bin size of 100 trajectories.

\begin{figure}[tbp]
  \begin{center}
    \begin{tabular}{cc}
    \includegraphics[width=80mm]{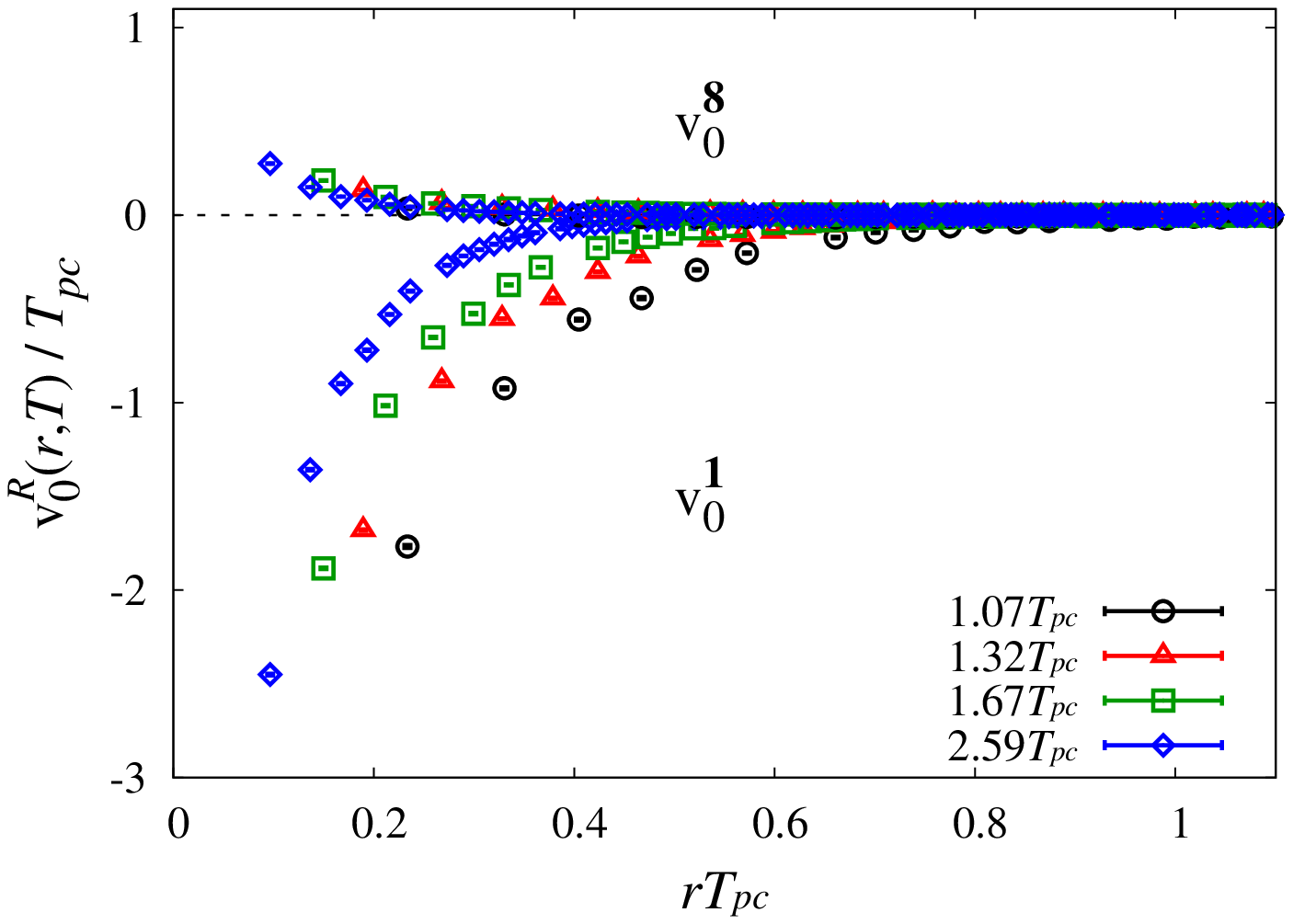} &
    \includegraphics[width=80mm]{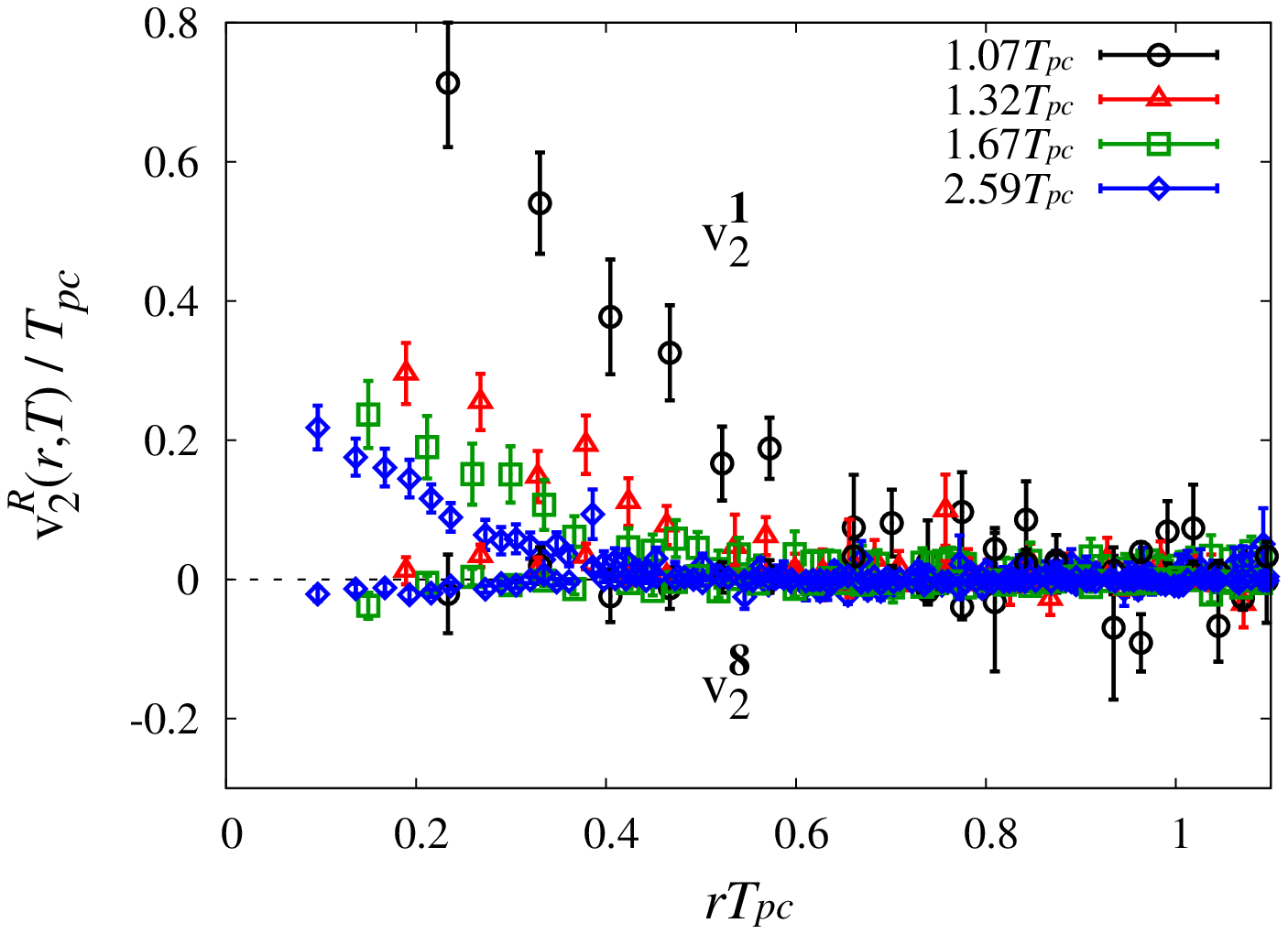}
    \end{tabular}
    \caption{$v_0^R$ (left) and $v_2^R$ (right)
    for color-singlet and octet $Q \bar{Q}$ channels above $T_{pc}$
    at $m_{\rm PS}/m_{\rm V} = 0.65$.
        }
    \label{fig:vqb065}
  \end{center}
\end{figure}

\begin{figure}[tbp]
  \begin{center}
    \begin{tabular}{cc}
    \includegraphics[width=80mm]{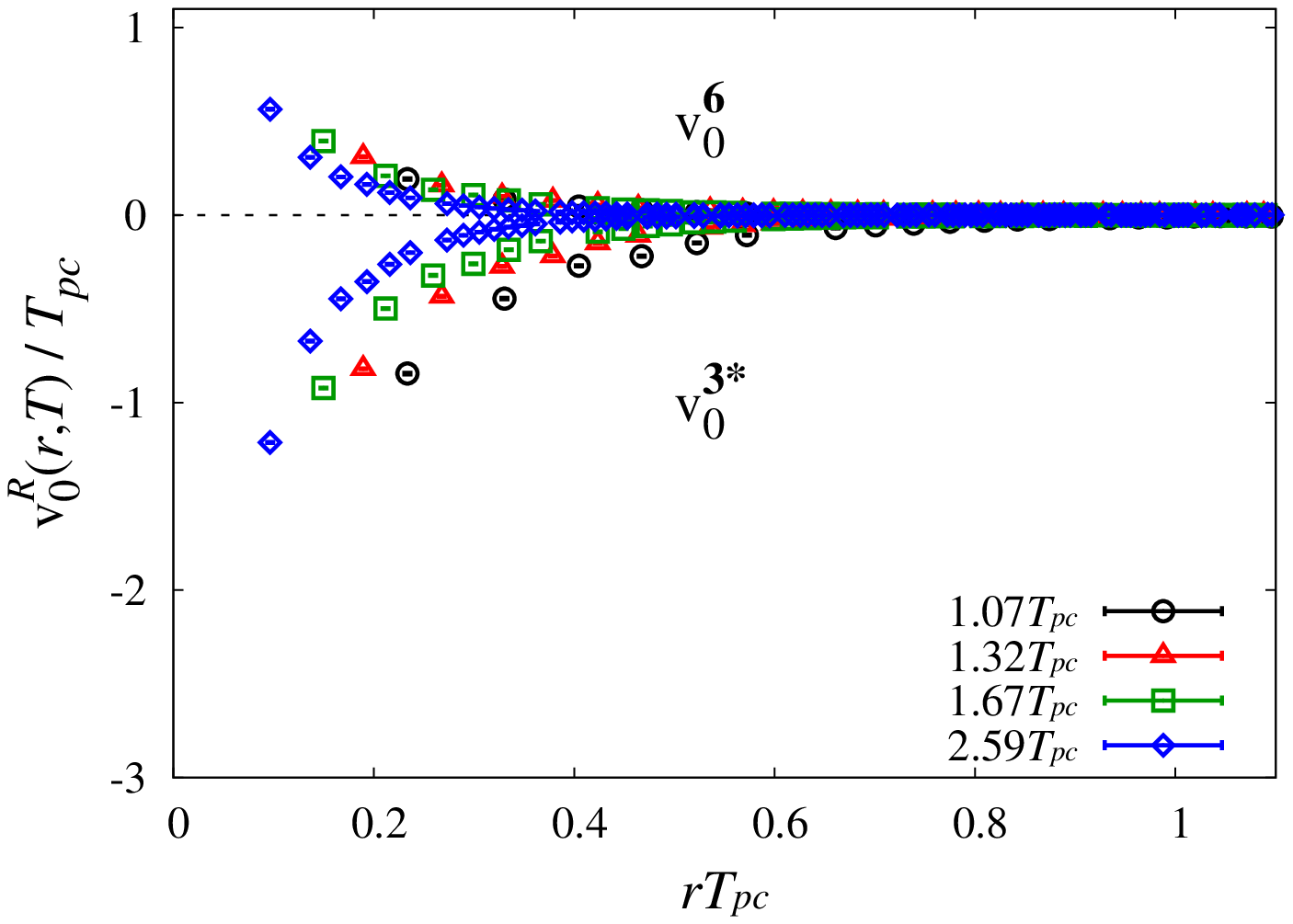} &
    \includegraphics[width=80mm]{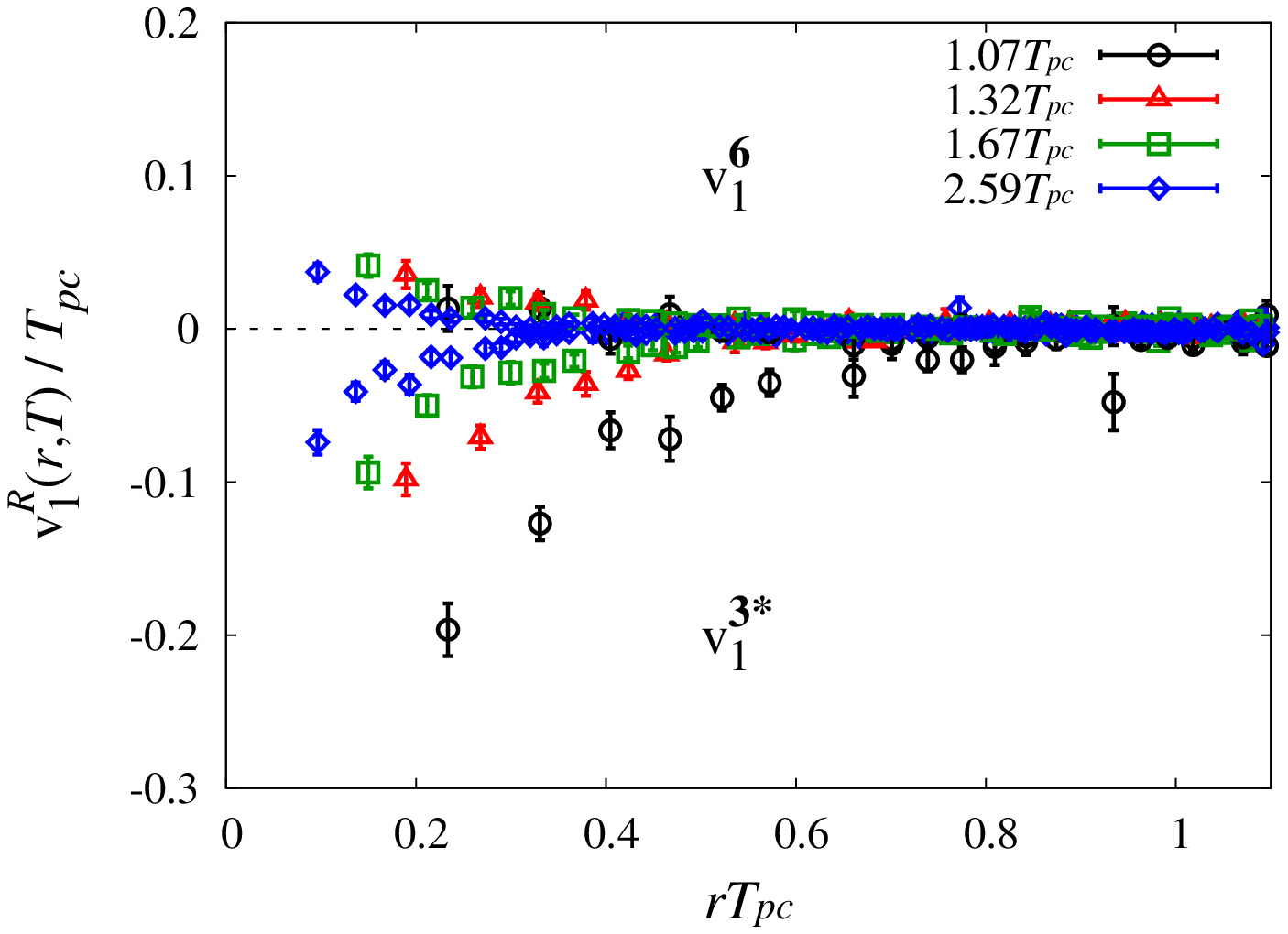}
    \end{tabular}
    \caption{$v_0^R$ (left) and $v_1^R$ (right)
    for color-sextet and antitriplet $Q Q$ channels above $T_{pc}$
    at $m_{\rm PS}/m_{\rm V} = 0.65$.
        }
    \label{fig:vqq1065}
  \end{center}
\end{figure}

\begin{figure}[tbp]
  \begin{center}
    \includegraphics[width=80mm]{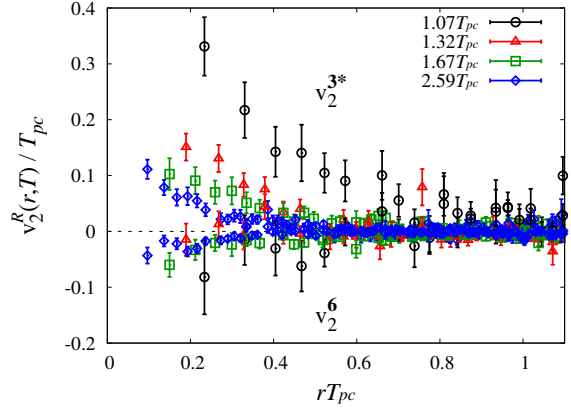}
    \caption{$v_2^R$
    for color-sextet and antitriplet $Q Q$ channels above $T_{pc}$
    at $m_{\rm PS}/m_{\rm V} = 0.65$.
        }
     \vspace{-0.5cm}
    \label{fig:vqq2065}
  \end{center}
\end{figure}

\begin{figure}[tbp]
  \begin{center}
    \begin{tabular}{cc}
    \includegraphics[width=80mm]{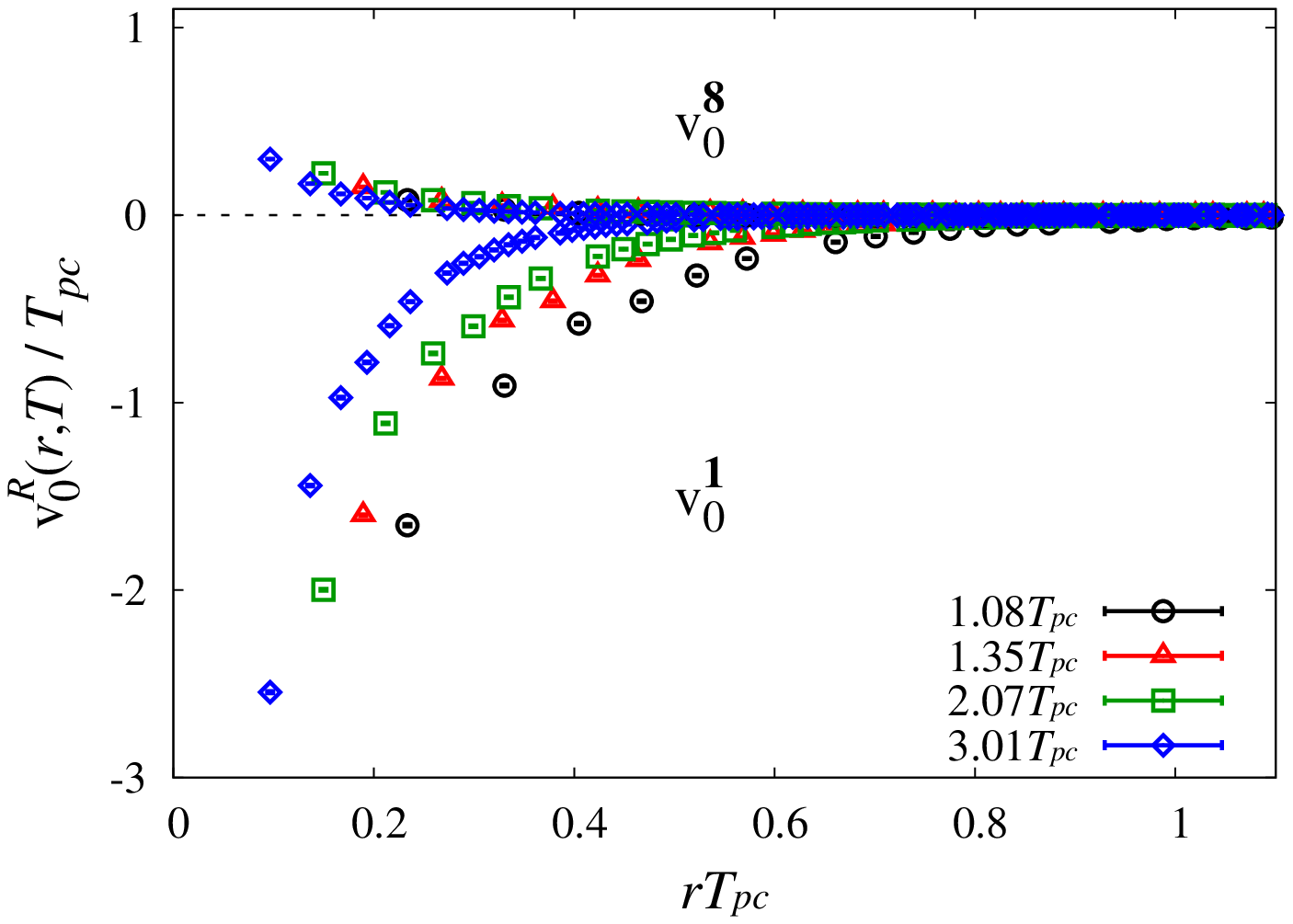} &
    \includegraphics[width=80mm]{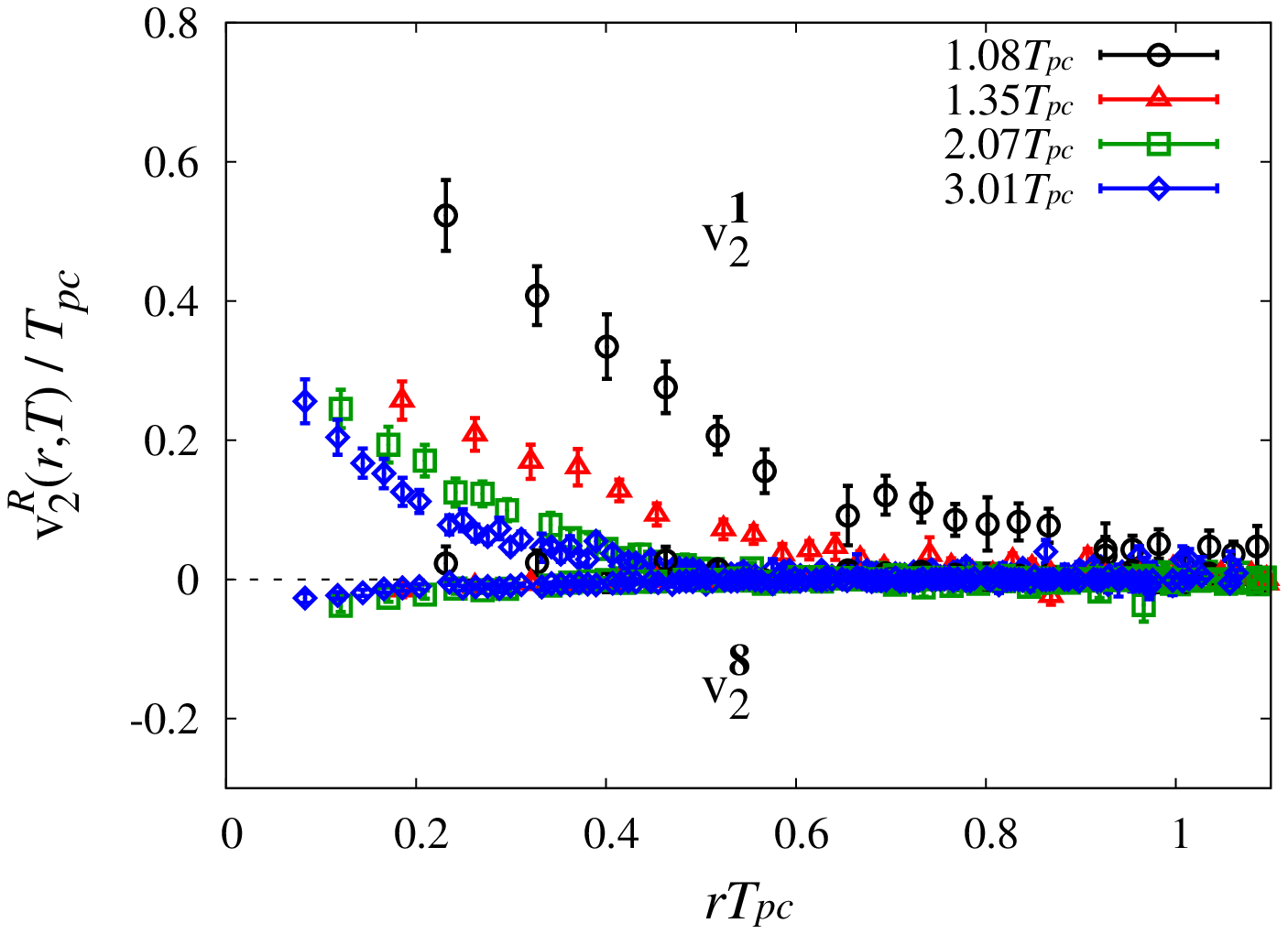}
    \end{tabular}
    \caption{The same figures as Fig.~\ref{fig:vqb065}
    at $m_{\rm PS}/m_{\rm V} = 0.80$.
        }
    \label{fig:vqb080}
  \end{center}
\end{figure}

\begin{figure}[tbp]
  \begin{center}
    \begin{tabular}{cc}
    \includegraphics[width=80mm]{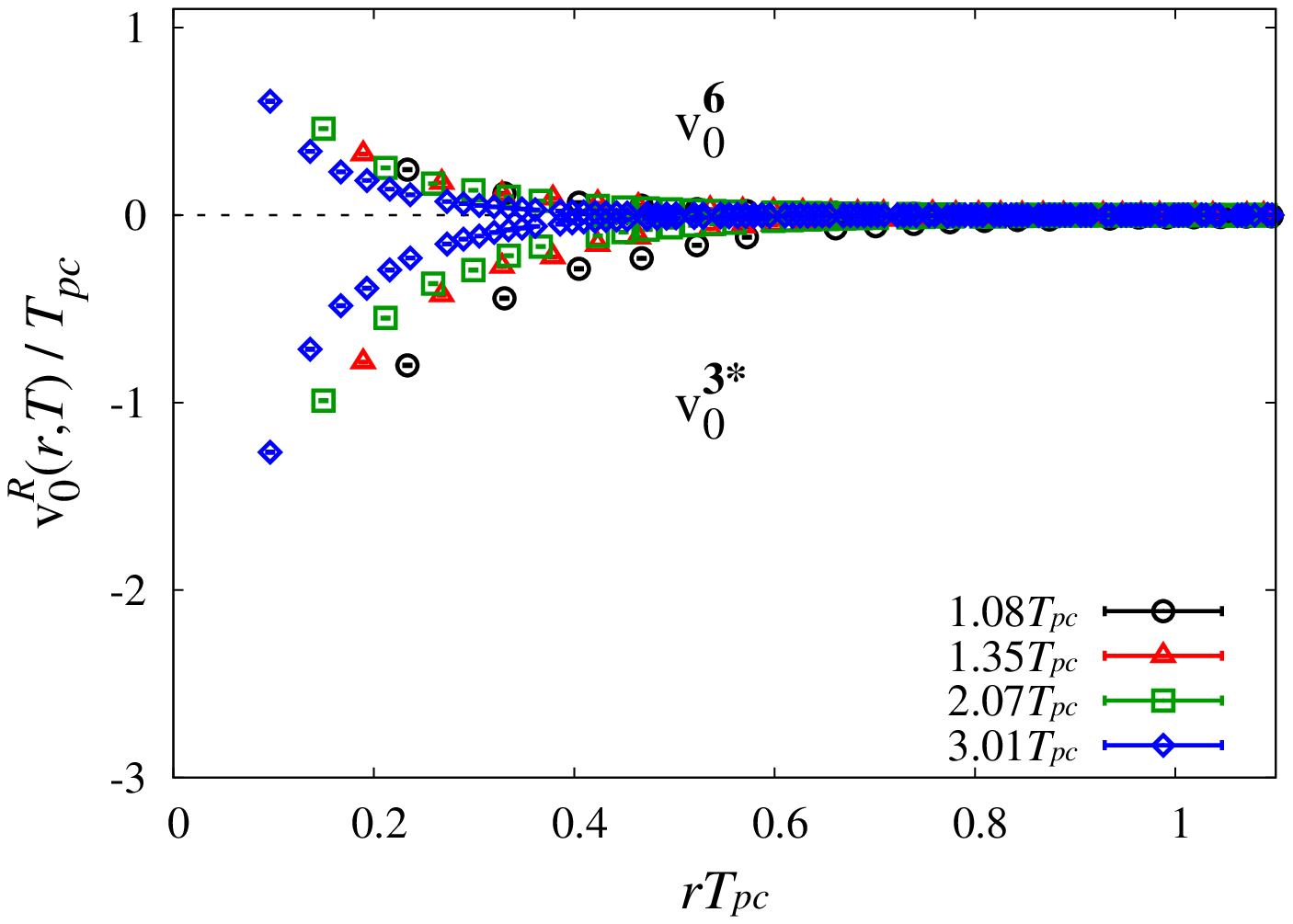} &
    \includegraphics[width=80mm]{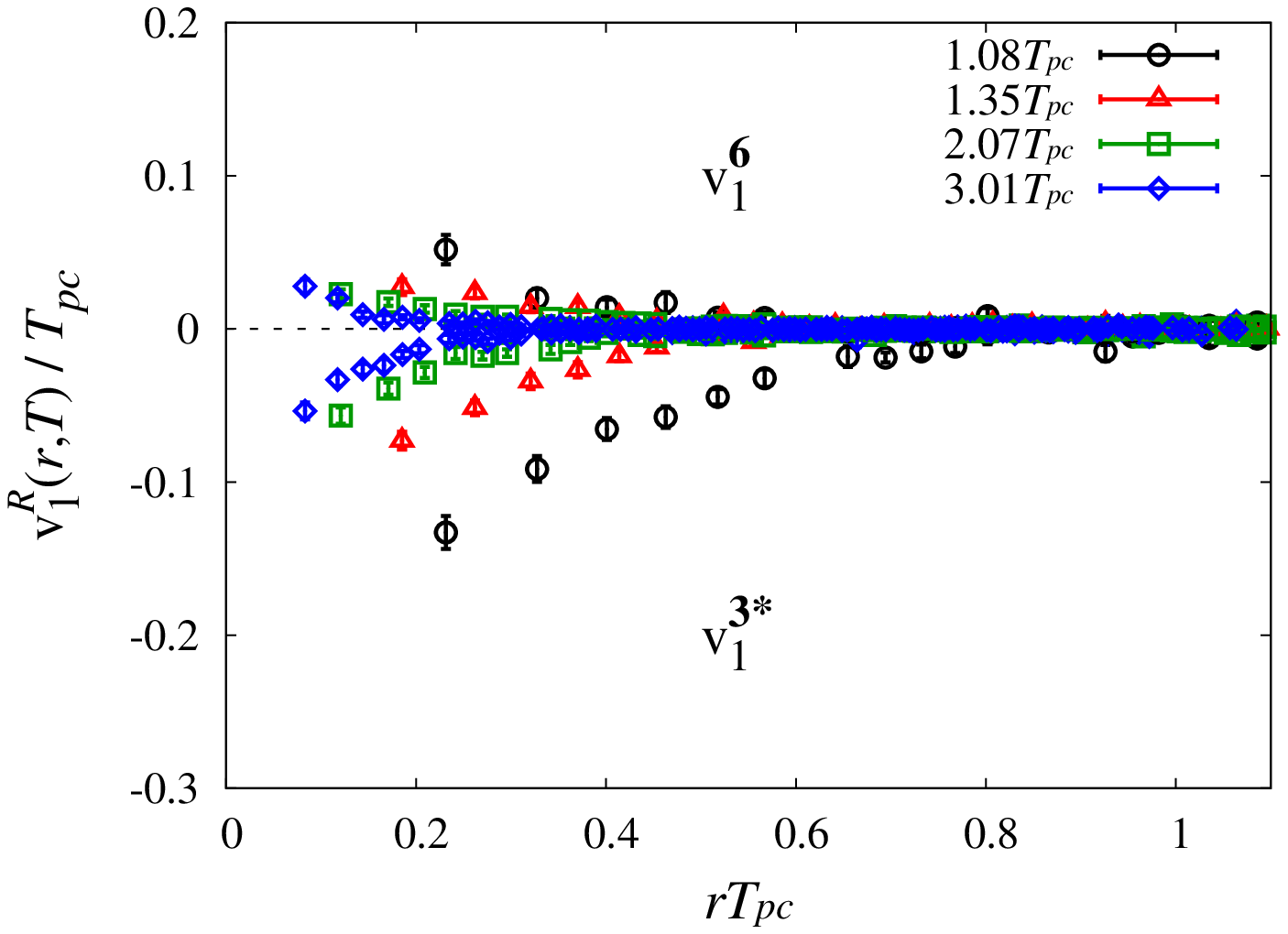}
    \end{tabular}
    \caption{The same figures as Fig.~\ref{fig:vqq1065}
    at $m_{\rm PS}/m_{\rm V} = 0.80$.
        }
    \label{fig:vqq1080}
  \end{center}
\end{figure}

\begin{figure}[tbp]
  \begin{center}
    \includegraphics[width=80mm]{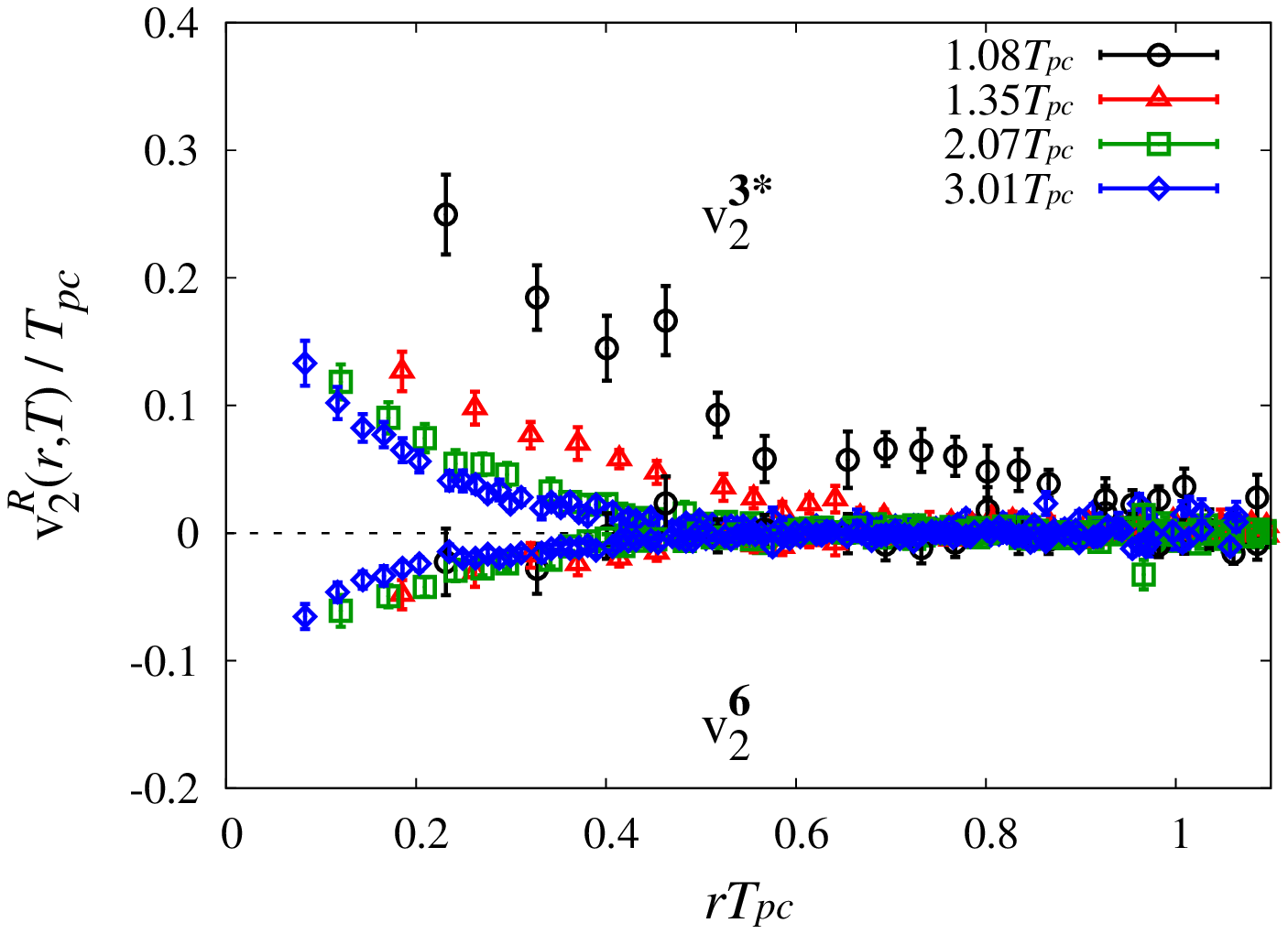}
    \caption{The same figures as Fig.~\ref{fig:vqq2065}
    at $m_{\rm PS}/m_{\rm V} = 0.80$.
        }
     \vspace{-0.5cm}
    \label{fig:vqq2080}
  \end{center}
\end{figure}

The results for the expansion coefficients of the normalized free energies at $m_{\rm PS} / m_{\rm V} = 0.65$ are  shown in Fig.~\ref{fig:vqb065} for the color singlet and octet $Q \bar{Q}$ channels, 
and in Figs.~\ref{fig:vqq1065} and \ref{fig:vqq2065} for the color sextet and antitriplet $QQ$ channels. 
Those obtained at $m_{\rm PS} / m_{\rm V} = 0.80$ are shown in Figs.~\ref{fig:vqb080}--\ref{fig:vqq2080}.

The $v_0^R$'s shown in Figs.~\ref{fig:vqb065}, \ref{fig:vqq1065}, \ref{fig:vqb080} and \ref{fig:vqq1080} 
are the normalized free energies at $\mu_q=0$.
The fact that, with increasing the distance $r$, $v_0^{\bf 1}$ and $v_0^{{\bf 3}^*}$ increase while $v_0^{\bf 8}$ and $v_0^{\bf 6}$ decrease represents the finding of our previous study \cite{Maezawa:2007fc} that,  at $\mu_q=0$, 
the inter-quark interaction is ``attractive'' in the color singlet and
antitriplet channels and is ``repulsive'' in the color octet and sextet channels.

From these Figures, we note that, both around $T_{pc}$ and at higher temperatures, 
the sign of $v_1^R$ is the same with that of $v_0^R$,
whereas the sign of a $v_2^R$ is the opposite of that of $v_0^R$:
\begin{eqnarray}
  v_1^R \cdot v_0^R &>& 0 \ \ \  \text{(only for $QQ$ free energies)} ,
\\
  v_2^R \cdot v_0^R &<& 0 .
\end{eqnarray}
Because $v_1^R$ is absent for $Q\bar{Q}$ free energies, this means that, in the leading-order of $\mu_{q}$,
the inter-quark interaction between $Q$ and $\bar{Q}$ becomes weak at finite $\mu_{q}$, 
while that between $Q$ and $Q$ becomes strong. 
In other words, $Q \bar{Q}$ ($QQ$) free energies are screened (anti-screened)
by the internal quarks induced at finite $\mu_{q}$.

\subsection{Screening properties at finite $T$ and $\mu_{q}$}

At $\mu_q=0$, the color channel dependence in the free energies was shown to be well absorbed in the kinematical Casimir factor at high temperatures \cite{Maezawa:2007fc}, 
as first noticed in quenched studies \cite{Nakamura1,Nakamura2}.
Therefore, %in order to study the screening effect in each color channel,
we fit the normalized free energies by a screened Coulomb form,
\begin{eqnarray}
\Vm (r,T,\mu_{q}) = C^R \, \frac{\alpha_{\rm eff}(T,\mu_{q})}{r} e^{- m_{D}(T,\mu_{q}) \, r}
,
\label{eq:scf}
\end{eqnarray}
where the Casimir factors 
 $C^R \equiv \langle \sum_{a=1}^{8} t_1^a\cdot t_2^a \rangle_R$
 for various color channels are given  by 
\begin{eqnarray}
C^{\bf 1}     = -\frac{4}{3}, \quad
C^{\bf 8}     =  \frac{1}{6}, \quad
C^{\bf 6}     =  \frac{1}{3}, \quad
C^{\bf 3^*}   = -\frac{2}{3}.
\end{eqnarray}
At small $\mu_q$, the effective running coupling $\alpha_{\rm eff}(T,\mu_{q})$ and the Debye screening mass $m_D(T,\mu_{q})$ 
 are expanded by powers of $\mu_{q}/T$:
\begin{eqnarray}
\alpha_{\rm eff} &=& 
\alpha_0 +
\alpha_1 \left( \frac{\mu_{q}}{T} \right) +
\alpha_2 \left( \frac{\mu_{q}}{T} \right)^2 +
O(\mu^3)
, \label{eq:alpha}
\\
m_D &=& 
m_{D,0} +
m_{D,2} \left( \frac{\mu_{q}}{T} \right)^2 +
O(\mu^4)
,
\label{eq:mD}
\end{eqnarray}
where we use the fact that the Debye screening mass does not
  have the odd powers in the Taylor expansion 
   because it corresponds to the self-energy of the two-point correlation 
  of the gauge field which is symmetric under $\mu_{q} \rightarrow - \mu_{q}$.
Properties of $\alpha_0(T)$ and $m_{D,0}(T)$ are discussed in \cite{Maezawa:2007fc}.

Expanding (\ref{eq:scf}) with respect to $\mu_{q}/T$ using (\ref{eq:alpha}) and (\ref{eq:mD}), and comparing with the expansion (\ref{eq:ENFE}) of the normalized free energies,
we obtain the following relations:
\begin{eqnarray}
v_0 (r,T) &=& C^R\, \frac{\alpha_0 (T)}{r}\, e^{- m_{D,0}(T)\, r}
, \\
\label{eq:0th}
\frac{v_1(r,T)}{v_0(r,T)} &=& \frac{\alpha_1(T)}{\alpha_0(T)}
\qquad \text{(only for $QQ$ free energies)}
\label{eq:1st}
,\\
\frac{v_2(r,T)}{v_0(r,T)} &=& \frac{\alpha_2(T)}{\alpha_0(T)} - m_{D,2}(T)\, r
.\label{eq:2nd}
\end{eqnarray}
Therefore, the expansion coefficients of $\alpha_{\rm eff}$ and $m_D$ for each $T$ 
can be calculated by fitting the normalized free energies for appropriate ranges of $r$.
We chose the fit ranges to be $0.5 \le rT \le 1.0$ for Eq.~(\ref{eq:1st})
 and $0.25 \le rT \le 1.0$ for Eq.~(\ref{eq:2nd}).
In Appendix \ref{ap:fit}, we study the fit range dependence of the fits, and find that the magnitude of systematic errors in the expansion coefficients due to the fit range are at most comparable with that of the statistical errors at $T \simge 1.2 T_{pc}$.

\begin{figure}[tbp]
  \begin{center}
    \begin{tabular}{cc}
    \includegraphics[width=84mm]{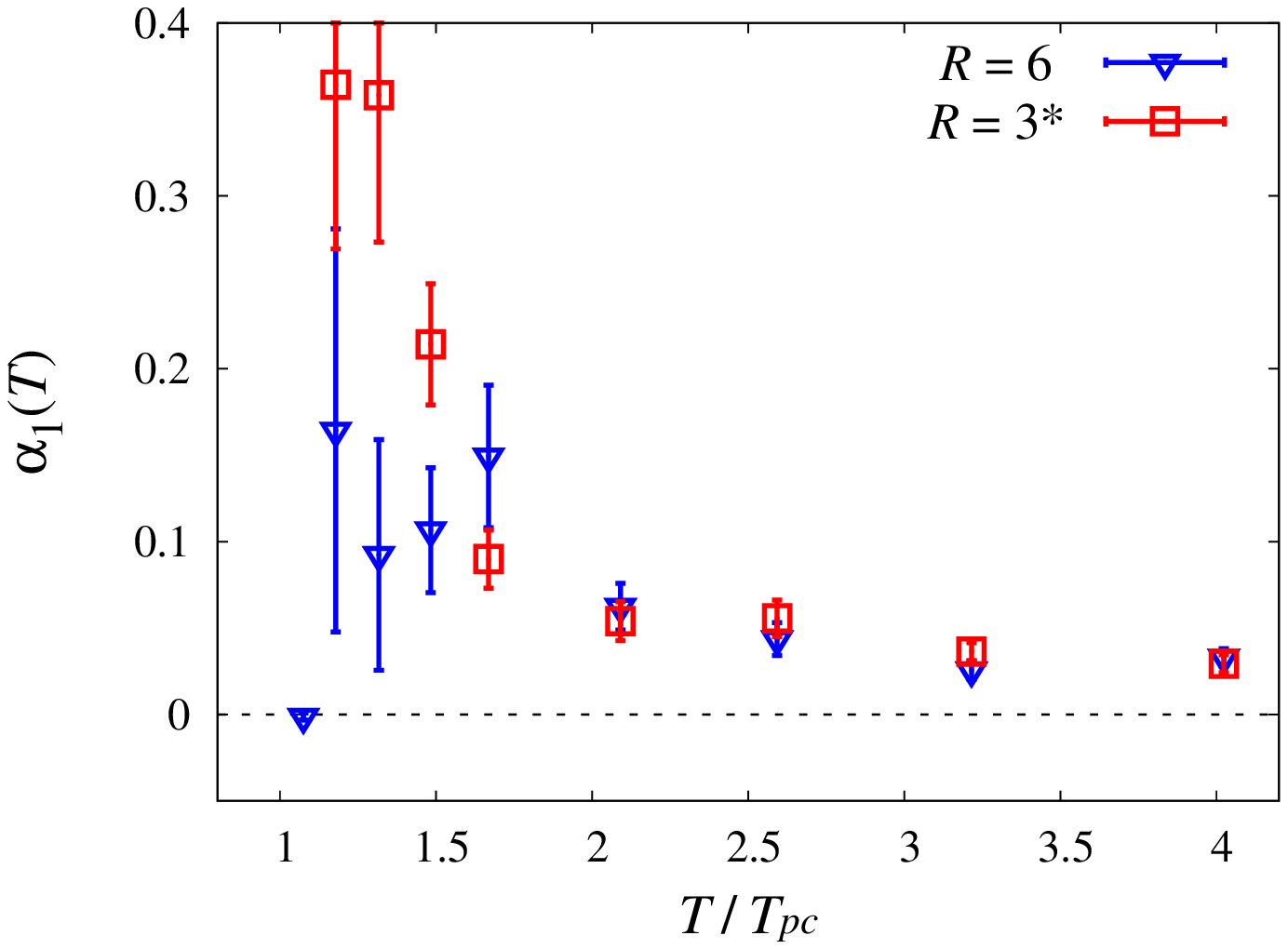} &
    \includegraphics[width=84mm]{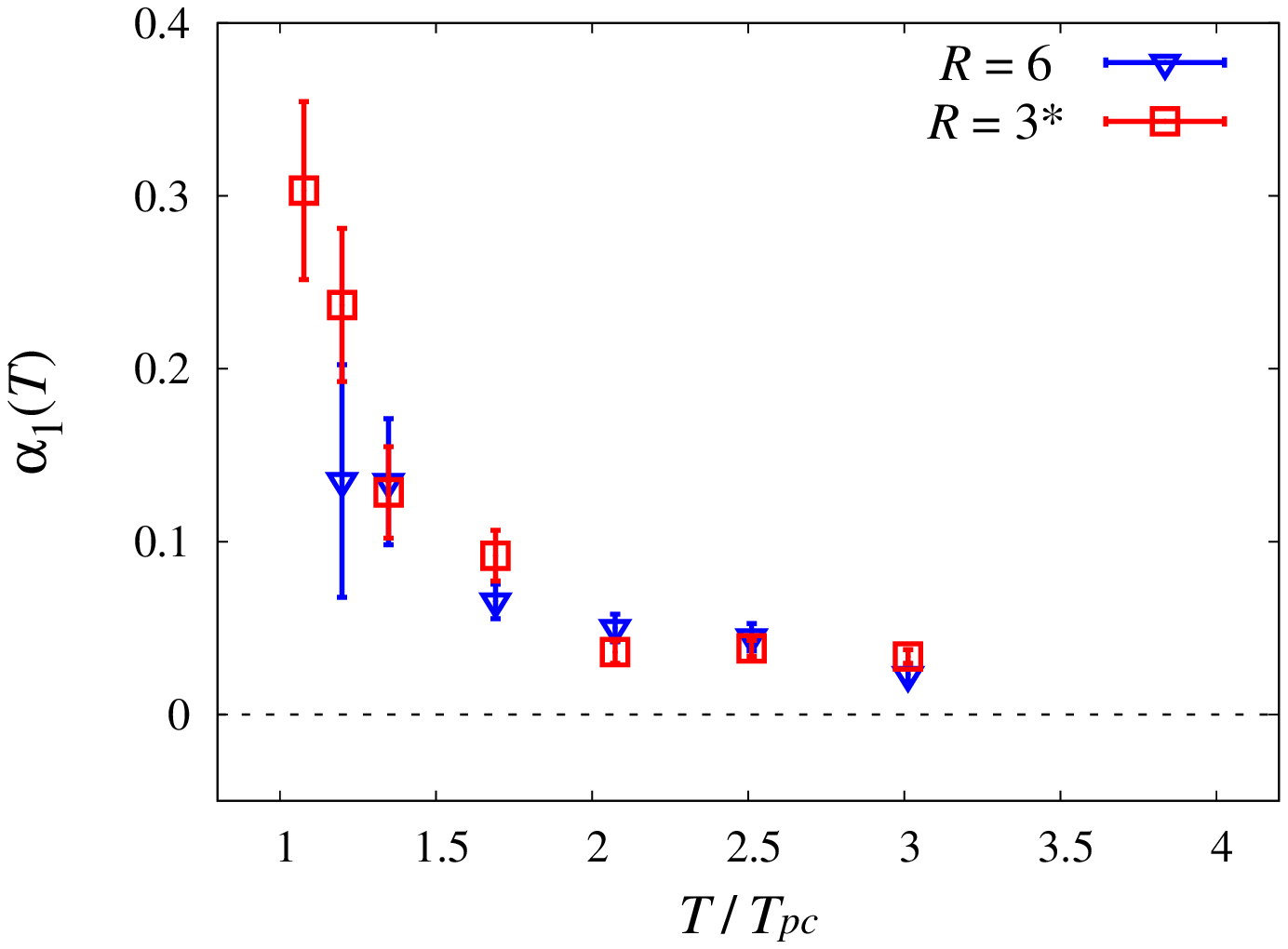}
    \end{tabular}
    \caption{$\alpha_1(T)$
     at $m_{\rm PS} / m_{\rm V} = 0.65$ (left) and 0.80 (right).
}
  \label{fig:AL1}
  \end{center}
\end{figure}

\begin{figure}[tbp]
  \begin{center}
    \begin{tabular}{cc}
    \includegraphics[width=84mm]{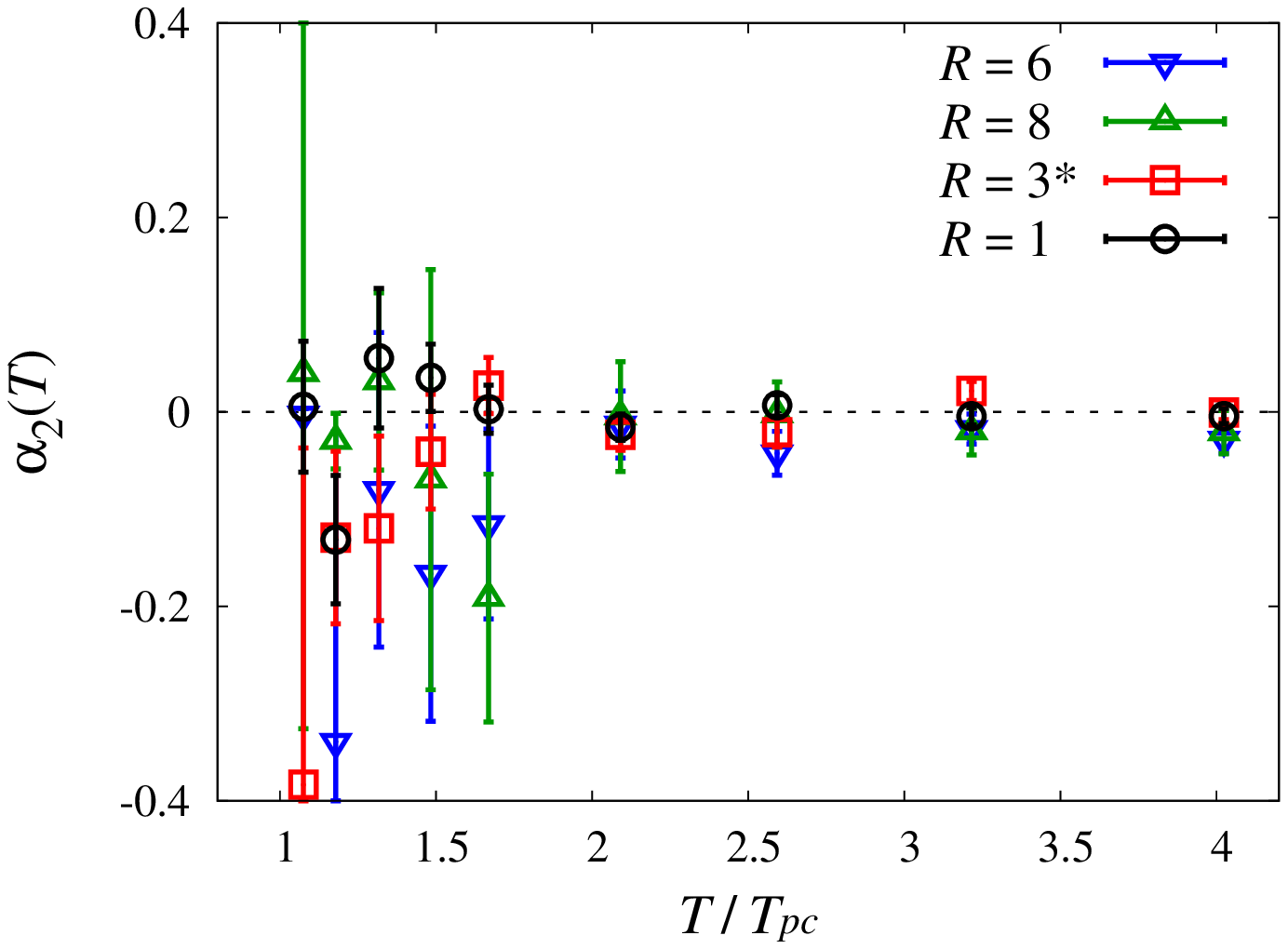} &
    \includegraphics[width=84mm]{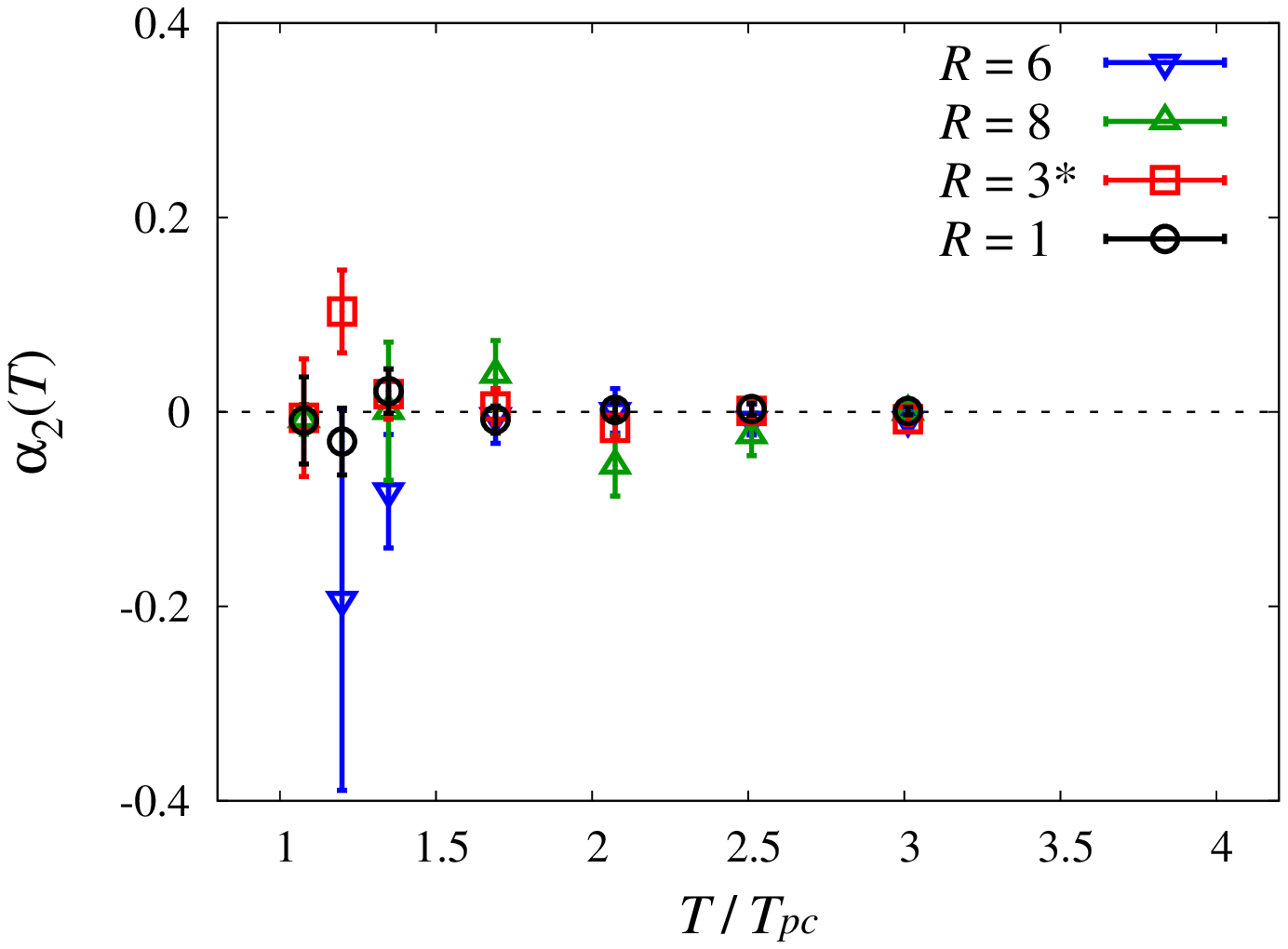}
    \end{tabular}
    \caption{$\alpha_2(T)$
     at $m_{\rm PS} / m_{\rm V} = 0.65$ (left) and 0.80 (right).
}
  \label{fig:AL2}
  \end{center}
\end{figure}

\begin{figure}[tbp]
  \begin{center}
    \begin{tabular}{cc}
    \includegraphics[width=84mm]{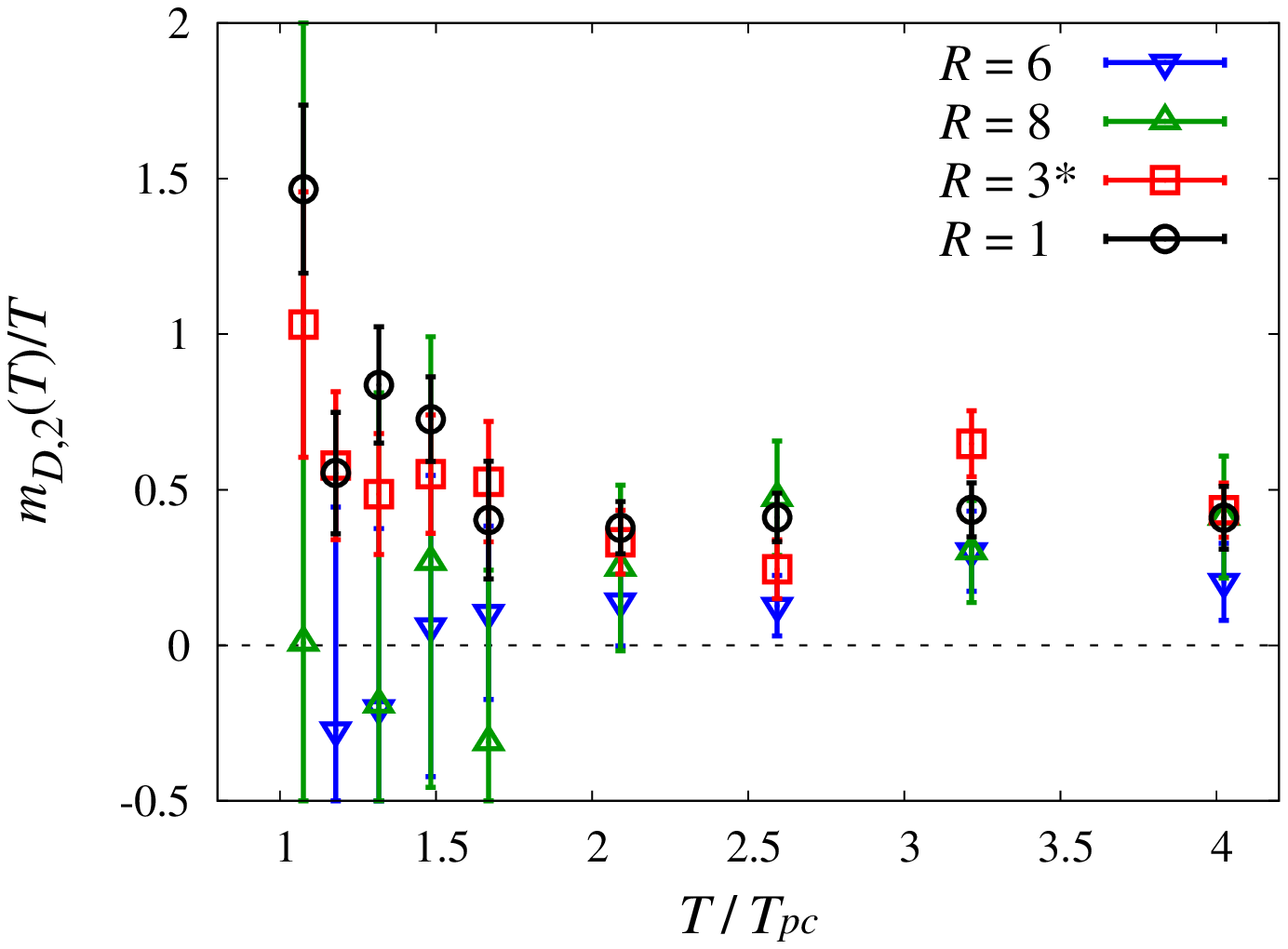} &
    \includegraphics[width=84mm]{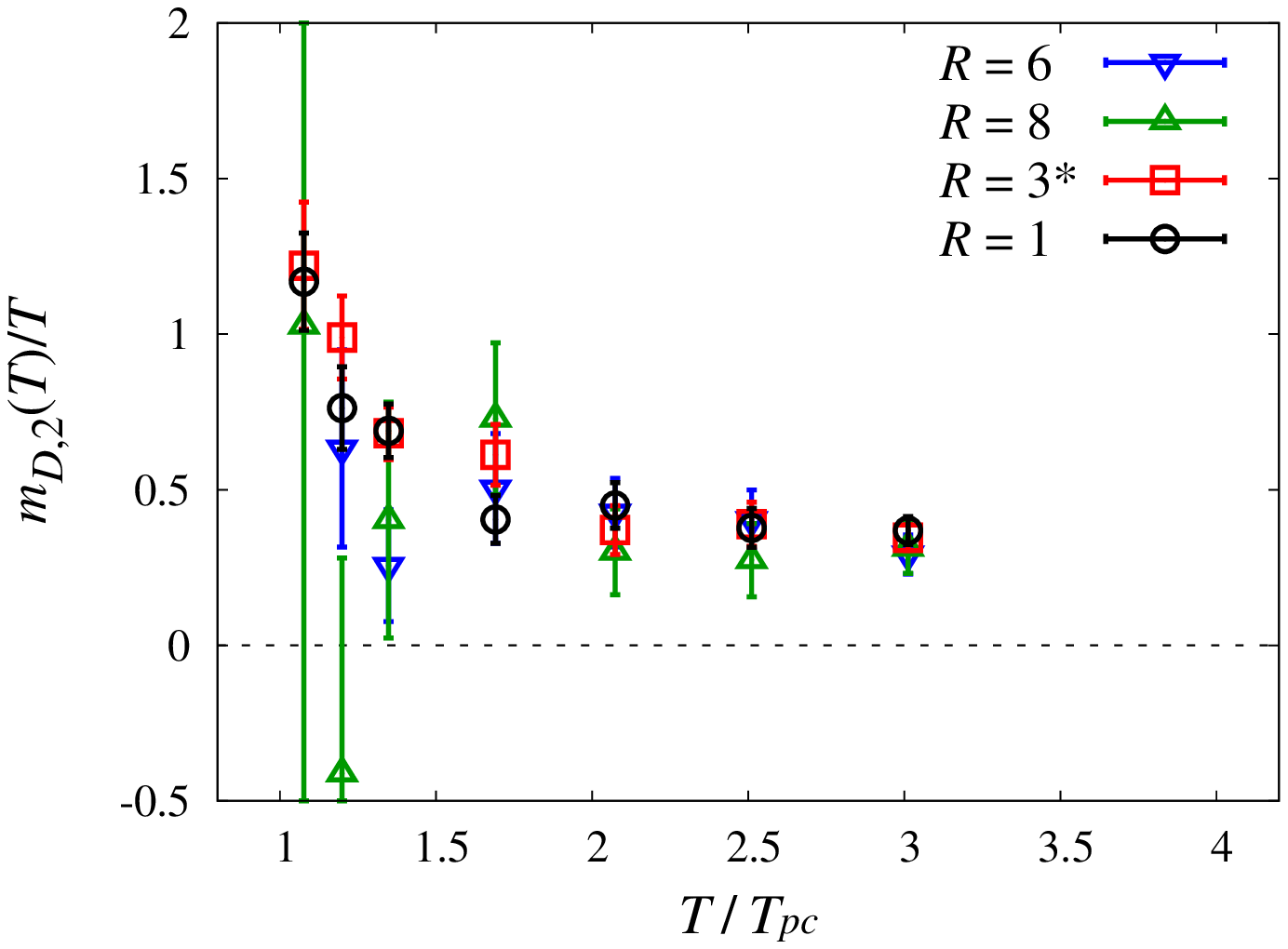}
    \end{tabular}
    \caption{$m_{D,2}(T)$
     at $m_{\rm PS} / m_{\rm V} = 0.65$ (left) and 0.80 (right).
}
  \label{fig:SM2}
  \end{center}
\end{figure}

The results for the first order coefficients $\alpha_1(T)$, which appear only for 
 the color sextet and antitriplet $QQ$ channels, are shown 
 in Fig.~\ref{fig:AL1} for $m_{\rm PS} / m_{\rm V} = 0.65$ (left) 
 and 0.80 (right).
The second order coefficients $\alpha_2(T)$ and $m_{D,2}(T)$ are shown in Figs.~\ref{fig:AL2}
 and \ref{fig:SM2} 
  at $m_{\rm PS} / m_{\rm V} = 0.65$ (left) and 0.80 (right), respectively.
Numerical values of these coefficients are summarized in Appendix \ref{ap:fit}.

\begin{figure}[tbp]
  \begin{center}
    \begin{tabular}{cc}
    \includegraphics[width=84mm]{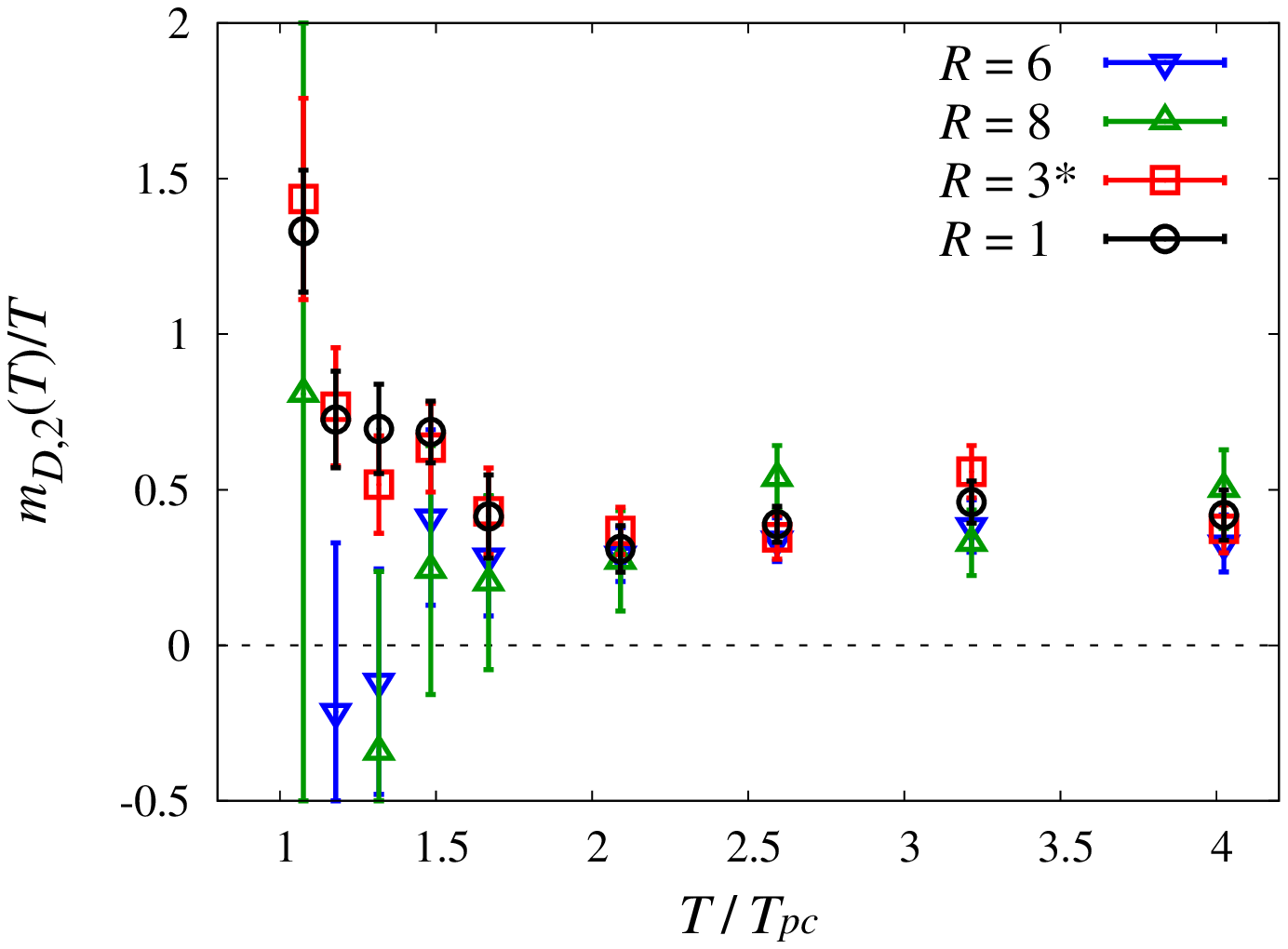} &
    \includegraphics[width=84mm]{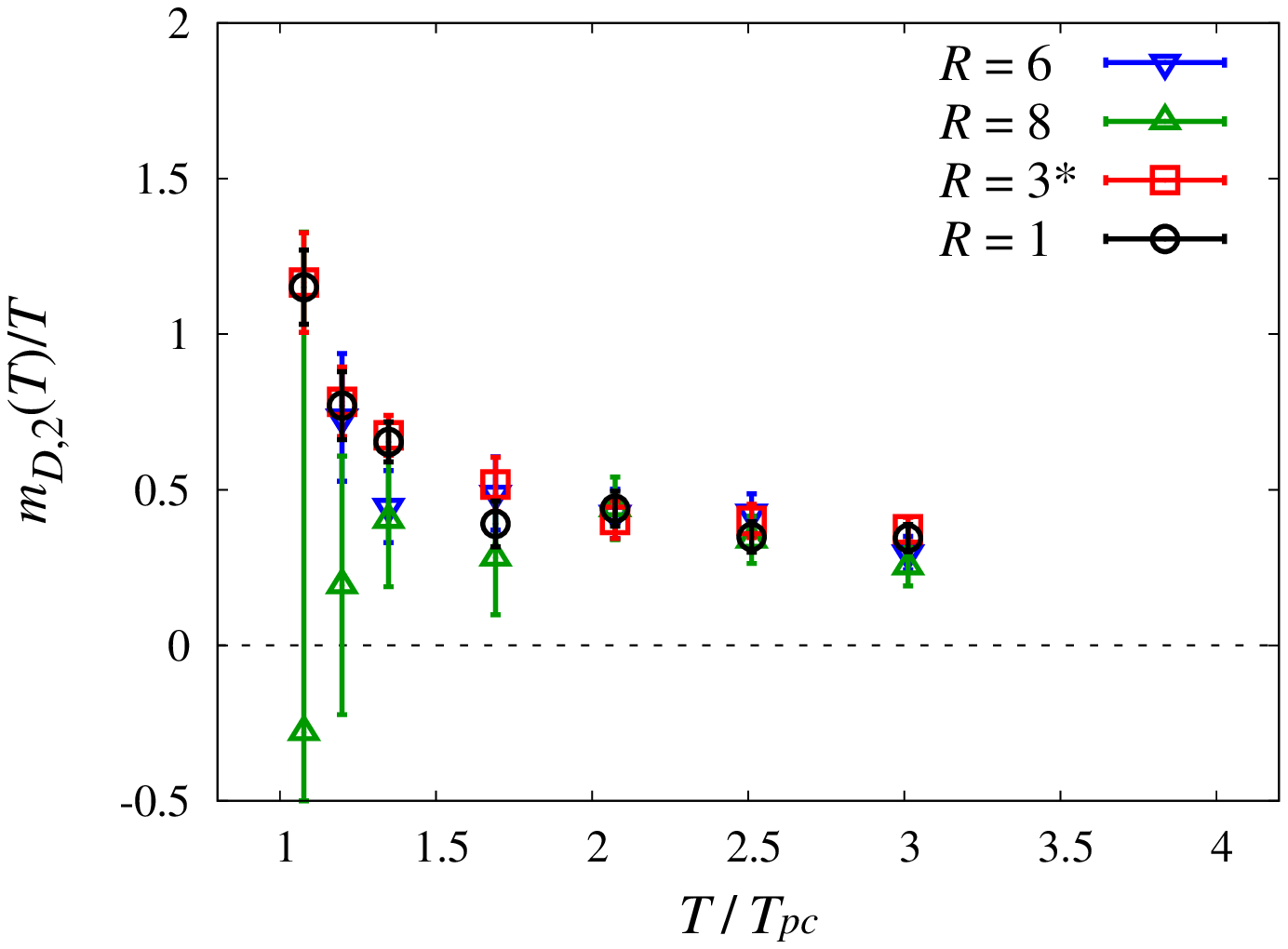}
    \end{tabular}
    \caption{Results of $m_{D,2} (T)$
     assuming $\alpha_2(T) =0$,
     at $m_{\rm PS} / m_{\rm V} = 0.65$ (left) and 0.80 (right).}
  \label{fig:SM2ex}
  \end{center}
\end{figure}

From these Figures, we find that 
 there is no significant channel dependence in these coefficients
  at high temperatures ($T \simge 2 T_{pc}$),
  similar to the case of $\alpha_0(T)$ and $m_{D,0}(T)$ studied in \cite{Maezawa:2007fc}.
We note that $m_{D,2}(T)$ is positive at $T \simge 1.5 T_{pc}$
 which means that magnitude of the Debye mass becomes larger at finite densities 
  in the leading-order of $\mu_{q}$.
This is qualitatively consistent with results calculated with an improved
 staggered quark action for the color-singlet channel \cite{Doring}.
We also find that,
  although $\alpha_1(T)$ remains finite even at $T \simeq 4 T_{pc}$,
 the magnitude of $\alpha_2(T)$ 
  is almost zero for all color channels at $T \simge 1.5 T_{pc}$. 
Therefore, to reduce statistical fluctuations in $m_{D,2}(T)$, we may assume $\alpha_2(T) = 0$ in the fit  
(\ref{eq:2nd}). 
%\begin{eqnarray}
%\frac{v_2}{v_0} &=& - m_{D,2} r.
%\label{eq:2ex}
%\end{eqnarray}
The results are shown in Fig.~\ref{fig:SM2ex}.
% for $m_{\rm PS} / m_{\rm V} = 0.65$ (left) and 0.80 (right) with the fit range of $0.5 \le rT \le 1.0$.
Smallness of the color-channel dependence became clearer.
Numerical values for $\alpha_1(T)$, $\alpha_2(T)$, $m_{D,2}(T)$
 and $m_{D,2} (T; \alpha_2=0)$ are summarized in Tables \ref{tab:1st}--\ref{tab:2ex}
 together with $\chi^2 / N_{\rm DF}$ for each fit.

%%%%%%%%%%%%%%%%%%%%%%%%%%%%%%%%%%%%%%%%%%%%%%%%%%%%%%%%%%%
\subsection{Comparison with the thermal perturbation theory}

%Let us compare the Debye screening mass at finite $T$ and $\mu_{q}$
%  on the lattice with that predicted in the thermal perturbation theory.
The 2-loop running coupling is given by,
\begin{eqnarray}
g^{-2}_{2 {\rm l}} (\kappa) 
=  \beta_0 \ln \left( \frac{\kappa}{\Lambda} \right)^2 + 
\frac{\beta_1}{\beta_0} 
\ln \ln \left( \frac{\kappa}{\Lambda} \right)^2
,
\end{eqnarray}
where $\kappa$ and $\Lambda$ are the renormalization point
 and the QCD scale parameter, respectively.
In the thermal perturbation theory the argument in the logarithms can be decomposed as 
$ \kappa / \Lambda = (\kappa/T)(T/T_{pc})(T_{pc}/\Lambda)$
  where we adopt 
$\Lambda= \Lambda_{\overline{\rm MS}}^{N_f=2} \simeq 261$ MeV
 \cite{Gockeler:2005rv} 
and $T_{pc} \simeq 171$ MeV \cite{cp1}.
We assume that the renormalization point $\kappa$ is 
in the range $\kappa = \pi T$ to $3 \pi T$.
Therefore, $g_{\rm 2l}$ can be viewed as a function of $T/T_{pc}$. 
In the leading order of the thermal perturbation theory,
 the Debye screening mass with $g_{\rm 2l}$ is given by
\begin{eqnarray}
m_D^{\rm LO} (T,\mu_{q}) = g_{2 {\rm l}} (\kappa) 
 \left\{ \left( 1+ \frac{N_f}{6} \right) T^2
+ \frac{N_f}{2 \pi^2} \mu_{q}^2 \right\}^{1/2}
.
\end{eqnarray}
Thus, the leading-order expansion coefficients are given by
\begin{eqnarray}
m_{D,0}^{\rm LO} =  \sqrt{1 + \frac{N_f}{6}} \, g_{2 {\rm l}}(\kappa)T
, \quad
m_{D,2}^{\rm LO} = \frac{1}{4 \pi^2} \frac{N_f}{\sqrt{1+N_f/6}} \, g_{2 {\rm l}}(\kappa) T
.
\label{eq:SMP}
\end{eqnarray}
Taking the ratio of these coefficients we find for $N_f=2$
\begin{eqnarray}
\frac{m_{D,2}^{\rm LO}}{m_{D,0}^{\rm LO}} = \frac{3}{8 \pi^2}
\label{eq:ECR}
.
\end{eqnarray}

\begin{figure}[tbp]
  \begin{center}
    \begin{tabular}{cc}
    \includegraphics[width=84mm]{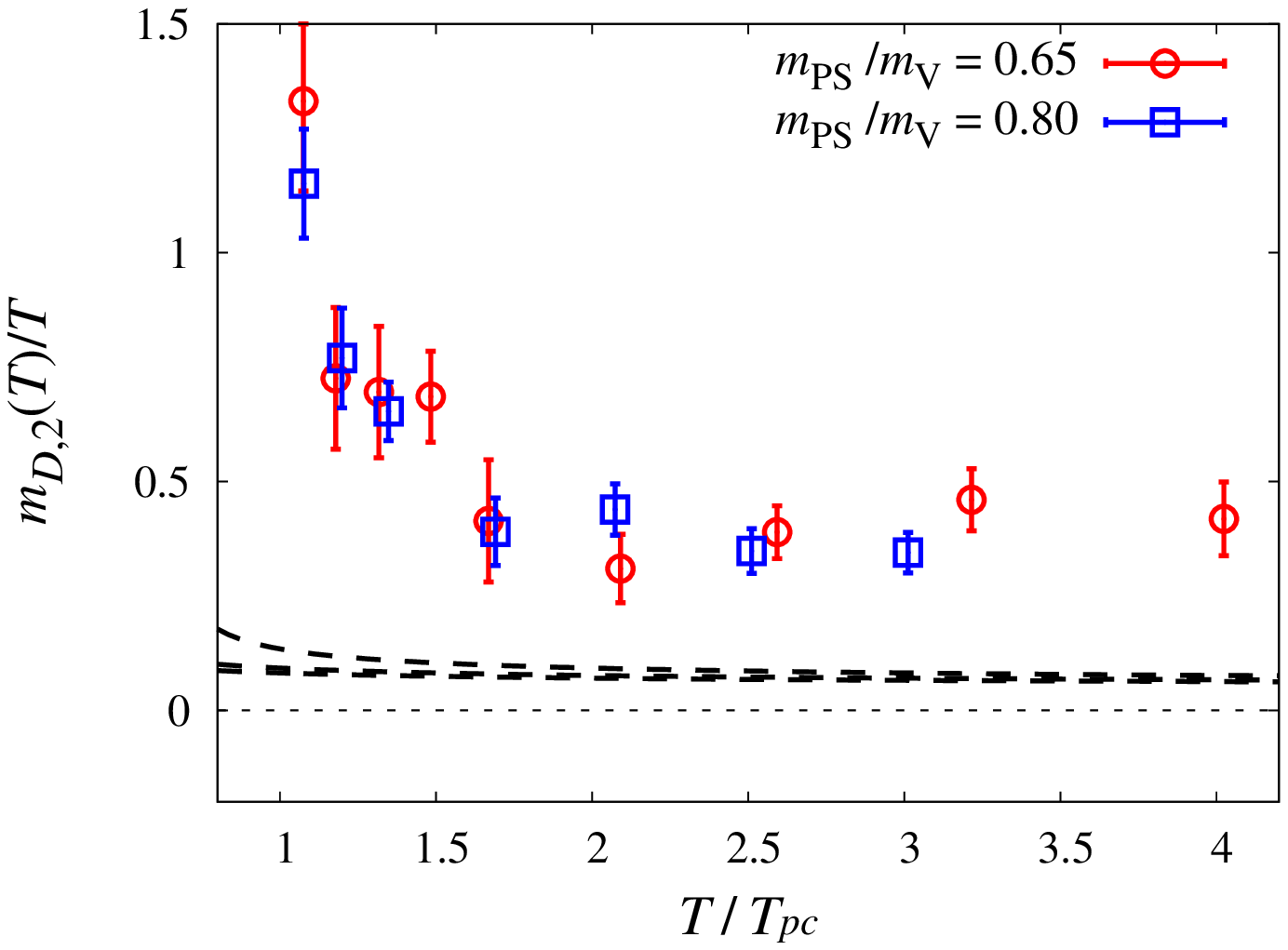} &
    \includegraphics[width=84mm]{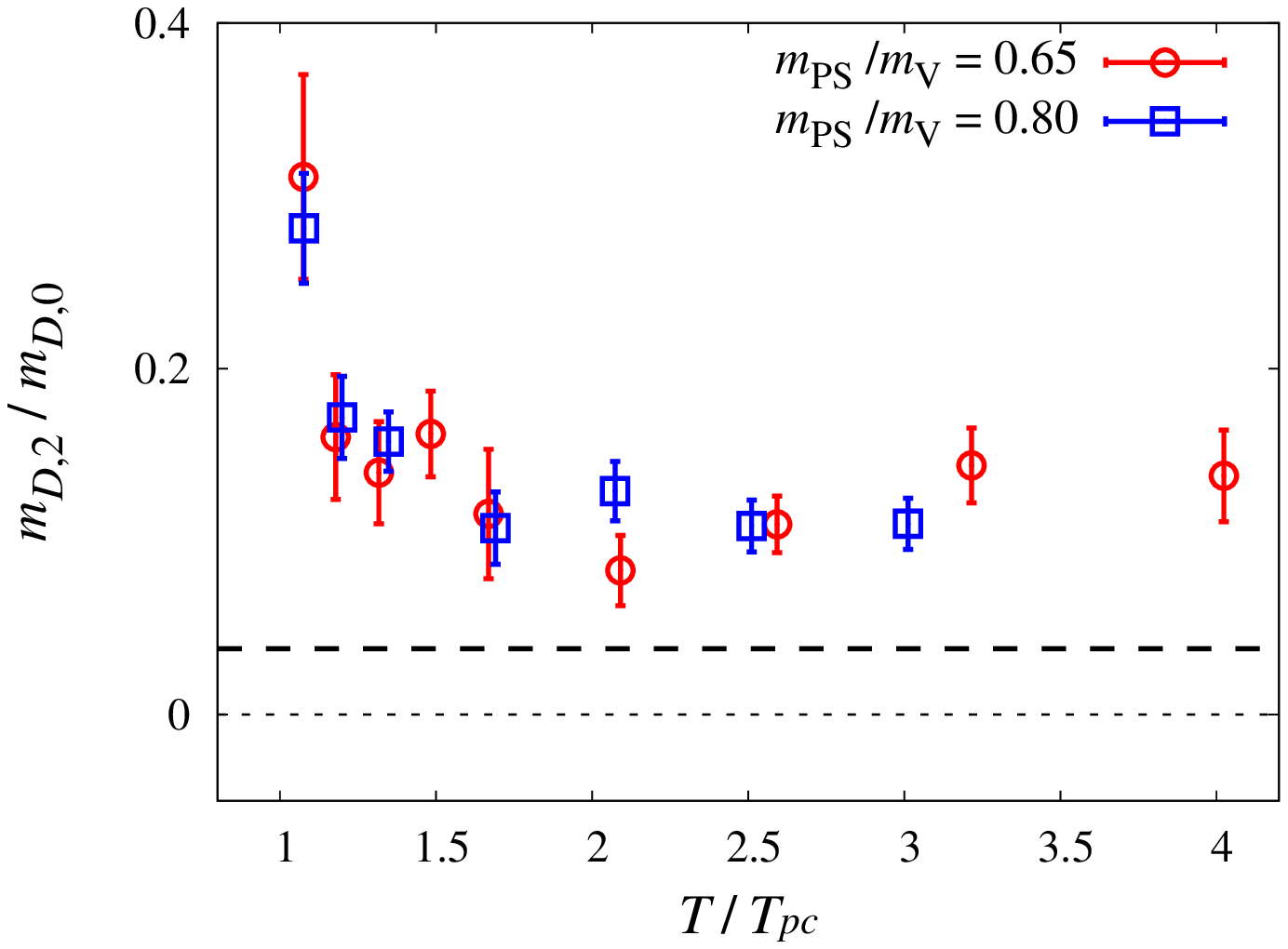}
    \end{tabular}
    \caption{(Left) $m_{D,2}(T)$ for the color singlet channel. Dashed lines represent 
    the prediction of the leading-order thermal perturbation theory 
    for $ \kappa = \pi T$, $2\pi T$ and $3 \pi T$ from above.
    (Right) $m_{D,2} / m_{D,0}$ for the color singlet channel. 
     The dashed line at  $m_{D,2} / m_{D,0} = 3/(8\pi^2)$ represents the prediction from the leading-order thermal perturbation theory.}
  \label{fig:SM2_pQCD}
  \end{center}
\end{figure}

In Ref.~\cite{Maezawa:2007fc}, we found that, at $\mu_q=0$, the leading-order thermal perturbation theory predicts much smaller values for $m_{D,0}(T)$ than the lattice results. 
In the left panel of Fig.~\ref{fig:SM2_pQCD}, we compare our results of $m_{D,2}(T)$ for the 
 color singlet channel with that of the leading-order thermal perturbation theory.  
Similar to the case of $m_{D,0}(T)$, we find that the lattice results of $m_{D,2}(T)$ are much larger than the prediction of the thermal perturbation theory at the leading-order. 

In Fig.~\ref{fig:SM2_pQCD} (right), we plot the lattice results for the ratio $m_{D,2}/ m_{D,0}$ and compare them with (\ref{eq:ECR}).
We find that this ratio also deviates from the prediction of the leading-order thermal perturbation theory.
We note that, with the p4-improved staggered quark action, the ratio $m_{D,2}/ m_{D,0}$ was reported to agree with  $3/8 \pi^2$ at $T \simge 1.5 T_{pc}$ \cite{Doring}.
Similar discrepancy between Wilson and staggered type quark actions has been already reported for Debye screening masses at $\mu_q=0$ \cite{Maezawa:2007fc}.
Further investigations at smaller lattice spacings etc. are required to clarify the origin of the discrepancy.
At $\mu_q=0$, it was shown that the discrepancy with the thermal perturbation theory is largely removed for $m_{D,0}(T)$ with the improved Wilson quark action when we include the next-to-leading-order contributions \cite{Maezawa:2007fc}. 
Thus, a higher order calculation of the thermal perturbation theory at finite $\mu_{q}$ will also be important to understand the results obtained on the lattice.

%%%%%%%%%%%%%%%%%%%%%%%%%%%%%%%%%%%%%%%%%%%%%%%%%%%%%%%%%%%%%
\section{Conclusions}
\label{sec:conc}

Comparison of results obtained by different lattice formulations is important to estimate theoretical uncertainties in lattice QCD calculations. 
Since most lattice QCD simulations  at finite temperatures and densities have been performed using staggered-type quark actions so far, studies with a different lattice quark action is particularly important.
In this paper, we carried out the first calculation of the equation of state at nonzero densities with two flavors of improved Wilson quarks.  
Simulations are performed on a $16^3 \times 4$ lattice along 
the lines of constant physics corresponding to $m_{\rm PS}/m_{\rm V} = 0.65$ and 0.80 
in the $(\beta,K)$ plane. 
With Wilson-type quarks, statistical fluctuations of physical observables at finite density are much severer than with staggered-type quarks.  
To tame the problem, we combined and developed several improvement techniques.

Adopting the Taylor expansion method, we calculated the derivatives of pressure with respect to the chemical potentials $\mu_q$ and $\mu_I$ up to the fourth order . 
Using these derivatives, we studied the fluctuations of quark number and isospin densities at finite chemical potentials.
A quantitative difference between the second derivatives of $\chi_q$ and $\chi_I$ 
was observed:
$\chi_q$ shows a peak near $T_{pc}$, whose height increases as $\mu_q$ increases, 
whereas $\chi_I$ does not show a clear peak near $T_{pc}$. 
These behaviors agree qualitatively with the results obtained using p4-improved 
staggered fermions, and are consistent with the expectation from the effective sigma model.

With the current statistics, the statistical errors in the results were not small with the simple Taylor expansion method.
To improve the calculation, we adopted a hybrid method of the Taylor expansion and the reweighting techniques combined with a Gaussian approximation for the distribution of the complex phase of the quark determinant. 
In a previous study with a staggered-type quark \cite{eji07}, this method was shown to be efficient to suppress statistical fluctuations at finite densities.
We found that the statistical errors in the quark number density and the susceptibility at finite densities are reduced with the new method.
Although the simulations at different temperatures are independent, the resulting $T$-dependence in the quark number density and the susceptibility turned out to be smooth, and the heap in $\chi_q$ near $T_{pc}$ became clearer.
These results suggest that the sign problem at finite densities is mildened by such improvements.

We also studied the heavy-quark free energies and the Debye screening mass at finite densities in the high temperature phase.
We calculated the Taylor expansion coefficients of the heavy-quark free energies 
in all color channels up to the second order in $\mu_{q}/T$.
We found a characteristic difference between $Q\bar{Q}$ and $QQ$ free energies:
The inter-quark interactions between $Q$ and $\bar{Q}$ become week, while those between $Q$ and $Q$ become strong, as $\mu_{q}$ increases.
We also calculated the effective running coupling and
   the Debye screening mass for each
 color channel up to the second order of $\mu_{q}$.
Both quantities show no significant color channel dependence at $T \simge 2 T_{pc}$.
The second order coefficient of Debye screening mass, $m_{D,2}(T)$, turned out to be positive, 
implying that the Debye mass becomes larger as $\mu_{q}$ increases.
We note that our $m_{D,2}(T)$ does not agree with the leading-order thermal perturbation theory.
Higher orders are required to explain the lattice results.

\section*{Acknowledgements}
We would like to thank K.-I.~Ishikawa and the members of the CP-PACS 
Collaboration for providing us with the basic code 
for generating the configurations.
This work is in part supported 
by Grants-in-Aid of the Japanese Ministry
of Education, Culture, Sports, Science and Technology, 
(Nos.~17340066, 18540253, 19549001, 20340047, 21340049). 
SE is supported by U.S.\ Department of Energy (DE-AC02-98CH10886). 
This work is in part supported 
also by the Large-Scale Numerical Simulation
Projects of CCS/ACCC, Univ.~of Tsukuba, 
and by the Large Scale Simulation Program of High Energy
Accelerator Research Organization (KEK) 
Nos.06-19, 07-18, 08-10 and 09-18.

%%%%%%%%%%%%%%%%%%%%%%%%%%%%%%%%%%%%%%%%%%%%%%%%%%%%%%%%%%%%%%%%%%%%%

\appendix

%\section*{Appendix A: Pressure and quark number susceptibility in the free gas limit} 
\section{Pressure and quark number susceptibility in the free gas limit} 
\label{ap:free}

In order to estimate the equation of state at nonzero quark 
chemical potential, $\mu=\mu_q a$ in the high temperature limit, 
we calculate the pressure and its derivatives with respect to $\mu$ 
in the free quark gas limit.
Because the effect of finite quark mass becomes negligible 
in the high temperature limit,
we discuss only the case of massless quarks.

The partition function for free Wilson quarks is give by 
\begin{eqnarray}
{\cal Z}(K, \mu) &=& (\det M)^{N_{\rm f}},
\label{eq:partition_free} \\
  M_{xy} &=& \delta_{x,y}
    -K \sum_{i} \left[(1-\gamma_i) \delta_{x+\hat{i},y}
    +(1+\gamma_i) \delta_{x-\hat{i},y}\right] \nonumber \\
  && -K \left[{\rm e}^{\mu} (1-\gamma_4) \delta_{x+\hat{4},y}
    +{\rm e}^{-\mu} (1+\gamma_4) \delta_{x-\hat{4},y} \right],
  \label{eq:fermact_free}
\end{eqnarray}
on an $N_s^3 \times N_t$ lattice.
Note that the clover term vanishes for free quarks. 
We perform a unitary transformation into momentum space 
(Fourier transformation).
\begin{eqnarray}
\tilde{M}_{kl} \equiv \frac{1}{N_s^3 N_t} \sum_{x,y}
e^{-ikx+ily} M_{xy} \equiv U^{\dag}_{kx} M_{xy} U_{yl}.
\end{eqnarray}
Here
\begin{eqnarray}
U_{yl} & \equiv & \frac{1}{\sqrt{N_s^3 N_t}} e^{ily}, \hspace{5mm}
U^{\dag}_{kx} \ \equiv \ \frac{1}{\sqrt{N_s^3 N_t}} e^{-ikx},
\nonumber \\ 
U^{\dag}_{kx} U_{xl} &=& \frac{1}{N_s^3 N_t} \sum_x e^{ix(l-k)} 
= \delta_{k,l}, \hspace{5mm} 
\det (U^{\dag} U) = \det U^{\dag} \det U = 1
\end{eqnarray}
We then calculate the partition function,
\begin{eqnarray}
{\cal Z}(K, \mu) &=&  (\det M)^{N_{\rm f}}
 = (\det \tilde{M})^{N_{\rm f}}, \\
&& \hspace{-18mm}
\tilde{M}_{kl} = \frac{1}{N_s^3 N_t} \sum_{x} \left[ 
e^{-ix(k-l)} \left[ 1-K \sum_{i=1}^3 \left((1-\gamma_i) e^{il_i}
    +(1+\gamma_i) e^{-il_i} \right) \right. \right. \nonumber \\
&& \left. \left. -K \left( e^{\mu} (1-\gamma_4) e^{il_4}
    +e^{-\mu} (1+\gamma_4) e^{-il_4} \right) \right] \right] 
\nonumber \\ && \hspace{-15mm}
= \delta_{k,l} \left[ 1-K \sum_{i=1}^3 \left( 2\cos k_i 
    -2i \gamma_i \sin k_i \right) -K \left( 2\cos(k_4 -i\mu) 
    -2i \gamma_4 \sin(k_4 -i\mu) \right) \right],
\end{eqnarray}
where
\begin{eqnarray}
k_{\mu} &=& \frac{2 \pi j_{\mu}}{N_s}, \ \ 
j_{\mu}=0, \pm1, \cdots, N_s/2 \hspace{8mm} {\rm for \ \mu=1,2,3} \\
k_{4} &=& \frac{2 \pi (j_{4}+ 1/2)}{N_t}, \ \ 
j_{4}=0, \pm1, \cdots, N_t/2.
\end{eqnarray}
Introducing a $4 \times 4$-matrix which is defined by 
$\tilde{M}_{kl} = \delta_{k,l} \tilde{M}(k)$, 
\begin{eqnarray}
{\cal Z}(K, \mu) &=& \left( \prod_k \det \tilde{M}(k) 
\right)^{3N_{\rm f}}, 
\nonumber \\
\det \tilde{M}(k) \hspace{-2mm}
&=& \hspace{-2mm} \det \left[ 1-K \sum_{i=1}^3 \left( 2\cos k_i 
    -2i \gamma_i \sin k_i \right) -K \left( 2\cos(k_4 -i\mu) 
    -2i \gamma_4 \sin(k_4 -i\mu) \right) \right] 
\nonumber \\ && \hspace{-19mm}
= \left[ \left( 1-2K \sum_{i=1}^3 \cos k_i -2 \cos(k_4 -i\mu) \right)^2
    +4K^2 \sum_{i=1}^3 \sin^2 k_i +4K^2 \sin^2 (k_4 -i\mu) \right]^2
\nonumber \\ && \hspace{-19mm}
= \left[ \left( 1-8K+4K \sum_{i=1}^3 \sin^2 \left( \frac{k_i}{2} \right) 
    +4K \sin^2 \left( \frac{k_4 -i\mu}{2} \right) \right)^2 \right.
\nonumber \\ &&  \left.
    +4K^2 \sum_{i=1}^3 \sin^2 k_i +4K^2 \sin^2 (k_4 -i\mu) \right]^2
\nonumber \\ && \hspace{-19mm}
= \left[ (1-8K)^2 +8K(1-8K) \left( 
\sum_{i=1}^3 \sin^2 \left( \frac{k_i}{2} \right) 
+ \sin^2 \left( \frac{k_4 -i\mu}{2} \right) \right) \right.
\nonumber \\ && \hspace{-19mm}
\left. +4K^2 \left[ \left( 2\sum_{i=1}^3 \sin^2 
\left( \frac{k_i}{2} \right) \right)^2 
+4 \left( 2\sum_{i=1}^3 \sin^2 \left( \frac{k_i}{2} \right) 
+1 \right) \sin^2 \left( \frac{k_4 -i\mu}{2} \right) 
+ \sum_{i=1}^3 \sin^2 k_i \right] \right]^2 .
\end{eqnarray}
$( \det (a_0I +a_1 i \gamma_1 +a_2 i \gamma_2 +a_3 i \gamma_3 
+a_4 i \gamma_4) = (a_0^2+a_1^2+a_2^2+a_3^2+a_4^2)^2)$
\vspace{5mm}

In the massless quark limit $K=1/8$,
\begin{eqnarray}
\det \tilde{M} (k)
= \frac{16}{8^4} \left[ A(k) + B^2(k) +4(B(k)+1) 
\sin^2 \left( \frac{k_4 -i\mu}{2} \right) \right]^2 , 
\end{eqnarray}
where
\begin{eqnarray}
A(k)=\sum_{i=1}^3 \sin^2 k_i, \hspace{5mm}
B(k)=2\sum_{i=1}^3 \sin^2 \left( \frac{k_i}{2} \right). 
\end{eqnarray}

We calculate the derivatives of pressure with respect to $\mu$
at $\mu=0, K=1/8$ numerically. 
\begin{eqnarray}
\frac{p}{T^4} &=& N_t^4 \left( \frac{1}{N_s^3 N_t} \ln {\cal Z}(T,\mu) 
-\frac{1}{N_s^4} \ln {\cal Z}(T=0, \mu=0) \right), \\
c_n &=& \left.
\frac{1}{n!} \frac{\partial^n (p/T^4)}{\partial (\mu_q/T)^n} \right|_{\mu=0}
=  \frac{N_t^{3-n}}{N_s^3} \left.
\frac{\partial^n \ln {\cal Z}(T)}{\partial \mu^n}  \right|_{\mu=0}. 
%\hspace{10mm} \left( 
%\frac{\partial \ln {\cal Z}(T=0, \mu=0)}{\partial \mu}=0 \right)
\end{eqnarray}
Here, ${\cal Z}(T, \mu)$ and ${\cal Z}(T=0, \mu)$ are the partition functions 
calculated on $N_s^3 \times N_t$ and $N_s^4$ lattices, respectively, 
The derivative of the normalization $\ln {\cal Z}(T=0, \mu=0)$ in $\mu$ is, 
of course, zero.
The derivatives of $\ln {\cal Z}$ at $\mu=0$ are given by  
\begin{eqnarray}
\frac{\partial \ln {\cal Z}}{\partial \mu} &=& 
3N_{\rm f} \frac{\partial}{\partial \mu} \sum_k \ln \det \tilde{M}(k)
=6N_{\rm f} \sum_k \left( \frac{{\cal D}_1(k)}{{\cal D}_0(k)} \right),
\\
\frac{\partial^2 \ln {\cal Z}}{\partial \mu^2} &=&
3N_{\rm f} \frac{\partial^2}{\partial \mu^2} \sum_k \ln \det \tilde{M}(k)
=6N_{\rm f} \sum_k \left( \frac{{\cal D}_2(k)}{{\cal D}_0(k)}
- \frac{{\cal D}_1^2(k)}{{\cal D}_0^2(k)} \right),
\\
\frac{\partial^3 \ln {\cal Z}}{\partial \mu^3} &=& 
3N_{\rm f} \frac{\partial^3}{\partial \mu^3} \sum_k \ln \det \tilde{M}(k)
=6N_{\rm f} \sum_k \left( \frac{{\cal D}_3(k)}{{\cal D}_0(k)}
-3 \frac{{\cal D}_2(k) {\cal D}_1(k)}{{\cal D}_0^2(k)}
+2 \frac{{\cal D}_1^3(k)}{{\cal D}_0^3(k)} \right),
\\
\frac{\partial^4 \ln {\cal Z}}{\partial \mu^4} &=& 
3N_{\rm f} \frac{\partial^4}{\partial \mu^4} \sum_k \ln \det \tilde{M}(k)
\nonumber \\ && \hspace{-8mm}
=6N_{\rm f} \sum_k \left( \frac{{\cal D}_4(k)}{{\cal D}_0(k)}
-4 \frac{{\cal D}_3(k) {\cal D}_1(k)}{{\cal D}_0^2(k)}
-3 \frac{{\cal D}_2^2(k)}{{\cal D}_0^2(k)} 
+12 \frac{{\cal D}_2(k) {\cal D}_1^2(k)}{{\cal D}_0^3(k)}
-6 \frac{{\cal D}_1^4(k)}{{\cal D}_0^4(k)} \right),
\end{eqnarray}
where
\begin{eqnarray}
{\cal D}_0 &=& A(k) + B^2(k) +4\,[B(k)+1] \sin^2 (k_4/2) , \\
{\cal D}_{n: {\rm odd}} &=& -2i \,[B(k)+1] \sin k_4 , \\ 
{\cal D}_{n: {\rm even}} &=& -2 \,[B(k)+1] \cos k_4 .
\end{eqnarray}

\begin{figure}[t]
\begin{center}
\begin{tabular}{cc}
\includegraphics[width=2.4in]{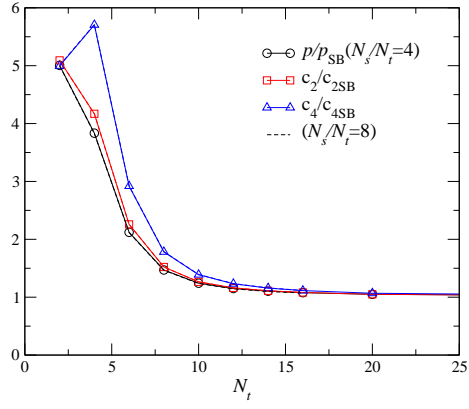}
\end{tabular}
\vspace*{-2mm}
\caption{
The results of $p(\mu=0)$ (circle), $c_2$ (square) and $c_4$ (triangle) 
normalized by the values of their continuum limit. 
}
\label{fig:freegas}
\end{center}
%\vspace{-0.5cm}
\end{figure}

The odd derivatives vanish as in the case of interacting quarks.
Since $c_n = 0 $ for $n>4$ in the continuum limit, we calculate $p$ at $\mu=0$ as well as $c_2$ and $c_4$.
The numerical results normalized by the values of their continuum Stephan-Boltzmann limit 
are plotted in Fig.~\ref{fig:freegas}.
Circle, square and triangle symbols are the results of $p(\mu=0)$, $c_2$ 
and $c_4$ for each $N_t$ with $N_s/N_t=4$, respectively. 
The results with $N_s/N_t=8$ are also shown by the dashed lines. 
The $N_s$ dependence is found to be negligible. 
However, the results are much larger than unity for small $N_t$, 
suggesting sizable lattice discretization effects for $N_t < 10$.

%%%%%%%%%%%%%%%%%%%%%%%%%%%%%%%%%%%%%%%%%%%%%%%%%%%%
\section{Derivatives of $\ln {\cal Z}$ in the Gaussian approximation} 
\label{ap:gaussian}

We discuss the error from the Gaussian approximation of 
a complex phase distribution function. 
We calculate the second and forth derivatives of $\ln {\cal Z}$ when 
the Gaussian approximation is applied, and compare with the exact results.

Denoting the derivative of $\ln \det M$ as 
\begin{eqnarray}
{\cal D}_n \equiv \left. N_{\rm f}
\frac{\partial^n \ln \det M(\mu)}{\partial \mu^n} \right|_{\mu=0},
\end{eqnarray}
the partition function with the Gaussian approximation Eq.~(\ref{eq:parap}) 
can be expanded in a power series, 
\begin{eqnarray}
\frac{{\cal Z}(\mu)}{{\cal Z}(0)} & \approx & \left\langle 
\exp \left[ F -\frac{\left\langle \theta^2 \right\rangle_F}{2} \right] 
\right\rangle_{(\mu=0)} 
= \int \exp \left[ -\frac{\left\langle \theta^2 \right\rangle_F}{2} \right] 
\langle \exp(F) \rangle_F \ w_0 (F) \ dF \nonumber \\
%&=& \int \exp \left[ -\frac{\left\langle \theta^2 \right\rangle_F}{2} \right] 
%\left\langle \exp \left[ \frac{{\cal D}_2 \mu^2}{2} + \frac{{\cal D}_4 \mu^4}{4!} 
%+ \cdots \right] \right\rangle_F w_0 (F) \ dF \nonumber \\
&=& \int \exp \left[ \frac{\left\langle {\cal D}_1^2 \right\rangle_F \mu^2}{2}
+\frac{\left\langle {\cal D}_1 {\cal D}_3 \right\rangle_F \mu^4}{3!}
+\frac{\left\langle {\cal D}_3^2 \right\rangle_F 
\mu^6}{2 \times (3!)^2} + \cdots \right] 
\nonumber \\ && \hspace{-5mm} \times 
\exp \left[ \frac{\left\langle {\cal D}_2 \right\rangle_F \mu^2}{2} 
+ \frac{\left\langle {\cal D}_4 \right\rangle_F \mu^4}{4!} 
+ \frac{\left\langle {\cal D}_2^2 \right\rangle_F \mu^4}{8}
- \frac{\left\langle {\cal D}_2 \right\rangle_F^2 \mu^4}{8}
+ \cdots \right] w_0 (F) \ dF .
\end{eqnarray}
We then obtain the second and forth derivatives of $\ln {\cal Z}$ at $\mu=0$.
The second derivative is 
\begin{eqnarray}
\left. \frac{\partial \ln {\cal Z}}{\partial (\mu^2)} \right|_{\mu=0} 
= \left. \frac{1}{2} \frac{\partial^2 \ln {\cal Z}}{\partial \mu^2} 
\right|_{\mu=0} 
= \int \left( \frac{\left\langle {\cal D}_1^2 \right\rangle_F}{2}
+ \frac{\left\langle {\cal D}_2 \right\rangle_F}{2} 
\right) w_0 (F) \ dF
=\frac{1}{2} \left( \left\langle {\cal D}_1^2 \right\rangle + 
\left\langle {\cal D}_2 \right\rangle \right).
\end{eqnarray}
This result is, of course, the same as the exact result.
Next, we calculate the forth derivative.
\begin{eqnarray}
\left. \frac{\partial^2 \ln {\cal Z}}{\partial (\mu^2)^2} \right|_{\mu=0} 
&=& \left. \frac{1}{12} 
\frac{\partial^4 \ln {\cal Z}}{\partial \mu^4} \right|_{\mu=0}
\nonumber \\
&& \hspace{-1cm} =  
\int 
\left( \frac{\left\langle {\cal D}_1 {\cal D}_3 \right\rangle_F}{3}
+ \frac{\left\langle {\cal D}_4 \right\rangle_F}{12} 
+ \frac{\left\langle {\cal D}_2^2 \right\rangle_F}{4}
- \frac{\left\langle {\cal D}_2 \right\rangle_F^2}{4}
+\left( \frac{\left\langle {\cal D}_1^2 \right\rangle_F}{2}
+ \frac{\left\langle {\cal D}_2 \right\rangle_F}{2} \right)^2 \right) 
w_0 (F) \ dF 
\nonumber \\ &&
- \left( \int 
\left( \frac{\left\langle {\cal D}_1^2 \right\rangle_F}{2}
+ \frac{\left\langle {\cal D}_2 \right\rangle_F}{2} \right) 
w_0 (F) \ dF \right)^2 
\nonumber \\ && \hspace{-1cm} = 
\frac{\left\langle {\cal D}_1 {\cal D}_3 \right\rangle}{3}
+ \frac{\left\langle {\cal D}_4 \right\rangle}{12} 
+ \frac{\left\langle {\cal D}_2^2 \right\rangle}{4}
+ \int \left(
\frac{\left\langle {\cal D}_1^2 \right\rangle_F^2}{4}
+\frac{\left\langle {\cal D}_1^2 \right\rangle_F
\left\langle {\cal D}_2 \right\rangle_F}{2}
\right) w_0 (F) \ dF
- \frac{1}{4} \left( \left\langle {\cal D}_1^2 \right\rangle
+ \left\langle {\cal D}_2 \right\rangle \right)^2 .
\end{eqnarray}
Because $F= {\cal D}_2(\mu^2/2) + O(\mu^4)$ and 
$ \int \left\langle \cdots \right\rangle_F {\cal O}[F]
w_0 (F) dF = \left\langle \cdots {\cal O}[F] \right\rangle $ 
for any function of $F$: ${\cal O}[F]$, 
\begin{eqnarray}
\int \left\langle {\cal D}_1^2 \right\rangle_F
\left\langle {\cal D}_2 \right\rangle_F w_0 (F) \ dF
= \left\langle {\cal D}_1^2 {\cal D}_2 \right\rangle + O(\mu^2).
\end{eqnarray}

Moreover,  as discussed in Sec.\ref{sec:gaussphase}, 
${\cal D}_1$ is given by a sum of the local number density 
operator $(\sim \bar{\psi} \gamma_0 \psi(x))$ at $\mu_q=0$. 
If the simulation is performed apart from a singular point, 
we may adopt the Gaussian approximation for the distribution of ${\cal D}_1$.
In such a case, ${\cal D}_1$ satisfies 
\begin{eqnarray}
\left\langle {\cal D}_1^2 \right\rangle_F^2
\approx \frac{1}{3} \left\langle {\cal D}_1^4 \right\rangle_F.
\end{eqnarray}

%Moreover, for the case that the distribution function $w(F, \theta)$ 
%is given by 
%\begin{eqnarray}
%w(|F|, \theta) \approx \sqrt{\frac{a_2(F)}{\pi}} 
%\left( 1-\frac{3a_4(F)}{4a_2^2 (F)} + \cdots \right)^{-1} w_0 (F) 
%e^{-[a_2(F) \theta^2 + a_4(F) \theta^4 + \cdots ]} 
%\label{eq:wga}
%\end{eqnarray}
%with small $a_4/a_2$. the ratio of $\langle \theta^4 \rangle_F$ and 
%$\langle \theta^2 \rangle_F^2$ can be evaluated by 
%\begin{eqnarray}
%\frac{\left\langle \theta^4 \right\rangle_F }{ 
%\left\langle \theta^2 \right\rangle_F^2 } 
%=3 - 3 \frac{2a_4 (F)}{a_2^2 (F)} 
%+ O \left[ \left( \frac{a_4}{a_2} \right)^2 \right]. 
%\label{eq:a2a4}
%\end{eqnarray}
%Because $1/a_2 \sim O(\mu^2)$, $a_4/a_2^2$ vanishes 
%in the $\mu \to 0$ limit unless 
%$a_4/a_2$ diverges.
%Hence, we obtain 
%$ \left\langle \theta^4 \right\rangle_F = 
%3 \left\langle \theta^2 \right\rangle_F^2$ in the $\mu \to 0$ limit.

Substituting this equation, the forth derivative becomes
\begin{eqnarray}
\frac{\partial^4 \ln {\cal Z}}{\partial \mu^4} 
= 4 \left\langle {\cal D}_1 {\cal D}_3 \right\rangle
+ \left\langle {\cal D}_4 \right\rangle 
+ 3 \left\langle {\cal D}_2^2 \right\rangle
+ \left\langle {\cal D}_1^4 \right\rangle
+ 6 \left\langle {\cal D}_1^2 {\cal D}_2 \right\rangle
- 3 \left( \left\langle {\cal D}_1^2 \right\rangle
+ \left\langle {\cal D}_2 \right\rangle \right)^2.
\end{eqnarray}
This is the same as the exact result. 
In this calculation, we assumed that 
the distribution function of the total quark number, ${\cal D}_1$, 
is of Gaussian at $\mu_q=0$.
%the spatial density correlation 
%is small in comparison with the size of the lattice. 
%If a simulation is performed at a point away from the critical point 
%with sufficiently large volume, this assumption is valid.
Within this condition, we find that the Gaussian approximation does not affect 
the calculation of the derivatives of $\ln {\cal Z}$ up to $O(\mu^4)$. 
Similar discussion is also possible for the higher order terms of 
$\mu$ and one can find out the condition in which the Gaussian 
approximation is valid for each order of $\mu$. 

%%%%%%%%%%%%%%%%%%%%%%%%%%%%%%%%%%%%%%%%%%%%%%%%%%%%
\section{Results of expansion coefficients for $\alpha_{\rm eff}$ and $m_D$}
\label{ap:fit}

To evaluate expansion coefficients 
 of $\alpha_{\rm eff}$ and $m_D$, 
 we fit the normalized free energies with (\ref{eq:1st})--(\ref{eq:2nd}).
Our results of the expansion coefficients together with the quality of the fits are summarized in Tables \ref{tab:1st}--\ref{tab:2ex}.
We adopt the fit ranges $0.5 \le rT \le 1.0$ for Eq.~(\ref{eq:1st})
 and $0.25 \le rT \le 1.0$ for Eq.~(\ref{eq:2nd}).
These fit ranges are chosen by examining the fit range dependence as follows.
Let us denote the fit range as $R_{\rm ini} \le rT \le R_{\rm fin}$.
We find that the fit results are insensitive to $R_{\rm fin}$ when $R_{\rm fin}$ is sufficiently large.
To evaluate the sensitivity on $R_{\rm ini}$, we introduce $R_{{\rm ini}+2}$ as the next-neighboring longer distance on the lattice. 
For example, when $R_{\rm ini} = 0.5$ at $N_t=4$, the lattice distance of the point is 2 and the next-neighboring longer distance is $\sqrt{1^2+1^2+2^2} = \sqrt{6}$, and thus $R_{{\rm ini}+2} = \sqrt{6}/4$.
Similarly, when $R_{\rm ini} = 0.25$ at $N_t=4$,  $R_{{\rm ini}+2} = \sqrt{3}/4$.
Then, we estimate the systematic error due to the fit range by the difference of the fit results
 between $R_{\rm ini}$ 
  and $R_{{\rm ini}+2}$ with fixed $R_{\rm fin}$.
The systematic errors are shown in the second parentheses
 for $\alpha_1(T)$ of color-antitriplet channel in Tab.~\ref{tab:1st},
 for $m_{D,2}(T)$ of color-singlet channel in Tabs.~\ref{tab:2nd_SM}
 and %for $m_{D,2}(T)$ of color singlet channel with the assumption $\alpha_2(T)=0$  in Tab.~
\ref{tab:2ex}.
We find that the systematic errors are almost comparable with the statistical errors,
  except very close to $T_{pc}$.

\begin{table}[tp]
 \begin{center}
 \caption{Results $\alpha_1 (T)$ and $\chi^2/N_{\rm DF}$ 
 for the fit of 1st order coefficients
  at $m_{\rm PS}/m_{\rm V}=0.65$ (left) and 0.80 (right).
  The 2nd parentheses in ${\bf 3}^\ast$ channel of $\alpha_1(T)$
  expresses the systematic errors due to difference of the fit range.}
 \label{tab:1st}
 {\renewcommand{\arraystretch}{1} \tabcolsep = 6.5mm %\footnotesize
 \newcolumntype{.}{D{.}{.}{6}}
 \begin{tabular}{c..|cc}
\hline \hline
 \multicolumn{5}{c}{$m_{\rm PS} / m_{\rm V} = 0.65$} \\
\hline
 \multicolumn{1}{c}{} & 
 \multicolumn{2}{c|}{$\alpha_1 (T) \times 10$} & 
 \multicolumn{2}{c} {$\chi^2 / N_{\rm DF}$} \\
\hline
 \multicolumn{1}{c} {$T/T_{pc}$} &
 \multicolumn{1}{c} {$R={\bf 6}$} & 
 \multicolumn{1}{c|}{${\bf 3}^*$} & 
 \multicolumn{1}{c} {${\bf 6}$} & 
 \multicolumn{1}{c} {${\bf 3}^*$} \\
\hline
 \multicolumn{1}{c}  { 1.07      } &
 \multicolumn{1}{.}  { -0.01(  2)} &
 \multicolumn{1}{.|} { 10.86(501)( 42)} &
 \multicolumn{1}{c}  { 0.36      } &
 \multicolumn{1}{c}  { 1.10      } \\
 1.18 &  1.64(116) &  3.64( 95)( 54) &  1.11 &  0.52  \\
 1.32 &  0.92( 66) &  3.58( 85)(  2) &  2.42 &  0.47  \\
 1.48 &  1.07( 36) &  2.14( 35)( 21) &  1.27 &  0.82  \\
 1.67 &  1.49( 41) &  0.90( 16)( 18) &  1.04 &  0.86  \\
 2.09 &  0.62( 13) &  0.54( 11)( 11) &  0.92 &  1.68  \\
 2.59 &  0.44(  9) &  0.56( 10)(  7) &  1.91 &  2.14  \\
 3.22 &  0.26(  5) &  0.36(  5)(  4) &  0.48 &  1.25  \\
 4.02 &  0.33(  5) &  0.29(  5)(  2) &  1.13 &  1.16  \\
\hline \hline
\\
\hline \hline
 \multicolumn{5}{c}{$m_{\rm PS} / m_{\rm V} = 0.80$} \\
\hline
 \multicolumn{1}{c}{} & 
 \multicolumn{2}{c|}{$\alpha_1 (T) \times 10$} & 
 \multicolumn{2}{c} {$\chi^2 / N_{\rm DF}$} \\
\hline
 \multicolumn{1}{c} {$T/T_{pc}$} &
 \multicolumn{1}{c} {$R={\bf 6}$} & 
 \multicolumn{1}{c|}{${\bf 3}^*$} & 
 \multicolumn{1}{c} {${\bf 6}$} & 
 \multicolumn{1}{c} {${\bf 3}^*$} \\
\hline
 \multicolumn{1}{c}  { 1.08      } &
 \multicolumn{1}{.}  { $-$       } &
 \multicolumn{1}{.|} { 3.03( 51)( 40) } &
 \multicolumn{1}{c}  { $-$       } &
 \multicolumn{1}{c}  { 0.71      } \\
1.20 &  1.35( 67) &  2.37( 44)( 68) &  0.98 &  1.27 \\
1.35 &  1.35( 36) &  1.28( 26)( 22) &  2.18 &  0.88 \\
1.69 &  0.65( 10) &  0.92( 14)( 13) &  1.05 &  1.21 \\
2.07 &  0.50(  7) &  0.36(  6)( 16) &  1.87 &  2.81 \\
2.51 &  0.45(  7) &  0.38(  4)(  1) &  0.70 &  0.38 \\
3.01 &  0.23(  3) &  0.34(  3)(  1) &  1.83 &  2.04 \\
\hline \hline
 \end{tabular}
 }
 \end{center}
\end{table}

\begin{table}[tp]
 \begin{center}
 \caption{$\chi^2 / N_{\rm DF}$ for the fit of 2nd order coefficients
  at $m_{\rm PS}/m_{\rm V}=0.65$ (left) and 0.80 (right).}
 \label{tab:2nd_chi}
 {\renewcommand{\arraystretch}{1} \tabcolsep = 3.0mm \footnotesize
 \begin{tabular}{ccccc|ccccc}
\hline \hline
 \multicolumn{5}{c|}{$m_{\rm PS} / m_{\rm V} = 0.65$} & 
 \multicolumn{5}{c} {$m_{\rm PS} / m_{\rm V} = 0.80$} \\
 \hline
 $T/T_{pc}$  & $R={\bf 1}$ & ${\bf 8}$ &
 ${\bf 6}$ & ${\bf 3}^*$ & 
 $T/T_{pc}$  & $R={\bf 1}$ & ${\bf 8}$ &
 ${\bf 6}$ & ${\bf 3}^*$  \\
\hline
 1.07 &  0.87 &  1.29 &  0.64 &  1.05 &  1.08 &  0.63 &  1.15 &  $-$  &  1.82 \\
 1.18 &  0.43 &  0.85 &  0.64 &  1.04 &  1.20 &  1.00 &  1.70 &  2.17 &  1.46 \\
 1.32 &  1.86 &  0.95 &  1.17 &  2.22 &  1.35 &  0.99 &  0.65 &  0.64 &  0.46 \\
 1.48 &  1.02 &  1.56 &  1.10 &  1.32 &  1.69 &  2.83 &  2.49 &  1.53 &  1.44 \\
 1.67 &  1.12 &  1.73 &  1.22 &  0.61 &  2.07 &  0.95 &  0.64 &  1.05 &  0.98 \\
 2.09 &  1.01 &  0.96 &  2.43 &  1.84 &  2.51 &  1.83 &  1.28 &  0.73 &  1.25 \\
 2.59 &  1.19 &  1.49 &  1.66 &  1.21 &  3.01 &  1.08 &  1.76 &  1.08 &  0.59 \\
 3.22 &  1.02 &  1.83 &  1.98 &  0.90 & & & & & \\
 4.02 &  1.61 &  0.72 &  1.10 &  1.86 & & & & & \\
\hline \hline
 \end{tabular}
 }
 \end{center}
\end{table}

\begin{table}[tp]
 \begin{center}
 \caption{Results of $\alpha_2 (T) \times 10$ 
  at $m_{\rm PS}/m_{\rm V}=0.65$ (left) and 0.80 (right).}
 \label{tab:2nd_AL}
 {\renewcommand{\arraystretch}{1} \tabcolsep = 2.0mm \footnotesize
 \newcolumntype{.}{D{.}{.}{6}}
 \begin{tabular}{c....|c....}
 \hline \hline
 \multicolumn{5}{c|}{$m_{\rm PS} / m_{\rm V} = 0.65$} & 
 \multicolumn{5}{c} {$m_{\rm PS} / m_{\rm V} = 0.80$} \\
 \hline
 \multicolumn{1}{c}{$T/T_{pc}$} &
 \multicolumn{1}{c} {$R={\bf 1}$} & 
 \multicolumn{1}{c} {${\bf 8}$} & 
 \multicolumn{1}{c} {${\bf 6}$} & 
 \multicolumn{1}{c|}{${\bf 3}^*$} & 
 \multicolumn{1}{c}{$T/T_{pc}$} &
 \multicolumn{1}{c} {$R={\bf 1}$} & 
 \multicolumn{1}{c} {${\bf 8}$} & 
 \multicolumn{1}{c} {${\bf 6}$} & 
 \multicolumn{1}{c}{${\bf 3}^*$} \\
 \hline
 \multicolumn{1}{c}  { 1.07      } &
 \multicolumn{1}{.}  { 0.05( 67) } &
 \multicolumn{1}{.}  { 0.40(365) } &
 \multicolumn{1}{.}  { -0.03(  5)} &
 \multicolumn{1}{.|} { -3.83(346)} &
 \multicolumn{1}{c}  { 1.08      } &
 \multicolumn{1}{.}  { -0.09( 44)} &
 \multicolumn{1}{.}  { -0.08( 15)} &
 \multicolumn{1}{c}  { $-$       } &
 \multicolumn{1}{.}  { -0.06( 60)} \\
 1.18 & -1.31( 65) & -0.30( 28) & -3.39(318) & -1.29( 88) &  1.20 & -0.30( 34) & -8.42(1002)& -1.93(196) &  1.03( 42) \\
 1.32 &  0.55( 71) &  0.31( 91) & -0.80(161) & -1.20( 94) &  1.35 &  0.21( 22) &  0.01( 71) & -0.81( 58) &  0.18( 25) \\
 1.48 &  0.35( 34) & -0.70(216) & -1.66(151) & -0.41( 59) &  1.69 & -0.08( 13) &  0.38( 35) & -0.04( 28) &  0.06( 18) \\
 1.67 &  0.03( 24) & -1.91(127) & -1.15( 97) &  0.27( 28) &  2.07 &  0.02(  7) & -0.55( 31) &  0.01( 22) & -0.16(  9) \\
 2.09 & -0.16( 13) & -0.05( 56) & -0.13( 34) & -0.24( 15) &  2.51 &  0.03(  6) & -0.24( 20) & -0.08( 15) &  0.01(  7) \\
 2.59 &  0.07( 11) & -0.03( 33) & -0.42( 22) & -0.21( 15) &  3.01 &  0.01(  3) &  0.00( 10) & -0.08(  7) & -0.07(  5) \\
 3.22 & -0.04(  8) & -0.20( 24) & -0.18( 15) &  0.22(  9) & & & & & \\
 4.02 & -0.04(  7) & -0.21( 22) & -0.28( 13) & -0.01(  7) & & & & & \\
\hline \hline
 \end{tabular}
 }
 \end{center}
\end{table}

\begin{table}[tp]
 \begin{center}
 \caption{Results of $m_{D,2}(T)$ 
  at $m_{\rm PS}/m_{\rm V}=0.65$ (left) and 0.80 (right).
  The 2nd parentheses in the color-singlet channel
  expresses the systematic errors due to difference of the fit range.}
 \label{tab:2nd_SM}
 {\renewcommand{\arraystretch}{1} \tabcolsep = 2.5mm %\footnotesize
 \newcolumntype{.}{D{.}{.}{6}}
 \begin{tabular}{c....|c....}
 \hline \hline
 \multicolumn{5}{c|}{$m_{\rm PS} / m_{\rm V} = 0.65$} & 
 \multicolumn{5}{c} {$m_{\rm PS} / m_{\rm V} = 0.80$} \\
 \hline
 \multicolumn{1}{c} {$T/T_{pc}$} &
 \multicolumn{1}{c} {$R={\bf 1}$} & 
 \multicolumn{1}{c} {${\bf 8}$} & 
 \multicolumn{1}{c} {${\bf 6}$} & 
 \multicolumn{1}{c|}{${\bf 3}^*$} & 
 \multicolumn{1}{c} {$T/T_{pc}$} &
 \multicolumn{1}{c} {$R={\bf 1}$} & 
 \multicolumn{1}{c} {${\bf 8}$} & 
 \multicolumn{1}{c} {${\bf 6}$} & 
 \multicolumn{1}{c} {${\bf 3}^*$} \\
 \hline
% 1.07 &  1.47( 26) &  0.01(348) &  2.17(229) &  1.03( 42) &  1.08 &  1.17( 15) &  1.03(221) & *.**(***) &  1.22( 20) \\
 \multicolumn{1}{c}  { 1.07      } &
 \multicolumn{1}{.}  { 1.47( 26)( 52) } &
 \multicolumn{1}{.}  { 0.01(348) } &
 \multicolumn{1}{.}  { 2.17(229) } &
 \multicolumn{1}{.|} { 1.03( 42) } &
 \multicolumn{1}{c}  { 1.08      } &
 \multicolumn{1}{.}  { 1.17( 15)( 10) } &
 \multicolumn{1}{.}  { 1.03(221) } &
 \multicolumn{1}{c}  { $-$       } &
 \multicolumn{1}{.}  { 1.22( 20) } \\
 1.18 &  0.55( 19)( 26) & -3.18(180) & -0.27( 71) &  0.58( 23) &  1.20 &  0.76( 13)( 12) & -0.41( 69) &  0.63( 31) &  0.99( 13) \\
 1.32 &  0.84( 18)( 28) & -0.19(100) & -0.20( 57) &  0.49( 19) &  1.35 &  0.69(  8)(  1) &  0.40( 37) &  0.26( 17) &  0.68(  8) \\
 1.48 &  0.73( 13)(  2) &  0.27( 72) &  0.06( 48) &  0.55( 19) &  1.69 &  0.40(  7)( 31) &  0.73( 24) &  0.50( 17) &  0.61(  9) \\
 1.67 &  0.40( 18)( 35) & -0.31( 55) &  0.10( 27) &  0.53( 19) &  2.07 &  0.45(  7)( 12) &  0.30( 13) &  0.43( 11) &  0.37(  7) \\
 2.09 &  0.38(  8)( 22) &  0.25( 26) &  0.14( 14) &  0.33( 10) &  2.51 &  0.38(  6)(  5) &  0.27( 11) &  0.40(  9) &  0.39(  7) \\
 2.59 &  0.41(  7)( 27) &  0.47( 18) &  0.13(  9) &  0.24(  9) &  3.01 &  0.37(  4)(  7) &  0.31(  8) &  0.29(  6) &  0.35(  4) \\
 3.22 &  0.44(  8)(  3) &  0.30( 16) &  0.30( 12) &  0.65( 10) & & & & & \\
 4.02 &  0.41( 10)( 28) &  0.41( 19) &  0.20( 12) &  0.43(  8) & & & & & \\
\hline \hline
 \end{tabular}
 }
 \end{center}
\end{table}

\begin{table}[tp]
 \begin{center}
 \caption{Results of $m_{D,2}(T)$
 determined with the assumption $\alpha_2(T) =0$
  at $m_{\rm PS}/m_{\rm V}=0.65$ (left) and 0.80 (right).
  The 2nd parentheses in the color-singlet channel
  expresses the systematic errors due to difference of the fit range.}
 \label{tab:2ex}
 {\renewcommand{\arraystretch}{1} \tabcolsep = 2.5mm %\footnotesize
 \newcolumntype{.}{D{.}{.}{6}}
 \begin{tabular}{c....|c....}
 \hline \hline
 \multicolumn{5}{c|}{$m_{\rm PS} / m_{\rm V} = 0.65$} & 
 \multicolumn{5}{c} {$m_{\rm PS} / m_{\rm V} = 0.80$} \\
 \hline
 \multicolumn{1}{c} {$T/T_{pc}$} &
 \multicolumn{1}{c} {$R={\bf 1}$} & 
 \multicolumn{1}{c} {${\bf 8}$} & 
 \multicolumn{1}{c} {${\bf 6}$} & 
 \multicolumn{1}{c|}{${\bf 3}^*$} & 
 \multicolumn{1}{c} {$T/T_{pc}$} &
 \multicolumn{1}{c} {$R={\bf 1}$} & 
 \multicolumn{1}{c} {${\bf 8}$} & 
 \multicolumn{1}{c} {${\bf 6}$} & 
 \multicolumn{1}{c} {${\bf 3}^*$} \\
 \hline
% 1.07 &  1.33( 19) &  0.81(153) &  3.09(166) &  1.43( 32) &  1.08 &  1.15( 11) & -0.28(160) &  *.**      &  1.17( 16) \\
 \multicolumn{1}{c}  { 1.07      } &
 \multicolumn{1}{.}  { 1.33( 19)( 10) } &
 \multicolumn{1}{.}  { 0.81(153) } &
 \multicolumn{1}{.}  { 3.09(166) } &
 \multicolumn{1}{.|} { 1.43( 32) } &
 \multicolumn{1}{c}  { 1.08      } &
 \multicolumn{1}{.}  { 1.15( 11)(  6) } &
 \multicolumn{1}{.}  { -0.28(160)} &
 \multicolumn{1}{c}  { $-$       } &
 \multicolumn{1}{.}  { 1.17( 16) } \\
 1.18 &  0.73( 15)(  1) & -1.84(123) & -0.21( 54) &  0.77( 18) &  1.20 &  0.77( 10)( 16) &  0.19( 41) &  0.73( 20) &  0.78( 11) \\
 1.32 &  0.70( 14)( 21) & -0.34( 58) & -0.12( 36) &  0.52( 15) &  1.35 &  0.65(  6)(  9) &  0.40( 21) &  0.45( 11) &  0.67(  6) \\
 1.48 &  0.69(  9)(  1) &  0.24( 40) &  0.41( 28) &  0.64( 14) &  1.69 &  0.39(  7)(  7) &  0.28( 18) &  0.49( 11) &  0.52(  8) \\
 1.67 &  0.41( 13)( 14) &  0.20( 27) &  0.29( 19) &  0.43( 13) &  2.07 &  0.44(  5)(  3) &  0.44( 10) &  0.42(  7) &  0.40(  6) \\
 2.09 &  0.31(  7)(  0) &  0.27( 16) &  0.29(  8) &  0.37(  7) &  2.51 &  0.35(  4)(  2) &  0.34(  7) &  0.43(  5) &  0.40(  5) \\
 2.59 &  0.39(  5)(  6) &  0.54( 10) &  0.34(  7) &  0.35(  7) &  3.01 &  0.34(  4)(  2) &  0.25(  6) &  0.30(  5) &  0.37(  3) \\
 3.22 &  0.46(  6)(  8) &  0.33( 10) &  0.38(  8) &  0.56(  8) & & & & & \\
 4.02 &  0.42(  8)(  1) &  0.50( 12) &  0.33(  9) &  0.37(  7) & & & & & \\
\hline \hline
 \end{tabular}
 }
 \end{center}
\end{table}

%%%%%%%%%%%%%%%%%%%%%%%%%%%%%%%%%%%%%%%%%%%%%%%%%%%%%%%%%%%%%%%

\end{document}